\documentclass[%
reprint,
twocolumn,
superscriptaddress,
preprintnumbers,
bibnotes,
amsmath,amssymb,
pra,
lengthcheck
]{revtex4-1}
\usepackage[english]{babel}
\usepackage{graphicx}% Include figure files
\usepackage{dcolumn}% Align table columns on decimal point
\usepackage{bm}% bold math
\usepackage{dsfont}% bold math
\usepackage{newtxtext,newtxmath}
\usepackage[colorlinks,urlcolor=blue,linkcolor=blue,citecolor=blue,anchorcolor=blue]{hyperref}
\usepackage{color}
\usepackage{xcolor}
\usepackage{float}

\begin{document}
\preprint{APS/123-QED}

\title{Non-Hermitian topology driven by an identity term: An exactly solvable paradigm}

\author{Lingfang Li}
\affiliation{Key Laboratory of Quantum Materials and Devices of Ministry of Education, School of Physics, Frontiers Science Center for Mobile Information Communication and Security, Southeast University, Nanjing 211189, China}

\author{Yating Wei}
\affiliation{Key Laboratory of Quantum Materials and Devices of Ministry of Education, School of Physics, Frontiers Science Center for Mobile Information Communication and Security, Southeast University, Nanjing 211189, China}

\author{Yang Ruan}
\affiliation{Key Laboratory of Quantum Materials and Devices of Ministry of Education, School of Physics, Frontiers Science Center for Mobile Information Communication and Security, Southeast University, Nanjing 211189, China}

\author{Gangzhou Wu}
\affiliation{Key Laboratory of Quantum Materials and Devices of Ministry of Education, School of Physics, Frontiers Science Center for Mobile Information Communication and Security, Southeast University, Nanjing 211189, China}

\author{Jun Wang}
\affiliation{Key Laboratory of Quantum Materials and Devices of Ministry of Education, School of Physics, Frontiers Science Center for Mobile Information Communication and Security, Southeast University, Nanjing 211189, China}

\author{Shihua Chen}
\email{cshua@seu.edu.cn}
\affiliation{Key Laboratory of Quantum Materials and Devices of Ministry of Education, School of Physics, Frontiers Science Center for Mobile Information Communication and Security, Southeast University, Nanjing 211189, China}
\affiliation{Purple Mountain Laboratories, Nanjing 211111, China}

\author{Tong Lin}
\affiliation{Advanced Photonics Center, School of Electronic Science and Engineering, Southeast University, Nanjing 210096, China}

\author{Ching Hua Lee}
\email{phylch@nus.edu.sg}
\affiliation{Department of Physics, National University of Singapore, Singapore 117551, Republic of Singapore}

\author{Zhenhua Ni}
\email{zhni@seu.edu.cn}
\affiliation{Key Laboratory of Quantum Materials and Devices of Ministry of Education, School of Physics, Frontiers Science Center for Mobile Information Communication and Security, Southeast University, Nanjing 211189, China}
\affiliation{Purple Mountain Laboratories, Nanjing 211111, China}
\affiliation{Advanced Photonics Center, School of Electronic Science and Engineering, Southeast University, Nanjing 210096, China}

\date{\today}
\begin{abstract}
An identity term in the Hamiltonian is conventionally regarded as spectrally inert---it shifts energies but does not alter eigenstate topology. We show that under non-Hermitian skin pumping, this paradigm fails: a momentum-dependent identity term actively deforms the generalized Brillouin zone, thereby challenging established topological criteria that rely on fixed complex contours. Here, by introducing spin-orbit coupling into a Hatano-Nelson chain, we present an exact analytical solution for the entire non-Hermitian eigensystem under open boundary conditions. Our solution reveals how inter-cell spin-orbit coupling, synergizing with this non-trivial identity term, induces topological edge states and robust zero modes in the complete absence of chiral symmetry. This work establishes an exactly solvable paradigm for non-Hermitian topology beyond symmetry protection, and provides a rigorous benchmark for testing topological invariants in systems with momentum-dependent identity terms.
\end{abstract}

\maketitle

%\tableofcontents

Non-Hermitian physics \cite{Ash20,Kawa19,Okuma23,El18,Cou21,Ber21,Ding22,Zhang23} has unveiled a host of intriguing phenomena beyond the Hermitian framework, with several cornerstone discoveries such as the non-Hermitian skin effect (NHSE) \cite{Yao18,LeeC19,Borgnia20,Zhao2019,Tai23,Longhi22,Okuma2020,Zha20,Sun21,Wei20,Hou24,LiC23,
LiL22,Lei24,Gli24,Yosh24,Shen25,LiQ25,Hu25,Yang25}, non-Bloch topological phases \cite{Gong18,Shen18,Longhi19,Yok19,Guo21,Hou23,Lee22,Ban2023,DaiT24,Wang24}, and exceptional point physics \cite{Zhen2015,Ding16,Peng16,Chen17,Ozd19,Ali2019,Kaw19,Zhao25,Kul25,Meng24,Arouca20,Xue26,Liu25}. The NHSE, where an extensive number of bulk eigenstates accumulate at boundaries, defies the conventional bulk-boundary correspondence \cite{Kun18,Zirn21,Zhu2020,Yang20,Hel20,Hou22,Xiao20,Naka24} and necessitates the use of the generalized Brillouin zone (GBZ) for a correct description of topology in one-dimensional (1D) non-Hermitian systems \cite{Yok19,Yang20,Verma24}. This non-Bloch band theory has successfully classified topological phases and predicted novel boundary phenomena in diverse non-Hermitian platforms such as cold atoms \cite{Li2019,Liang22,Zhou22,Shen23}, electric circuits \cite{Lee18,Ste21,Sahin25,Zhang24,YangH24,Zou24}, mechanical metamaterials \cite{Bender13,Gha20,Xue22,Zhang26}, photonic lattices or arrays \cite{Ozawa2019,Peng14,Feng2017,Par18,Zhou18,Slootman24}, and quantum systems \cite{Rud09,Xiao17,FSong19,ShenR25,Shen24}. However, a significant portion of this progress has been anchored in models possessing chiral or particle-hole symmetry \cite{Gong18,Kawa19}, such as the celebrated Su-Schrieffer-Heeger (SSH) model \cite{Hel20,Par18,Zhan21}, where the energy spectra are symmetric about zero, hence greatly simplifying the analysis of topological edge modes \cite{Kawa19,Yok19,Guo21}.

Yet most realistic systems lack such protecting symmetries, falling into the broad class of non-chiral models \cite{Kawa19,YangK24,Jia25,Jiang23}. Existing analytical progress in this direction has primarily addressed systems where the Hamiltonian can be expressed in or built upon the standard two-band form $\mathcal{H}(\beta)=\boldsymbol{\mathrm{d}}(\beta)\cdot \boldsymbol{\sigma}$ \cite{Jiang2018,Li24,Yang24,Zhang25,ZhangY25}. In such cases, the non-chirality arises solely from the non-planarity of $\boldsymbol{\mathrm{d}}(\beta)$, but non-Bloch topological invariants have still been established \cite{Li24,ZhangY25}.

A more formidable challenge arises, however, when the Hamiltonian contains a non-vanishing, momentum-dependent identity term, i.e., $\mathcal{H}(\beta)=d_0(\beta)\mathds{1}+\boldsymbol{\mathrm{d}}(\beta)\cdot \boldsymbol{\sigma}$ with $d_0(\beta)\neq0$ \cite{Jiao21,Wu2023,Zhong25}. This identity term actively deforms the GBZ, rendering inapplicable topological criteria that rely on a fixed GBZ contour, including those based on winding-number differences \cite{Yok19,Wei26} or geometric constructions \cite{LeeC19,Li19,Hu24}, even for non-chiral systems \cite{Li24,ZhangY25}. While a distinct topological invariant defined on the Riemann surface for such generic models has recently been proposed \cite{Zhong25}, an exact analytical solution yielding explicit edge-state energies and phase boundaries for any specific model has remained elusive. Consequently, probing the topological physics in this regime still largely relies on case-by-case numerical GBZ computations \cite{Wu2023}, an approach that is demanding and can become intractable for complex GBZ morphologies \cite{Meng25,Li2020,Qin23,Kazuki21,Qin26}, leaving a gap between the topological classification and quantitative prediction.

In this Letter, we resolve this challenge by presenting an exactly solvable paradigm for non-Hermitian topology driven by a non-trivial identity term. Concretely, we introduce spin-orbit (SO) coupling \cite{Mura06,Sou16,Sch16,Man15} into the non-Hermitian Hatano-Nelson (HN) chain---a prototypical non-chiral system that exhibits the NHSE \cite{Hatano96,Liu2022,LiZ2024,Longhi17,Zhang22,Chen25,Ors25} and captures the essential topological physics of the most general symmetry-free two-band model solved in Supplemental Material \cite{SuppMat}. We demonstrate that the inter-cell SO coupling $\delta_2$, in synergy with this identity term, gives rise to a rich landscape of topological phenomena, including topological islands carved out of a broad trivial phase, robust zero modes that require neither chiral symmetry nor $(E,-E)$ spectral pairing, as well as a topological phase transition driven purely by the GBZ deformation (see End Matter), in an otherwise topologically trivial HN chain. Crucially, we achieve a complete analytical solution for the entire eigensystem under open boundary conditions (OBC), culminating in closed-form expressions for the edge-state energy and its existence condition. Our paradigmatic solution thus transcends symmetry-based classifications \cite{Gong18,Kawa19,Li25}, establishing inter-cell SO coupling as a generic mechanism for non-chiral topology and providing an analytical benchmark for testing topological invariants \cite{Zhong25} in identity-term-driven systems with complex GBZ morphologies.

\begin{figure}[ht]
\begin{center}
	\includegraphics[width=8.7cm]{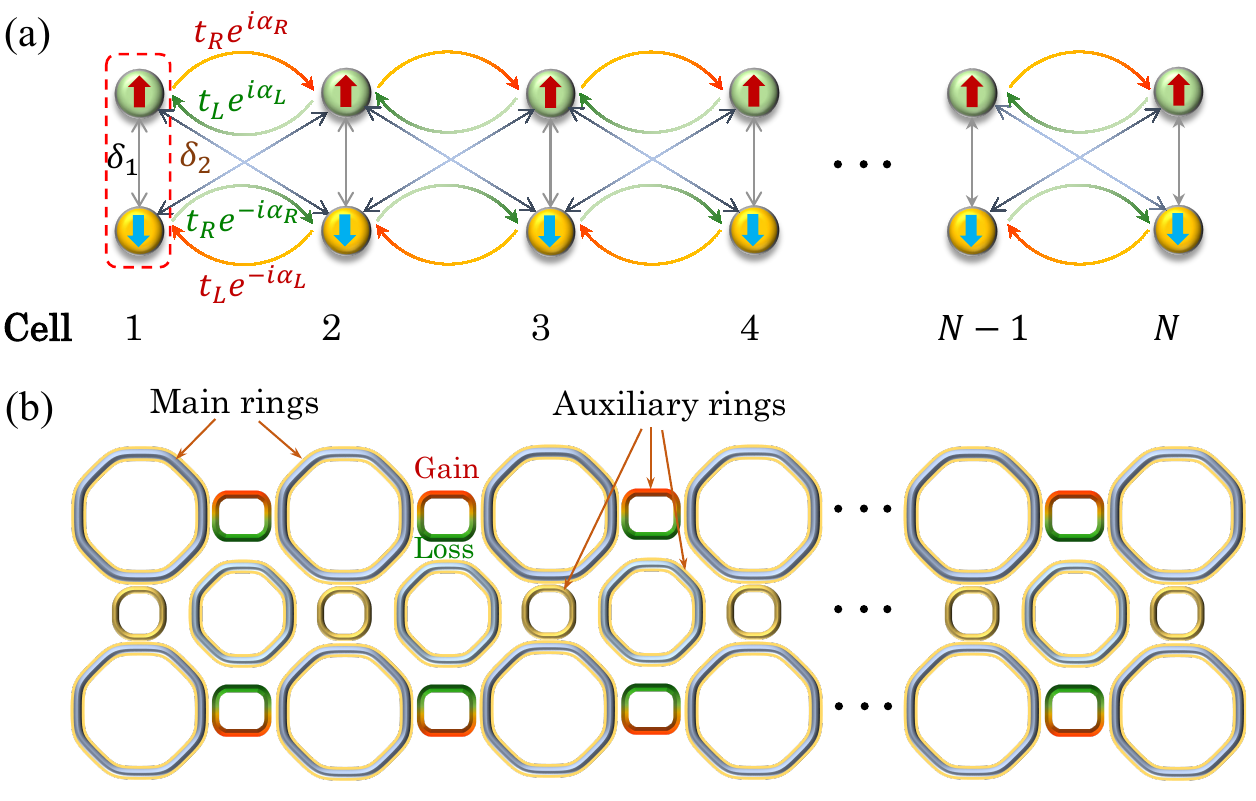}\\
    \includegraphics[width=8.7cm]{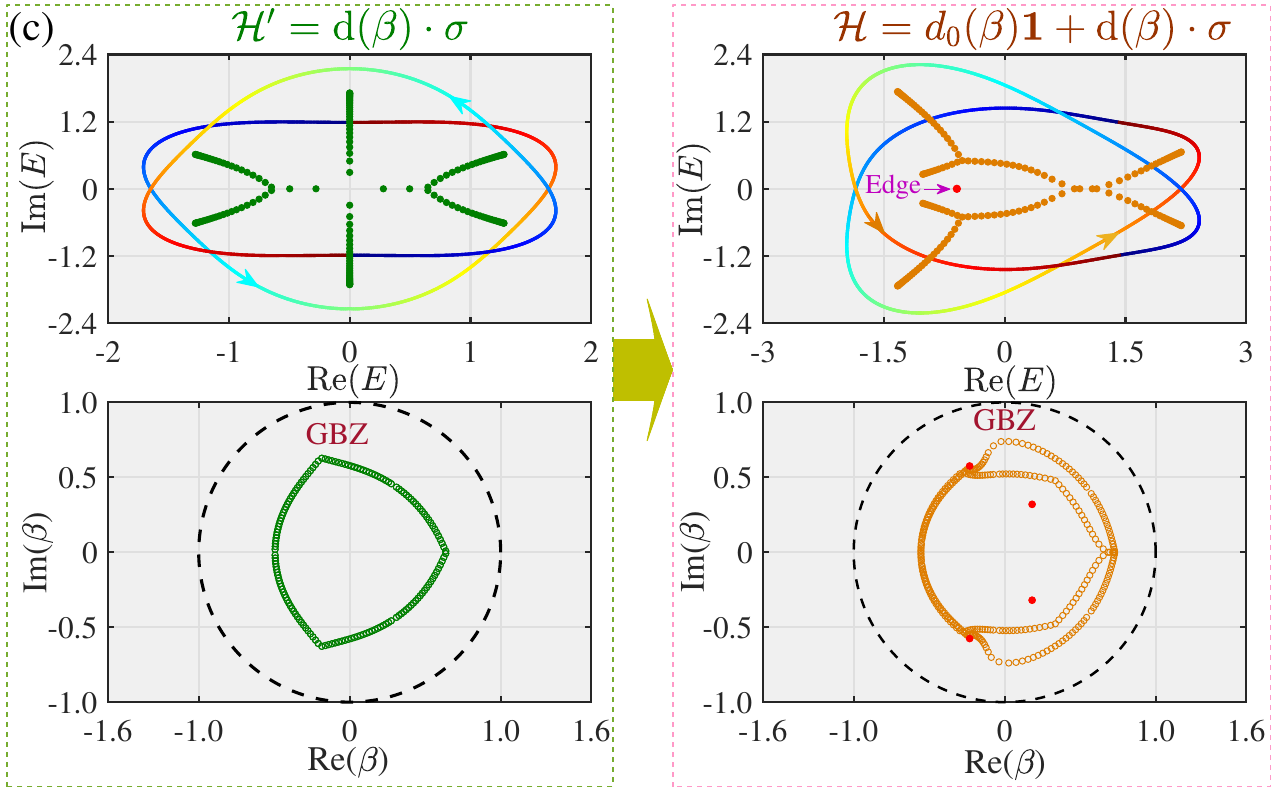}%%
	\caption{Topological phase transition induced by an identity term. (a) Schematic of a 1D non-Hermitian HN chain with SO coupling. (b) Optical implementation of such a chain using a microring resonator array. (c) Demonstration of how the identity term $d_0(\beta)\mathds{1}$ actively deforms the GBZ and induces topological edge states marked by red dots, for $t_L =2$, $t_R =0.6$, $\alpha_L=\pi/3$, $\alpha_R= \pi/4$, $\delta_1=1$, $\delta_2=0.4$, and $N =80$. Supplemental Fig. 1 gives further examples and Fig. \ref{fig5} in End Matter shows the full evolution of spectrum and GBZ with increasing $d_0(\beta)$ strength.} \label{fig1}
\end{center}
\end{figure}

Our 1D non-Hermitian HN model with SO coupling, depicted in Fig.~\ref{fig1}{(a)}, generalizes those in Refs. \cite{Long15,Li2020,Qin23,San25}. Under  OBC, its Hamiltonian reads
\begin{eqnarray}
\hat{H}&=&\sum_{n=1}^{N}\left\{t_{L}e^{i\alpha_L\sigma_z}\hat{c}_{n}^{\dagger}\hat{c}_{n+1}
+t_{R}e^{i\alpha_R\sigma_z}\hat{c}_{n+1}^{\dagger}\hat{c}_{n}\right.\nonumber\\
&+&\left.\delta_1\hat{c}_{n}^{\dagger}(\boldsymbol{b}\cdot \boldsymbol{\sigma})\hat{c}_{n}+\left[\delta_2\hat{c}_{n}^{\dagger}(\boldsymbol{b}\cdot \boldsymbol{\sigma})\hat{c}_{n+1}+h.c.\right]\right\},\label{Eq1}
\end{eqnarray}
where $\hat{c}_{n}^{\dagger}=(\hat{c}_{n,\uparrow}^{\dagger},\hat{c}_{n,\downarrow}^{\dagger})$ creates an electron with spin up ($\uparrow$) or  spin down ($\downarrow$) on the $n$-th unit cell (adopting the convention $\hat{c}_{N+1}\equiv0$ and $h.c.$ denotes Hermitian conjugate). Here, $\boldsymbol{\sigma}=(\sigma_x,\sigma_y,\sigma_z)$ is the vector of Pauli matrices, $\boldsymbol{b}=(1,0,0)$ is a unit vector along the $x$ direction, setting the magnetic field axis, and $\delta_1$ and $\delta_2$ denote the intra-cell and inter-cell SO coupling strengths, respectively. The two SU(2) phase factors $e^{i\alpha_{R,L}\sigma_z}$ are embedded in the hopping amplitudes, with $t_{R,L}\in\mathbb{R}$ and $\alpha_{R,L}\in \mathbb{R}$.

The corresponding non-Bloch Hamiltonian $\mathcal{H}(\beta\equiv e^{ik})$ is obtained, via Fourier transform, as
\begin{equation}
\mathcal{H}(\beta)=d_0(\beta)\mathds{1}+\boldsymbol{\mathrm{d}}(\beta)\cdot \boldsymbol{\sigma}=\left[\begin{array}{cc}G_{\uparrow}(\beta) &\Delta(\beta)\\\Delta(\beta)&G_{\downarrow}(\beta)
	\end{array}\right],  \label{Eq2}
\end{equation}
where $\mathds{1}$ is the identity matrix, $d_0(\beta)=[G_{\uparrow}(\beta)+G_{\downarrow}(\beta)]/2=t_L\cos(\alpha_L)\beta+t_R\cos(\alpha_R)/\beta$, $\boldsymbol{\mathrm{d}}(\beta)=(\Delta(\beta), ~0,~[G_{\uparrow}(\beta)-G_{\downarrow}(\beta)]/2)$, $\Delta(\beta)=\delta_{1}+\delta_{2}(\beta+1/\beta)$, $G_{\uparrow}(\beta) =t_{L}e^{i\alpha_L}\beta+t_{R}e^{i\alpha_R}/\beta$, and $G_{\downarrow}(\beta) = t_{L}e^{-i\alpha_L}\beta+t_{R}e^{-i\alpha_R}/\beta$. The $\beta$-dependent identity term $d_0(\beta)\mathds{1}$ represents a diagonal shift that actively modifies the GBZ and hence the topology. As illustrated in Fig.~\ref{fig1}(c), for a fixed set of parameters where the chain is topologically trivial without this term, introducing $d_0(\beta)\mathds{1}$ can deform the GBZ from a single contour into separate or twisted contours and concomitantly induce topological edge states \cite{Meng25} (see Supplemental Fig.~1 for further examples \cite{SuppMat}). We note that the familiar non-Hermitian SSH model with next-nearest-neighbor hoppings \cite{Yok19,Guo21,Hou23} corresponds to the special case $\alpha_L=\alpha_R=\pi/2$, where $d_0(\beta)$ vanishes.

In the Bloch representation ($\beta = e^{ik}$), the Hamiltonian (\ref{Eq2}) possesses time-reversal symmetry, $\sigma_x \mathcal{H}^*(k) \sigma_x = \mathcal{H}(-k)$,  ensuring a spectrum symmetric about the real axis, yet generally respects neither Hermiticity nor chiral symmetry. At $\delta_1=0$, it additionally satisfies a $\pi$-translation symmetry, $\mathcal{H}(k+\pi) = -\mathcal{H}(k)$, which enforces a spectral symmetry about the origin [eigenvalue pairs ($E, -E$)]. The combination of an actively modified GBZ (driven by the identity term $d_0(\beta)\mathds{1}$) and the absence of chiral symmetry places our model beyond the scope of previous generic topological criteria, such as those relying on a fixed GBZ or winding-number differences on a predetermined contour.  It is precisely this scenario that makes the exact analytical solution developed here both essential and enabling.

The above SO-coupled HN chain can be realized across a variety of non-Hermitian setups such as photonic resonator arrays \cite{Chai20,Mittal21,Xin23,Wong25}, topolectrical circuits \cite{Ste21,Sahin25,Zhang24,YangH24}, and acoustic metamaterials \cite{Gha20,Xue22,Zhang26}. As an example, Fig. \ref{fig1}{(b)} shows an optical implementation based on a silicon-nitride microring resonator array, where the two spin states are mapped to the clockwise and counter-clockwise modes of each ring. The required non-reciprocal hoppings and spin-flip couplings are implemented via asymmetric inter-ring couplings and auxiliary linking rings, respectively \cite{Ors25}. Non-Hermiticity is introduced through balanced gain and loss in selected rings. This photonic setup offers in-situ tunability of all essential parameters such as hopping amplitudes, phases, and SO strengths, and hence is appropriate for observing the topological edge states predicted here.

\begin{figure*}
	[ht!]
	\begin{center}
		\includegraphics[width=16cm]%
		{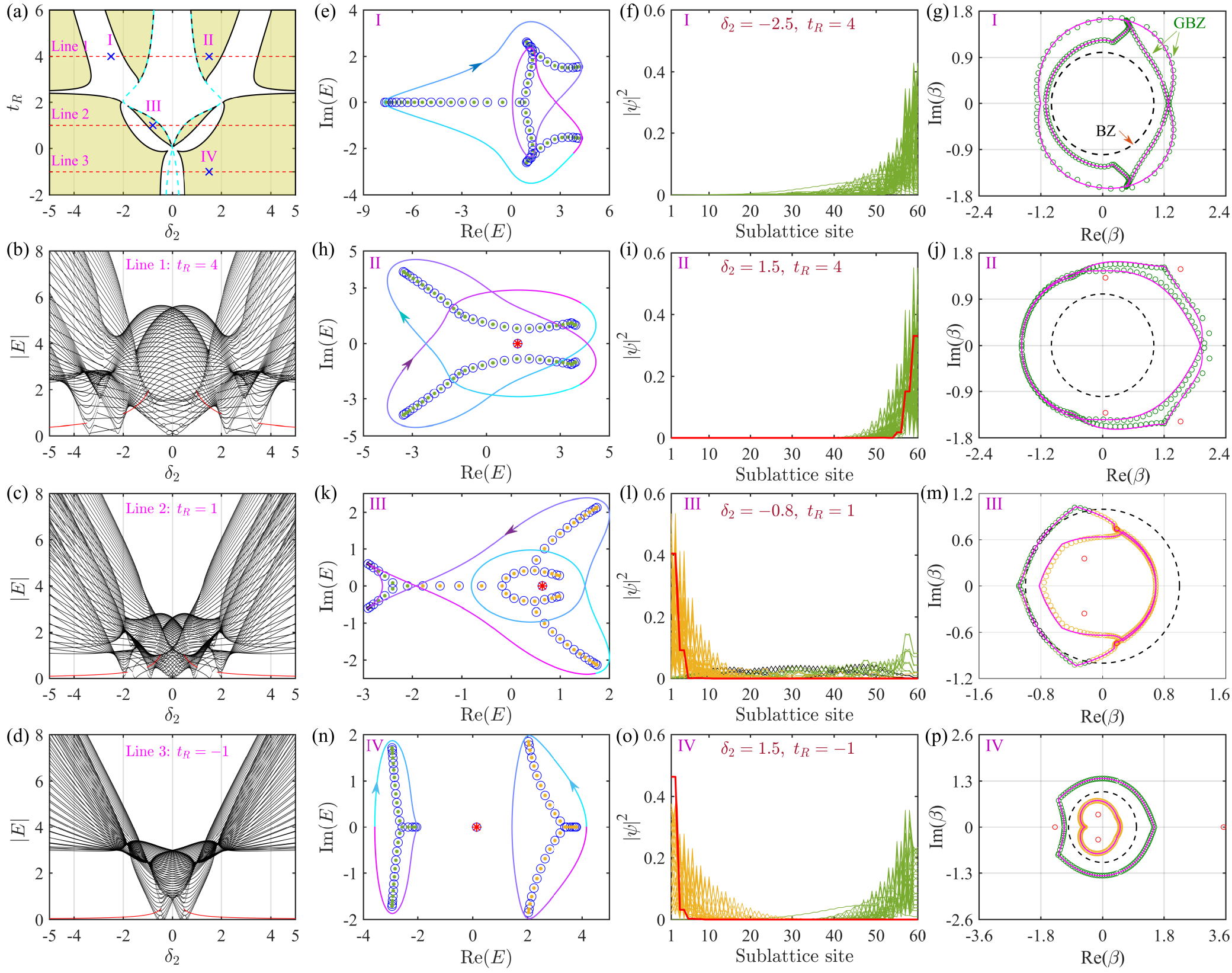}%%
		\caption{Analytically accessible phase diagram and topological edge states featuring distinct spectral morphologies. (a) Non-trivial phase diagram obtained directly from Eq. (\ref{Eq10}) in the $(\delta_2, t_R)$ plane for $t_L = 2$, $\alpha_L = \pi/3$, $\alpha_R = \pi/4$, $\delta_1 = 1$, with the topological regimes (yellow) emerging only when $\delta_2 \neq 0$. (b)-(d) Energy bands along the cut lines 1-3 in (a), where topological edge states are highlighted in red. Panels (e)-(p) validate the phase diagram at points I--IV. Point I (e-g) lies outside the topological regime and shows no edge states, whereas points II (h-j), III (k-m), and IV (n-p) are inside the regime and all host topological edge states. (e,h,k,n) Complex energy spectrum ($N=30$), where solid (open) circles mark analytical (numerical) OBC eigenvalues and the red asterisk marks the edge-state energy, with the spectrum under periodic boundary conditions (PBC) indicated by the colored loop. (f,i,l,o) Spatial profiles of all eigenstates, where green (yellow) curves denote bulk states localized near the right (left) boundary and red curves mark topological edge states. (g,j,m,p) Actively modified GBZ, with open circles for numerical GBZ ($N=60$), magenta curves for exact GBZ, and red circles marking the four $\beta$-roots of the topological edge state.} \label{fig2}
	\end{center}
\end{figure*}

We now present the exact solution for the open-boundary eigensystem $\hat{H}|\psi\rangle=E|\psi\rangle$. The starting point is the characteristic equation $\mathrm{det}[\mathcal{H}(\beta)-E]=0$, which reduces to
\begin{equation}
\left[G_{\uparrow}(\beta)-E\right]\left[G_{\downarrow}(\beta)-E\right]=[\delta_{1}+\delta_{2}(\beta+1/\beta)]^2. \label{Eq3}
\end{equation}
For a candidate energy $E$, Eq.~(\ref{Eq3}) yields four roots $\beta_{i}$ ($i=1,\dots,4$) in the complex plane. The corresponding wavefunctions are constructed from linear combinations of these $\beta_{i}$ values, with the coefficients fixed by the OBC (see Sec. I in Supplemental Material \cite{SuppMat} for details).

The eigenenergies $E$ under OBC are uniquely determined by a polynomial equation of $2N$ degree:
\begin{equation}
p_1\lambda_{N+1}^{[1]}+p_2\lambda_{N+1}^{[2]}=0, \label{Eq4}
\end{equation}
where $p_1=\sqrt{(t_R^2-\delta_{2}^2)(t_L^2-\delta_{2}^2)}$ and $p_2=\delta_{2}^2-t_Lt_R\cos(\alpha_L-\alpha_R)$.  The polynomials $\lambda_{N+1}^{[i]}$ ($i=1,2$) are generated by the coupled recurrence relations
\begin{eqnarray}
\lambda_{n+1}^{[i]}&=&2A\lambda_n^{[i]}+2\mu_n^{[i]}-\lambda_{n-1}^{[i]},\nonumber\\
\mu_{n+1}^{[i]}&=&-2B\lambda_n^{[i]}+2\gamma_n^{[i]}-\mu_{n-1}^{[i]},\\  \label{Eq5}
\gamma_{n+1}^{[i]}&=&2C\lambda_n^{[i]}-\gamma_{n-1}^{[i]}, \nonumber
\end{eqnarray}
with initial conditions $\lambda_{0}^{[1]}=0$, $\mu_{0}^{[1]}=1$, $\gamma_{0}^{[1]}=0$, $\lambda_{1}^{[1]}=1$, $\mu_{1}^{[1]}=0$, $\gamma_{1}^{[1]}=0$, and $\lambda_{0}^{[2]}=0$, $\mu_{0}^{[2]}=0$, $\gamma_{0}^{[2]}=1$, $\lambda_{1}^{[2]}=0$, $\mu_{1}^{[2]}=1$, and $\gamma_{1}^{[2]}=0$. Here the coefficients are defined by
\begin{equation}
A=\frac{E^2-\delta_1^2-2p_2}{2p_1},  \label{Eq6}
\end{equation}
\begin{equation}
B=\frac{(\delta_{1}\delta_{2}+Et_L\cos\alpha_L)(\delta_{1}\delta_{2}+Et_R\cos\alpha_R)}{p_1^2}-1, \label{Eq7}
\end{equation}
\begin{equation}
C=\frac{(\kappa_2-1)E^2}{2p_1}+\frac{\kappa_1\delta_{1}\delta_{2}E}{p_1}+\frac{2p_2+\delta_1^2+\kappa_0\delta_{1}^2\delta_{2}^2}{2p_1}, \label{Eq8}
\end{equation}
with $\kappa_m=\frac{t_L^m\cos^m\alpha_L}{t_L^2-\delta^2_{2}}+ \frac{t_R^m\cos^m\alpha_R}{t_R^2-\delta^2_{2}}$ ($m=0,1,2$).
We point out that Eq.~(\ref{Eq4}) provides an exact algebraic scheme for obtaining the complete OBC spectrum without numerical diagonalization. Each solution $E$ of Eq.~(\ref{Eq4}) in turn determines $\beta_i(E)$ via Eq.~(\ref{Eq3}), thereby fully specifying the eigenstate. This exact solvability originates from a hidden algebraic structure: the quartic Eq.~(\ref{Eq3}) reduces to a cubic under a symmetric parametrization of the four $\beta$-roots, and the OBC simplifies to a linear combination of Chebyshev polynomials---a generalized reciprocity that persists even without chiral symmetry and despite the presence of the identity term (see Secs. II and VIII in Ref. \cite{SuppMat}).

A particularly intriguing consequence of the SO coupling is the emergence of topological edge states, which requires $\delta_2\neq0$ as a necessary condition. In the thermodynamic limit  ($N \to \infty$), these topological states appear as isolated eigenvalues in the complex energy plane. Their energy is given by the compact expression
\begin{equation}
E_e=- \frac{\delta_{1}t_Lt_R\sin(\alpha_L-\alpha_R)}{\delta_{2}(t_L\sin\alpha_L-t_R\sin\alpha_R)}, \label{Eq9}
\end{equation}
which constitutes one of our central results. It is worth emphasizing that while the finite-energy edge states ($E_e\neq 0$) themselves are of great interest, the topological zero-mode conditions ($\delta_{1} = 0$ or $\alpha_L=\alpha_R$, which leads to $E_e= 0$) are particularly striking. They dictate that robust zero modes can emerge in a system that lacks chiral symmetry---a regime where traditional symmetry-based topological classifications would not predict their existence. We note that a general expression for the edge-state energy was also obtained independently in Ref.~\cite{Zhong25} via the original OBC determinant (see Supplemental Material Sec. III~\cite{SuppMat} for details). Our solution~(\ref{Eq9}) is derived via a distinct OBC reduction \cite{Li24} and serves as the explicit realization for the present Hamiltonian. This explicit form is essential for deriving the exact phase boundary and zero-mode conditions absent in Ref.~\cite{Zhong25}.

The existence of topological edge states is governed by our following analytical parameter condition \cite{SuppMat}:
\begin{equation}
\max(|W(Y_+)|, |W(Y_-)|)\leqslant|W(X_e)|, \label{Eq10}
\end{equation}
where $W(x)=\mathrm{max}(|x+ \sqrt{x^{2}-1}|,|x- \sqrt{x^{2}-1}|)$, $X_e=-p_2/p_1$, and $Y_\pm=[BX_e-C\pm\sqrt{ (BX_e-C)^2-4CX_e^3}]/(2X_e^2)$. Here the coefficients $B$ and $C$ are evaluated at the edge-state energy $E=E_e$. The inequality in Eq.~(\ref{Eq10}) delineates the parameter region supporting topological edge states, which is a subset of the broader region $p_2^2 \geqslant p_1^2$ (see Supplemental Fig.~6 \cite{SuppMat}); the latter defines the phase condition for topological zero modes when $\delta_1=0$. These analytical conditions arise from the wavefunction topology on the GBZ and are fully consistent with the symmetry-free topological invariant of Ref.~\cite{Zhong25} (see Supplemental Fig.~7 \cite{SuppMat}). They provide direct access to the topological phase diagram even when the GBZ is strongly deformed by the identity term.

\begin{figure}
	[ht]
	\begin{center}
		\includegraphics[width=8.7cm]%
		{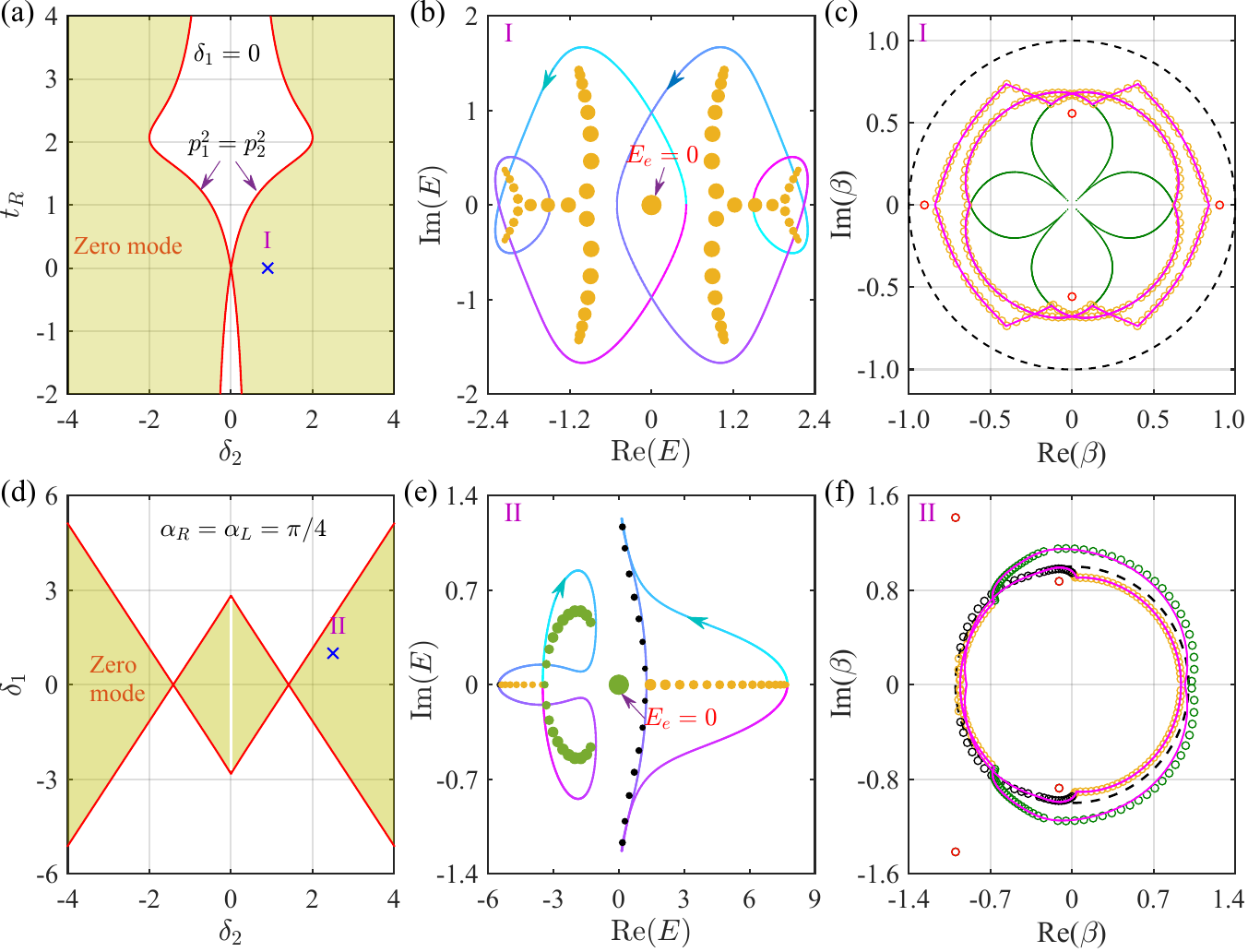}%%
		\caption{Exact topological zero modes under contrasting symmetries from Eq.~(\ref{Eq9}). Top row: The $\delta_1 = 0$ case, where a $\pi$-translation symmetry is present. (a) Phase diagram in the $(\delta_2, t_R)$ plane for $t_L = 2$, $\alpha_L = \pi/3$, $\alpha_R = \pi/4$, with the topological regime shaded in yellow. (b) Energy spectrum at the marked point ($\delta_2=0.9$, $t_R=0$) (blue cross) in (a), where solid circles denote OBC eigenvalues ($N=30$), with diameters proportional to eigenstate intensity and colors indicating localization direction (green: right-localized; yellow: left-localized). (c) GBZ at the same point, displayed as in Fig. \ref{fig2} (here green curves denote the auxiliary GBZ lines). Bottom row: The $\alpha_R = \alpha_L$ case, where both chiral symmetry and $\pi$-translation symmetry are broken. (d)--(f) show the phase diagram in the $(\delta_2, \delta_1)$ plane for $t_L = 2$, $t_R = 1$, $\alpha_L = \alpha_R = \pi/4$, the energy spectrum at the marked point ($\delta_2=2.5$, $\delta_1=1$), and the GBZ, respectively.} \label{fig3} 	\end{center}
\end{figure}

For illustration, we demonstrate in Fig.~\ref{fig2} the phase diagram and direct evidence for topological edge states. Figure~\ref{fig2}(a) shows a non-trivial phase diagram in the $(\delta_2, t_R)$ plane, where a topological regime (shaded yellow) emerges only when $\delta_2 \neq 0$, highlighting the indispensable role of inter-cell SO coupling. Crucially, the synergy between the identity term \(d_0(\beta)\mathds{1}\) and the SO couplings (\(\delta_{1,2}\)) actively sculpts this phase diagram, leading to the formation of topological islands (see Supplemental Figs. 3 and 6 \cite{SuppMat}). These islands generally host multi-branch spectral structures (which can evolve into vertically separated arcs under specific parameters), distinct from the horizontally separated arcs characterizing the non-island topological regimes, as illustrated in Supplemental Fig. 3 \cite{SuppMat} and Fig. \ref{fig6} in End Matter. Such structures do not arise in simpler models lacking this interplay \cite{Li24}. Remarkably, even within this complex landscape, the energy bands along three cuts [Figs.~\ref{fig2}(b)--\ref{fig2}(d)] clearly reveal topological edge states (red curves) within the bulk gap, with their existence intervals matching the topological domains predicted by the phase diagram. A detailed examination at the four marked points (I--IV) further confirms the bulk-boundary correspondence: points II, III, and IV, all inside the topological regime, consistently exhibit isolated eigenvalues (red asterisk) in the complex spectrum [Figs.~\ref{fig2}(h,k,n)] and edge-localized wavefunctions (red curves) [Figs.~\ref{fig2}(i,l,o)], whereas point I, outside the topological regime, shows no edge states  [Figs.~\ref{fig2}(e,f,g)]. Owing to the presence of the $\beta$-dependent identity term, the associated GBZ appears either as a single twisted contour [Figs.~\ref{fig2}(g) and \ref{fig2}(m); see also Supplemental Fig. 9 \cite{SuppMat} for enlarged views] or as two separate loops [Figs.~\ref{fig2}(j) and \ref{fig2}(p)], reflecting the distinct topology of the PBC spectrum. This spectral winding topology is, in principle, linked to the emergence of the NHSE \cite{Okuma2020,Zha20} (see Sec. IV in Ref.~\cite{SuppMat} for further discussion).

\begin{figure}
	[ht!]
	\begin{center}
		\includegraphics[width=8.7cm]%
		{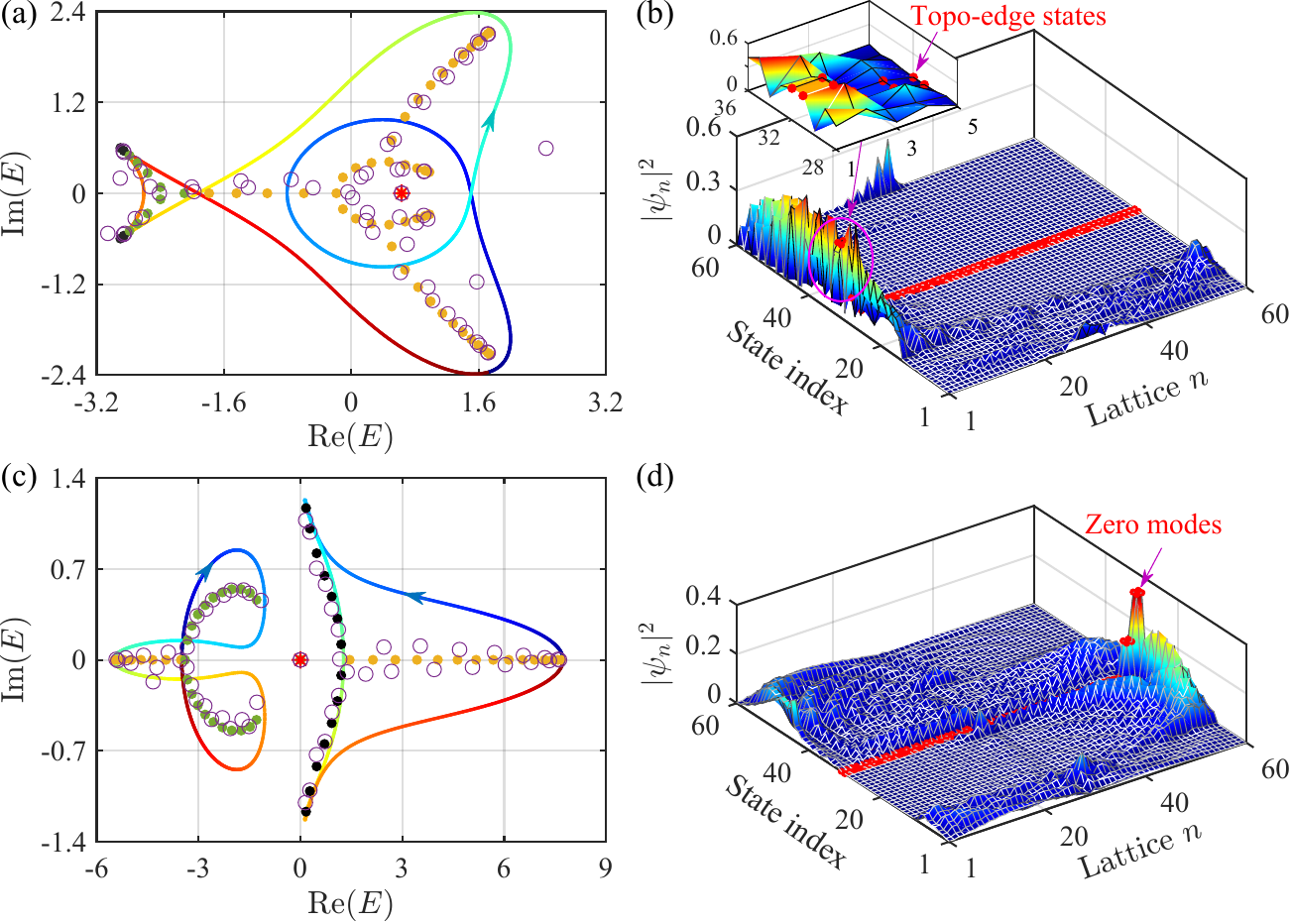}%%
		\caption{Robustness of topological edge states against local perturbations. (a, b) Response of a finite-energy topological edge state [from Fig.~\ref{fig2}, point III] to a local disorder at the 15th unit cell, where only the $\alpha_R$ value is changed to $\pi/2$. (a) Energy spectrum: solid (open) circles denote the unperturbed (perturbed) case. (b) Spatial distribution of all eigenstates, where the analytical edge-state solution is marked by red circles (see inset). (c, d) Response of a zero mode [from Fig.~\ref{fig3}(e)] to a different local disorder ($\alpha_L \to \pi/3$, $\alpha_R \to \pi/6$ at the 15th cell), displayed as in (a, b).} \label{fig4}
	\end{center}
\end{figure}

Figure~\ref{fig3} shows the emergence of topological zero modes under two distinct symmetry conditions. The top row (the $\delta_1 = 0$ case) explores a regime that preserves $\pi$-translation symmetry but lacks chiral symmetry. Its phase diagram [Fig.~\ref{fig3}(a)] defines a topological region supporting zero modes, and the spectrum at a chosen point (blue cross) confirms a topological edge state pinned at zero energy [Fig.~\ref{fig3}(b)], accompanied by a symmetric GBZ [Fig.~\ref{fig3}(c)]. In stark contrast, the $\alpha_R = \alpha_L$ case shown in the bottom row breaks both chiral symmetry and $\pi$-translation symmetry via $\delta_2 \neq 0$. Strikingly, a topological zero mode persists even in this fully non-chiral setting, as evidenced by its phase diagram [Fig.~\ref{fig3}(d)], asymmetric spectrum hosting a zero-energy state [Fig.~\ref{fig3}(e)], and distinctive GBZ [Fig.~\ref{fig3}(f)]. These findings, corroborated by further cases in Supplemental Fig.~4 \cite{SuppMat}, establish that robust zero modes can emerge in systems devoid of any conventional symmetry that mandates $(E,-E)$ pairing. This highlights a broader mechanism for topological zero modes in non-Hermitian systems, distinct from the SSH paradigm (see Fig.~\ref{fig6} in End Matter for topological phase transition).

Finally, we demonstrate in Fig.~\ref{fig4} the robustness of topological edge states against local perturbations in the non-chiral regime. A finite-energy topological edge state from Fig.~\ref{fig2} (point III) and a topological zero mode from Fig.~\ref{fig3}(e) are each subjected to a local change of the hopping phases at the $15$th unit cell. The perturbed energy spectra [Figs.~\ref{fig4}(a) and \ref{fig4}(c)] show that, while the bulk states shift noticeably, both the finite-energy edge state and the zero mode remain pinned at their original energies. Correspondingly, their spatial profiles [Figs.~\ref{fig4}(b) and \ref{fig4}(d)] are essentially unchanged and coincide with the unperturbed analytical solutions (red circles). Moreover, as shown in Supplemental Fig.~8, the edge states also survive under random on-site and hopping disorders. This insensitivity to local or random disorders underscores the intrinsic robustness of these edge modes, even when chiral symmetry is absent.

In conclusion, we have established an exactly solvable paradigm for non-Hermitian topology driven by a momentum-dependent identity term, realized via an SO-coupled non-Hermitian HN chain. Our solution reveals that this identity term, synergizing with inter-cell SO coupling, actively deforms the GBZ and thereby can induce topological edge states, including robust zero modes that require neither chiral symmetry nor the $(E,-E)$ spectral pairing. This paradigmatic solution transcends standard symmetry-classified topology and offers a pathway to topological protection under relaxed symmetries. Moreover, our analytical approach, which directly addresses the challenge posed by a GBZ-modifying identity term, readily extends to the most general two-band models without any protecting symmetry (see Sec.~VIII in Ref. \cite{SuppMat}), enabling exact studies in broader settings \cite{Wu2023,Jiao21,Zhang25,ZhangY25,Zhong25}.  These findings open new avenues for designing robust non-Hermitian topological states \cite{Ber21,Ding22,Zhang23}, with implications for topological frequency combs \cite{Mittal21,Flower24,Wu25}, protected entanglement sources \cite{Dai22}, and quantum simulation circuits \cite{Mei20,Dai24,Zhang24}.

\begin{acknowledgments}
{\it Acknowledgments}---This work was supported by the National Natural Science Foundation of China (Grants No. 12374301, No. 11974075, and No. 62105061), the Jiangsu Provincial Frontier Technology Research and Development Program (Grant No. BF2025065). C. H. L. acknowledges support from the National Research Foundation, Singapore under its QEP2.0 programme (NRF2021-QEP2-02-P09), and the Ministry of Education, Singapore (MOE Award No. MOE-T2EP50222-0003).
\end{acknowledgments}

{\it Data availability}---The data that support the findings of this article are openly available \cite{DataCode}.

\bibliographystyle{myprl}
\bibliography{PRL_BibFile}

\begin{thebibliography}{130}%
\makeatletter
\providecommand \@ifxundefined [1]{%
 \@ifx{#1\undefined}
}%
\providecommand \@ifnum [1]{%
 \ifnum #1\expandafter \@firstoftwo
 \else \expandafter \@secondoftwo
 \fi
}%
\providecommand \@ifx [1]{%
 \ifx #1\expandafter \@firstoftwo
 \else \expandafter \@secondoftwo
 \fi
}%
\providecommand \natexlab [1]{#1}%
\providecommand \enquote  [1]{``#1''}%
\providecommand \bibnamefont  [1]{#1}%
\providecommand \bibfnamefont [1]{#1}%
\providecommand \citenamefont [1]{#1}%
\providecommand \href@noop [0]{\@secondoftwo}%
\providecommand \href [0]{\begingroup \@sanitize@url \@href}%
\providecommand \@href[1]{\@@startlink{#1}\@@href}%
\providecommand \@@href[1]{\endgroup#1\@@endlink}%
\providecommand \@sanitize@url [0]{\catcode `\\12\catcode `\$12\catcode
  `\&12\catcode `\#12\catcode `\^12\catcode `\_12\catcode `\%12\relax}%
\providecommand \@@startlink[1]{}%
\providecommand \@@endlink[0]{}%
\providecommand \url  [0]{\begingroup\@sanitize@url \@url }%
\providecommand \@url [1]{\endgroup\@href {#1}{\urlprefix }}%
\providecommand \urlprefix  [0]{URL }%
\providecommand \Eprint [0]{\href }%
\providecommand \doibase [0]{http://dx.doi.org/}%
\providecommand \selectlanguage[0]{\@gobble}%
\providecommand \bibinfo  [0]{\@secondoftwo}%
\providecommand \bibfield  [0]{\@secondoftwo}%
\providecommand \translation [1]{[#1]}%
\providecommand \BibitemOpen [0]{}%
\providecommand \bibitemStop [0]{}%
\providecommand \bibitemNoStop [0]{.\EOS\space}%
\providecommand \EOS [0]{\spacefactor3000\relax}%
\providecommand \BibitemShut  [1]{\csname bibitem#1\endcsname}%
\let\auto@bib@innerbib\@empty
%</preamble>
\bibitem [{\citenamefont {Ashida}\ \emph {et~al.}(2020)\citenamefont {Ashida},
  \citenamefont {Gong},\ and\ \citenamefont {Ueda}}]{Ash20}%
  \BibitemOpen
  \bibfield  {author} {\bibinfo {author} {\bibfnamefont {Y.}~\bibnamefont
  {Ashida}}, \bibinfo {author} {\bibfnamefont {Z.}~\bibnamefont {Gong}}, \ and\
  \bibinfo {author} {\bibfnamefont {M.}~\bibnamefont {Ueda}},\ }\enquote
  {\bibinfo {title} {Non-{Hermitian} physics},}\ \href {\doibase
  10.1080/00018732.2021.1876991} {\bibfield  {journal} {\bibinfo  {journal}
  {Adv. Phys.}\ }\textbf {\bibinfo {volume} {69}},\ \bibinfo {pages} {249}
  (\bibinfo {year} {2020})}\BibitemShut {NoStop}%
\bibitem [{\citenamefont {Kawabata}\ \emph
  {et~al.}(2019{\natexlab{a}})\citenamefont {Kawabata}, \citenamefont
  {Shiozaki}, \citenamefont {Ueda},\ and\ \citenamefont {Sato}}]{Kawa19}%
  \BibitemOpen
  \bibfield  {author} {\bibinfo {author} {\bibfnamefont {K.}~\bibnamefont
  {Kawabata}}, \bibinfo {author} {\bibfnamefont {K.}~\bibnamefont {Shiozaki}},
  \bibinfo {author} {\bibfnamefont {M.}~\bibnamefont {Ueda}}, \ and\ \bibinfo
  {author} {\bibfnamefont {M.}~\bibnamefont {Sato}},\ }\enquote {\bibinfo
  {title} {Symmetry and topology in non-{Hermitian} physics},}\ \href {\doibase
  10.1103/PhysRevX.9.041015} {\bibfield  {journal} {\bibinfo  {journal} {Phys.
  Rev. X}\ }\textbf {\bibinfo {volume} {9}},\ \bibinfo {pages} {041015}
  (\bibinfo {year} {2019}{\natexlab{a}})}\BibitemShut {NoStop}%
\bibitem [{\citenamefont {Okuma}\ and\ \citenamefont {Sato}(2023)}]{Okuma23}%
  \BibitemOpen
  \bibfield  {author} {\bibinfo {author} {\bibfnamefont {N.}~\bibnamefont
  {Okuma}}\ and\ \bibinfo {author} {\bibfnamefont {M.}~\bibnamefont {Sato}},\
  }\enquote {\bibinfo {title} {Non-{Hermitian} topological phenomena: {A}
  review},}\ \href {\doibase 10.1146/annurev-conmatphys-040521-033133}
  {\bibfield  {journal} {\bibinfo  {journal} {Annu. Rev. Condens. Matter
  Phys.}\ }\textbf {\bibinfo {volume} {14}},\ \bibinfo {pages} {83} (\bibinfo
  {year} {2023})}\BibitemShut {NoStop}%
\bibitem [{\citenamefont {El-Ganainy}\ \emph {et~al.}(2018)\citenamefont
  {El-Ganainy}, \citenamefont {Makris}, \citenamefont {Khajavikhan},
  \citenamefont {Musslimani}, \citenamefont {Rotter},\ and\ \citenamefont
  {Christodoulides}}]{El18}%
  \BibitemOpen
  \bibfield  {author} {\bibinfo {author} {\bibfnamefont {R.}~\bibnamefont
  {El-Ganainy}}, \bibinfo {author} {\bibfnamefont {K.~G.}\ \bibnamefont
  {Makris}}, \bibinfo {author} {\bibfnamefont {M.}~\bibnamefont {Khajavikhan}},
  \bibinfo {author} {\bibfnamefont {Z.~H.}\ \bibnamefont {Musslimani}},
  \bibinfo {author} {\bibfnamefont {S.}~\bibnamefont {Rotter}}, \ and\ \bibinfo
  {author} {\bibfnamefont {D.~N.}\ \bibnamefont {Christodoulides}},\ }\enquote
  {\bibinfo {title} {Non-{Hermitian} physics and {PT} symmetry},}\ \href
  {https://doi.org/10.1038/nphys4323} {\bibfield  {journal} {\bibinfo
  {journal} {Nat. Phys.}\ }\textbf {\bibinfo {volume} {14}},\ \bibinfo {pages}
  {11} (\bibinfo {year} {2018})}\BibitemShut {NoStop}%
\bibitem [{\citenamefont {Coulais}\ \emph {et~al.}(2021)\citenamefont
  {Coulais}, \citenamefont {Fleury},\ and\ \citenamefont {van Wezel}}]{Cou21}%
  \BibitemOpen
  \bibfield  {author} {\bibinfo {author} {\bibfnamefont {C.}~\bibnamefont
  {Coulais}}, \bibinfo {author} {\bibfnamefont {R.}~\bibnamefont {Fleury}}, \
  and\ \bibinfo {author} {\bibfnamefont {J.}~\bibnamefont {van Wezel}},\
  }\enquote {\bibinfo {title} {Topology and broken {Hermiticity}},}\ \href
  {\doibase 10.1038/s41567-020-01093-z} {\bibfield  {journal} {\bibinfo
  {journal} {Nat. Phys.}\ }\textbf {\bibinfo {volume} {17}},\ \bibinfo {pages}
  {9} (\bibinfo {year} {2021})}\BibitemShut {NoStop}%
\bibitem [{\citenamefont {Bergholtz}\ \emph {et~al.}(2021)\citenamefont
  {Bergholtz}, \citenamefont {Budich},\ and\ \citenamefont {Kunst}}]{Ber21}%
  \BibitemOpen
  \bibfield  {author} {\bibinfo {author} {\bibfnamefont {E.~J.}\ \bibnamefont
  {Bergholtz}}, \bibinfo {author} {\bibfnamefont {J.~C.}\ \bibnamefont
  {Budich}}, \ and\ \bibinfo {author} {\bibfnamefont {F.~K.}\ \bibnamefont
  {Kunst}},\ }\enquote {\bibinfo {title} {Exceptional topology of
  non-{Hermitian} systems},}\ \href {\doibase 10.1103/RevModPhys.93.015005}
  {\bibfield  {journal} {\bibinfo  {journal} {Rev. Mod. Phys.}\ }\textbf
  {\bibinfo {volume} {93}},\ \bibinfo {pages} {015005} (\bibinfo {year}
  {2021})}\BibitemShut {NoStop}%
\bibitem [{\citenamefont {Ding}\ \emph {et~al.}(2022)\citenamefont {Ding},
  \citenamefont {Fang},\ and\ \citenamefont {Ma}}]{Ding22}%
  \BibitemOpen
  \bibfield  {author} {\bibinfo {author} {\bibfnamefont {K.}~\bibnamefont
  {Ding}}, \bibinfo {author} {\bibfnamefont {C.}~\bibnamefont {Fang}}, \ and\
  \bibinfo {author} {\bibfnamefont {G.}~\bibnamefont {Ma}},\ }\enquote
  {\bibinfo {title} {Non-{Hermitian} topology and exceptional-point
  geometries},}\ \href {\doibase 10.1038/s42254-022-00516-5} {\bibfield
  {journal} {\bibinfo  {journal} {Nat. Rev. Phys.}\ }\textbf {\bibinfo {volume}
  {4}},\ \bibinfo {pages} {745} (\bibinfo {year} {2022})}\BibitemShut {NoStop}%
\bibitem [{\citenamefont {Zhang}\ \emph {et~al.}(2023)\citenamefont {Zhang},
  \citenamefont {Zangeneh-Nejad}, \citenamefont {Chen}, \citenamefont {Lu},\
  and\ \citenamefont {Christensen}}]{Zhang23}%
  \BibitemOpen
  \bibfield  {author} {\bibinfo {author} {\bibfnamefont {X.}~\bibnamefont
  {Zhang}}, \bibinfo {author} {\bibfnamefont {F.}~\bibnamefont
  {Zangeneh-Nejad}}, \bibinfo {author} {\bibfnamefont {Z.-G.}\ \bibnamefont
  {Chen}}, \bibinfo {author} {\bibfnamefont {M.-H.}\ \bibnamefont {Lu}}, \ and\
  \bibinfo {author} {\bibfnamefont {J.}~\bibnamefont {Christensen}},\ }\enquote
  {\bibinfo {title} {A second wave of topological phenomena in photonics and
  acoustics},}\ \href {\doibase 10.1038/s41586-023-06163-9} {\bibfield
  {journal} {\bibinfo  {journal} {Nature}\ }\textbf {\bibinfo {volume} {618}},\
  \bibinfo {pages} {687} (\bibinfo {year} {2023})}\BibitemShut {NoStop}%
\bibitem [{\citenamefont {Yao}\ and\ \citenamefont {Wang}(2018)}]{Yao18}%
  \BibitemOpen
  \bibfield  {author} {\bibinfo {author} {\bibfnamefont {S.}~\bibnamefont
  {Yao}}\ and\ \bibinfo {author} {\bibfnamefont {Z.}~\bibnamefont {Wang}},\
  }\enquote {\bibinfo {title} {Edge states and topological invariants of
  non-{Hermitian} systems},}\ \href {\doibase 10.1103/PhysRevLett.121.086803}
  {\bibfield  {journal} {\bibinfo  {journal} {Phys. Rev. Lett.}\ }\textbf
  {\bibinfo {volume} {121}},\ \bibinfo {pages} {086803} (\bibinfo {year}
  {2018})}\BibitemShut {NoStop}%
\bibitem [{\citenamefont {Lee}\ and\ \citenamefont {Thomale}(2019)}]{LeeC19}%
  \BibitemOpen
  \bibfield  {author} {\bibinfo {author} {\bibfnamefont {C.~H.}\ \bibnamefont
  {Lee}}\ and\ \bibinfo {author} {\bibfnamefont {R.}~\bibnamefont {Thomale}},\
  }\enquote {\bibinfo {title} {Anatomy of skin modes and topology in
  non-{Hermitian} systems},}\ \href {\doibase 10.1103/PhysRevB.99.201103}
  {\bibfield  {journal} {\bibinfo  {journal} {Phys. Rev. B}\ }\textbf {\bibinfo
  {volume} {99}},\ \bibinfo {pages} {201103(R)} (\bibinfo {year}
  {2019})}\BibitemShut {NoStop}%
\bibitem [{\citenamefont {Borgnia}\ \emph {et~al.}(2020)\citenamefont
  {Borgnia}, \citenamefont {Kruchkov},\ and\ \citenamefont
  {Slager}}]{Borgnia20}%
  \BibitemOpen
  \bibfield  {author} {\bibinfo {author} {\bibfnamefont {D.~S.}\ \bibnamefont
  {Borgnia}}, \bibinfo {author} {\bibfnamefont {A.~J.}\ \bibnamefont
  {Kruchkov}}, \ and\ \bibinfo {author} {\bibfnamefont {R.-J.}\ \bibnamefont
  {Slager}},\ }\enquote {\bibinfo {title} {Non-{Hermitian} boundary modes and
  topology},}\ \href {\doibase 10.1103/PhysRevLett.124.056802} {\bibfield
  {journal} {\bibinfo  {journal} {Phys. Rev. Lett.}\ }\textbf {\bibinfo
  {volume} {124}},\ \bibinfo {pages} {056802} (\bibinfo {year}
  {2020})}\BibitemShut {NoStop}%
\bibitem [{\citenamefont {Zhao}\ \emph {et~al.}(2019)\citenamefont {Zhao},
  \citenamefont {Qiao}, \citenamefont {Wu}, \citenamefont {Midya},
  \citenamefont {Longhi},\ and\ \citenamefont {Feng}}]{Zhao2019}%
  \BibitemOpen
  \bibfield  {author} {\bibinfo {author} {\bibfnamefont {H.}~\bibnamefont
  {Zhao}}, \bibinfo {author} {\bibfnamefont {X.}~\bibnamefont {Qiao}}, \bibinfo
  {author} {\bibfnamefont {T.}~\bibnamefont {Wu}}, \bibinfo {author}
  {\bibfnamefont {B.}~\bibnamefont {Midya}}, \bibinfo {author} {\bibfnamefont
  {S.}~\bibnamefont {Longhi}}, \ and\ \bibinfo {author} {\bibfnamefont
  {L.}~\bibnamefont {Feng}},\ }\enquote {\bibinfo {title} {Non-{Hermitian}
  topological light steering},}\ \href {\doibase 10.1126/science.aay1064}
  {\bibfield  {journal} {\bibinfo  {journal} {Science}\ }\textbf {\bibinfo
  {volume} {365}},\ \bibinfo {pages} {1163} (\bibinfo {year}
  {2019})}\BibitemShut {NoStop}%
\bibitem [{\citenamefont {Tai}\ and\ \citenamefont {Lee}(2023)}]{Tai23}%
  \BibitemOpen
  \bibfield  {author} {\bibinfo {author} {\bibfnamefont {T.}~\bibnamefont
  {Tai}}\ and\ \bibinfo {author} {\bibfnamefont {C.~H.}\ \bibnamefont {Lee}},\
  }\enquote {\bibinfo {title} {Zoology of non-{Hermitian} spectra and their
  graph topology},}\ \href {\doibase 10.1103/PhysRevB.107.L220301} {\bibfield
  {journal} {\bibinfo  {journal} {Phys. Rev. B}\ }\textbf {\bibinfo {volume}
  {107}},\ \bibinfo {pages} {L220301} (\bibinfo {year} {2023})}\BibitemShut
  {NoStop}%
\bibitem [{\citenamefont {Longhi}(2022)}]{Longhi22}%
  \BibitemOpen
  \bibfield  {author} {\bibinfo {author} {\bibfnamefont {S.}~\bibnamefont
  {Longhi}},\ }\enquote {\bibinfo {title} {Self-{Healing} of non-{Hermitian}
  topological skin modes},}\ \href {\doibase 10.1103/PhysRevLett.128.157601}
  {\bibfield  {journal} {\bibinfo  {journal} {Phys. Rev. Lett.}\ }\textbf
  {\bibinfo {volume} {128}},\ \bibinfo {pages} {157601} (\bibinfo {year}
  {2022})}\BibitemShut {NoStop}%
\bibitem [{\citenamefont {Okuma}\ \emph {et~al.}(2020)\citenamefont {Okuma},
  \citenamefont {Kawabata}, \citenamefont {Shiozaki},\ and\ \citenamefont
  {Sato}}]{Okuma2020}%
  \BibitemOpen
  \bibfield  {author} {\bibinfo {author} {\bibfnamefont {N.}~\bibnamefont
  {Okuma}}, \bibinfo {author} {\bibfnamefont {K.}~\bibnamefont {Kawabata}},
  \bibinfo {author} {\bibfnamefont {K.}~\bibnamefont {Shiozaki}}, \ and\
  \bibinfo {author} {\bibfnamefont {M.}~\bibnamefont {Sato}},\ }\enquote
  {\bibinfo {title} {Topological origin of non-{Hermitian} skin effects},}\
  \href {\doibase 10.1103/PhysRevLett.124.086801} {\bibfield  {journal}
  {\bibinfo  {journal} {Phys. Rev. Lett.}\ }\textbf {\bibinfo {volume} {124}},\
  \bibinfo {pages} {086801} (\bibinfo {year} {2020})}\BibitemShut {NoStop}%
\bibitem [{\citenamefont {Zhang}\ \emph {et~al.}(2020)\citenamefont {Zhang},
  \citenamefont {Yang},\ and\ \citenamefont {Fang}}]{Zha20}%
  \BibitemOpen
  \bibfield  {author} {\bibinfo {author} {\bibfnamefont {K.}~\bibnamefont
  {Zhang}}, \bibinfo {author} {\bibfnamefont {Z.}~\bibnamefont {Yang}}, \ and\
  \bibinfo {author} {\bibfnamefont {C.}~\bibnamefont {Fang}},\ }\enquote
  {\bibinfo {title} {Correspondence between winding numbers and skin modes in
  non-{Hermitian} systems},}\ \href {\doibase 10.1103/PhysRevLett.125.126402}
  {\bibfield  {journal} {\bibinfo  {journal} {Phys. Rev. Lett.}\ }\textbf
  {\bibinfo {volume} {125}},\ \bibinfo {pages} {126402} (\bibinfo {year}
  {2020})}\BibitemShut {NoStop}%
\bibitem [{\citenamefont {Sun}\ \emph {et~al.}(2021)\citenamefont {Sun},
  \citenamefont {Zhu},\ and\ \citenamefont {Hughes}}]{Sun21}%
  \BibitemOpen
  \bibfield  {author} {\bibinfo {author} {\bibfnamefont {X.-Q.}\ \bibnamefont
  {Sun}}, \bibinfo {author} {\bibfnamefont {P.}~\bibnamefont {Zhu}}, \ and\
  \bibinfo {author} {\bibfnamefont {T.~L.}\ \bibnamefont {Hughes}},\ }\enquote
  {\bibinfo {title} {Geometric response and disclination-induced skin effects
  in non-{Hermitian} systems},}\ \href {\doibase
  10.1103/PhysRevLett.127.066401} {\bibfield  {journal} {\bibinfo  {journal}
  {Phys. Rev. Lett.}\ }\textbf {\bibinfo {volume} {127}},\ \bibinfo {pages}
  {066401} (\bibinfo {year} {2021})}\BibitemShut {NoStop}%
\bibitem [{\citenamefont {Weidemann}\ \emph {et~al.}(2020)\citenamefont
  {Weidemann}, \citenamefont {Kremer}, \citenamefont {Helbig}, \citenamefont
  {Hofmann}, \citenamefont {Stegmaier}, \citenamefont {Greiter}, \citenamefont
  {Thomale},\ and\ \citenamefont {Szameit}}]{Wei20}%
  \BibitemOpen
  \bibfield  {author} {\bibinfo {author} {\bibfnamefont {S.}~\bibnamefont
  {Weidemann}}, \bibinfo {author} {\bibfnamefont {M.}~\bibnamefont {Kremer}},
  \bibinfo {author} {\bibfnamefont {T.}~\bibnamefont {Helbig}}, \bibinfo
  {author} {\bibfnamefont {T.}~\bibnamefont {Hofmann}}, \bibinfo {author}
  {\bibfnamefont {A.}~\bibnamefont {Stegmaier}}, \bibinfo {author}
  {\bibfnamefont {M.}~\bibnamefont {Greiter}}, \bibinfo {author} {\bibfnamefont
  {R.}~\bibnamefont {Thomale}}, \ and\ \bibinfo {author} {\bibfnamefont
  {A.}~\bibnamefont {Szameit}},\ }\enquote {\bibinfo {title} {Topological
  funneling of light},}\ \href {\doibase 10.1126/science.aaz8727} {\bibfield
  {journal} {\bibinfo  {journal} {Science}\ }\textbf {\bibinfo {volume}
  {368}},\ \bibinfo {pages} {311} (\bibinfo {year} {2020})}\BibitemShut
  {NoStop}%
\bibitem [{\citenamefont {Hou}\ \emph {et~al.}(2024)\citenamefont {Hou},
  \citenamefont {Wu}, \citenamefont {Li}, \citenamefont {Basit}, \citenamefont
  {Wei}, \citenamefont {Chen}, \citenamefont {Grelu},\ and\ \citenamefont
  {Ni}}]{Hou24}%
  \BibitemOpen
  \bibfield  {author} {\bibinfo {author} {\bibfnamefont {C.}~\bibnamefont
  {Hou}}, \bibinfo {author} {\bibfnamefont {G.}~\bibnamefont {Wu}}, \bibinfo
  {author} {\bibfnamefont {L.}~\bibnamefont {Li}}, \bibinfo {author}
  {\bibfnamefont {A.}~\bibnamefont {Basit}}, \bibinfo {author} {\bibfnamefont
  {Y.}~\bibnamefont {Wei}}, \bibinfo {author} {\bibfnamefont {S.}~\bibnamefont
  {Chen}}, \bibinfo {author} {\bibfnamefont {P.}~\bibnamefont {Grelu}}, \ and\
  \bibinfo {author} {\bibfnamefont {Z.}~\bibnamefont {Ni}},\ }\enquote
  {\bibinfo {title} {Non-{Hermitian} skin effects in two- and three-dimensional
  intertwined tight-binding lattices},}\ \href {\doibase
  10.1103/PhysRevB.109.205135} {\bibfield  {journal} {\bibinfo  {journal}
  {Phys. Rev. B}\ }\textbf {\bibinfo {volume} {109}},\ \bibinfo {pages}
  {205135} (\bibinfo {year} {2024})}\BibitemShut {NoStop}%
\bibitem [{\citenamefont {Li}\ \emph {et~al.}(2023)\citenamefont {Li},
  \citenamefont {Trauzettel}, \citenamefont {Neupert},\ and\ \citenamefont
  {Zhang}}]{LiC23}%
  \BibitemOpen
  \bibfield  {author} {\bibinfo {author} {\bibfnamefont {C.-A.}\ \bibnamefont
  {Li}}, \bibinfo {author} {\bibfnamefont {B.}~\bibnamefont {Trauzettel}},
  \bibinfo {author} {\bibfnamefont {T.}~\bibnamefont {Neupert}}, \ and\
  \bibinfo {author} {\bibfnamefont {S.-B.}\ \bibnamefont {Zhang}},\ }\enquote
  {\bibinfo {title} {Enhancement of second-order non-{Hermitian} skin effect by
  magnetic fields},}\ \href {\doibase 10.1103/PhysRevLett.131.116601}
  {\bibfield  {journal} {\bibinfo  {journal} {Phys. Rev. Lett.}\ }\textbf
  {\bibinfo {volume} {131}},\ \bibinfo {pages} {116601} (\bibinfo {year}
  {2023})}\BibitemShut {NoStop}%
\bibitem [{\citenamefont {Li}\ and\ \citenamefont {Lee}(2022)}]{LiL22}%
  \BibitemOpen
  \bibfield  {author} {\bibinfo {author} {\bibfnamefont {L.}~\bibnamefont
  {Li}}\ and\ \bibinfo {author} {\bibfnamefont {C.~H.}\ \bibnamefont {Lee}},\
  }\enquote {\bibinfo {title} {Non-{Hermitian} pseudo-gaps},}\ \href {\doibase
  https://doi.org/10.1016/j.scib.2022.01.017} {\bibfield  {journal} {\bibinfo
  {journal} {Sci. Bull.}\ }\textbf {\bibinfo {volume} {67}},\ \bibinfo {pages}
  {685} (\bibinfo {year} {2022})}\BibitemShut {NoStop}%
\bibitem [{\citenamefont {Lei}\ \emph {et~al.}(2024)\citenamefont {Lei},
  \citenamefont {Lee},\ and\ \citenamefont {Li}}]{Lei24}%
  \BibitemOpen
  \bibfield  {author} {\bibinfo {author} {\bibfnamefont {Z.}~\bibnamefont
  {Lei}}, \bibinfo {author} {\bibfnamefont {C.~H.}\ \bibnamefont {Lee}}, \ and\
  \bibinfo {author} {\bibfnamefont {L.}~\bibnamefont {Li}},\ }\enquote
  {\bibinfo {title} {Activating non-{Hermitian} skin modes by parity-time
  symmetry breaking},}\ \href {\doibase 10.1038/s42005-024-01591-z} {\bibfield
  {journal} {\bibinfo  {journal} {Commun. Phys.}\ }\textbf {\bibinfo {volume}
  {7}},\ \bibinfo {pages} {100} (\bibinfo {year} {2024})}\BibitemShut {NoStop}%
\bibitem [{\citenamefont {Gliozzi}\ \emph {et~al.}(2024)\citenamefont
  {Gliozzi}, \citenamefont {Tomasi},\ and\ \citenamefont {Hughes}}]{Gli24}%
  \BibitemOpen
  \bibfield  {author} {\bibinfo {author} {\bibfnamefont {J.}~\bibnamefont
  {Gliozzi}}, \bibinfo {author} {\bibfnamefont {G.~D.}\ \bibnamefont {Tomasi}},
  \ and\ \bibinfo {author} {\bibfnamefont {T.~L.}\ \bibnamefont {Hughes}},\
  }\enquote {\bibinfo {title} {Many-body non-{Hermitian} skin effect for
  multipoles},}\ \href {\doibase 10.1103/PhysRevLett.133.136503} {\bibfield
  {journal} {\bibinfo  {journal} {Phys. Rev. Lett.}\ }\textbf {\bibinfo
  {volume} {133}},\ \bibinfo {pages} {136503} (\bibinfo {year}
  {2024})}\BibitemShut {NoStop}%
\bibitem [{\citenamefont {Yoshida}\ \emph {et~al.}(2024)\citenamefont
  {Yoshida}, \citenamefont {Zhang}, \citenamefont {Neupert},\ and\
  \citenamefont {Kawakami}}]{Yosh24}%
  \BibitemOpen
  \bibfield  {author} {\bibinfo {author} {\bibfnamefont {T.}~\bibnamefont
  {Yoshida}}, \bibinfo {author} {\bibfnamefont {S.-B.}\ \bibnamefont {Zhang}},
  \bibinfo {author} {\bibfnamefont {T.}~\bibnamefont {Neupert}}, \ and\
  \bibinfo {author} {\bibfnamefont {N.}~\bibnamefont {Kawakami}},\ }\enquote
  {\bibinfo {title} {Non-{Hermitian Mott} skin effect},}\ \href {\doibase
  10.1103/PhysRevLett.133.076502} {\bibfield  {journal} {\bibinfo  {journal}
  {Phys. Rev. Lett.}\ }\textbf {\bibinfo {volume} {133}},\ \bibinfo {pages}
  {076502} (\bibinfo {year} {2024})}\BibitemShut {NoStop}%
\bibitem [{\citenamefont {Shen}\ \emph
  {et~al.}(2025{\natexlab{a}})\citenamefont {Shen}, \citenamefont {Chan},\ and\
  \citenamefont {Lee}}]{Shen25}%
  \BibitemOpen
  \bibfield  {author} {\bibinfo {author} {\bibfnamefont {R.}~\bibnamefont
  {Shen}}, \bibinfo {author} {\bibfnamefont {W.~J.}\ \bibnamefont {Chan}}, \
  and\ \bibinfo {author} {\bibfnamefont {C.~H.}\ \bibnamefont {Lee}},\
  }\enquote {\bibinfo {title} {Non-{Hermitian} skin effect along hyperbolic
  geodesics},}\ \href {\doibase 10.1103/PhysRevB.111.045420} {\bibfield
  {journal} {\bibinfo  {journal} {Phys. Rev. B}\ }\textbf {\bibinfo {volume}
  {111}},\ \bibinfo {pages} {045420} (\bibinfo {year}
  {2025}{\natexlab{a}})}\BibitemShut {NoStop}%
\bibitem [{\citenamefont {Li}\ \emph {et~al.}(2025{\natexlab{a}})\citenamefont
  {Li}, \citenamefont {Jiang},\ and\ \citenamefont {Lee}}]{LiQ25}%
  \BibitemOpen
  \bibfield  {author} {\bibinfo {author} {\bibfnamefont {Q.}~\bibnamefont
  {Li}}, \bibinfo {author} {\bibfnamefont {H.}~\bibnamefont {Jiang}}, \ and\
  \bibinfo {author} {\bibfnamefont {C.~H.}\ \bibnamefont {Lee}},\ }\enquote
  {\bibinfo {title} {Phase-space generalized {Brillouin} zone for spatially
  inhomogeneous non-{Hermitian} systems},}\ \href {\doibase
  https://doi.org/10.1002/advs.202508047} {\bibfield  {journal} {\bibinfo
  {journal} {Adv. Sci.}\ }\textbf {\bibinfo {volume} {12}},\ \bibinfo {pages}
  {e08047} (\bibinfo {year} {2025}{\natexlab{a}})}\BibitemShut {NoStop}%
\bibitem [{\citenamefont {Hu}\ \emph {et~al.}(2025)\citenamefont {Hu},
  \citenamefont {Wang}, \citenamefont {Lian},\ and\ \citenamefont
  {Wang}}]{Hu25}%
  \BibitemOpen
  \bibfield  {author} {\bibinfo {author} {\bibfnamefont {Y.-M.}\ \bibnamefont
  {Hu}}, \bibinfo {author} {\bibfnamefont {Z.}~\bibnamefont {Wang}}, \bibinfo
  {author} {\bibfnamefont {B.}~\bibnamefont {Lian}}, \ and\ \bibinfo {author}
  {\bibfnamefont {Z.}~\bibnamefont {Wang}},\ }\enquote {\bibinfo {title}
  {Many-body non-{Hermitian} skin effect with exact steady states in the
  dissipative quantum link model},}\ \href {\doibase 10.1103/wztw-l8wg}
  {\bibfield  {journal} {\bibinfo  {journal} {Phys. Rev. Lett.}\ }\textbf
  {\bibinfo {volume} {135}},\ \bibinfo {pages} {260401} (\bibinfo {year}
  {2025})}\BibitemShut {NoStop}%
\bibitem [{\citenamefont {Yang}\ and\ \citenamefont {Fang}(2025)}]{Yang25}%
  \BibitemOpen
  \bibfield  {author} {\bibinfo {author} {\bibfnamefont {T.-H.}\ \bibnamefont
  {Yang}}\ and\ \bibinfo {author} {\bibfnamefont {C.}~\bibnamefont {Fang}},\
  }\enquote {\bibinfo {title} {Real-time edge dynamics of non-{Hermitian}
  lattices},}\ \href {\doibase 10.1103/llbb-pcgk} {\bibfield  {journal}
  {\bibinfo  {journal} {Phys. Rev. Lett.}\ }\textbf {\bibinfo {volume} {135}},\
  \bibinfo {pages} {186401} (\bibinfo {year} {2025})}\BibitemShut {NoStop}%
\bibitem [{\citenamefont {Gong}\ \emph {et~al.}(2018)\citenamefont {Gong},
  \citenamefont {Ashida}, \citenamefont {Kawabata}, \citenamefont {Takasan},
  \citenamefont {Higashikawa},\ and\ \citenamefont {Ueda}}]{Gong18}%
  \BibitemOpen
  \bibfield  {author} {\bibinfo {author} {\bibfnamefont {Z.}~\bibnamefont
  {Gong}}, \bibinfo {author} {\bibfnamefont {Y.}~\bibnamefont {Ashida}},
  \bibinfo {author} {\bibfnamefont {K.}~\bibnamefont {Kawabata}}, \bibinfo
  {author} {\bibfnamefont {K.}~\bibnamefont {Takasan}}, \bibinfo {author}
  {\bibfnamefont {S.}~\bibnamefont {Higashikawa}}, \ and\ \bibinfo {author}
  {\bibfnamefont {M.}~\bibnamefont {Ueda}},\ }\enquote {\bibinfo {title}
  {Topological phases of non-{Hermitian} systems},}\ \href {\doibase
  10.1103/PhysRevX.8.031079} {\bibfield  {journal} {\bibinfo  {journal} {Phys.
  Rev. X}\ }\textbf {\bibinfo {volume} {8}},\ \bibinfo {pages} {031079}
  (\bibinfo {year} {2018})}\BibitemShut {NoStop}%
\bibitem [{\citenamefont {Shen}\ \emph {et~al.}(2018)\citenamefont {Shen},
  \citenamefont {Zhen},\ and\ \citenamefont {Fu}}]{Shen18}%
  \BibitemOpen
  \bibfield  {author} {\bibinfo {author} {\bibfnamefont {H.}~\bibnamefont
  {Shen}}, \bibinfo {author} {\bibfnamefont {B.}~\bibnamefont {Zhen}}, \ and\
  \bibinfo {author} {\bibfnamefont {L.}~\bibnamefont {Fu}},\ }\enquote
  {\bibinfo {title} {Topological band theory for non-{Hermitian}
  {Hamiltonians}},}\ \href {\doibase 10.1103/PhysRevLett.120.146402} {\bibfield
   {journal} {\bibinfo  {journal} {Phys. Rev. Lett.}\ }\textbf {\bibinfo
  {volume} {120}},\ \bibinfo {pages} {146402} (\bibinfo {year}
  {2018})}\BibitemShut {NoStop}%
\bibitem [{\citenamefont {Longhi}(2019)}]{Longhi19}%
  \BibitemOpen
  \bibfield  {author} {\bibinfo {author} {\bibfnamefont {S.}~\bibnamefont
  {Longhi}},\ }\enquote {\bibinfo {title} {Probing non-{Hermitian} skin effect
  and non-{Bloch} phase transitions},}\ \href {\doibase
  10.1103/PhysRevResearch.1.023013} {\bibfield  {journal} {\bibinfo  {journal}
  {Phys. Rev. Research}\ }\textbf {\bibinfo {volume} {1}},\ \bibinfo {pages}
  {023013} (\bibinfo {year} {2019})}\BibitemShut {NoStop}%
\bibitem [{\citenamefont {Yokomizo}\ and\ \citenamefont
  {Murakami}(2019)}]{Yok19}%
  \BibitemOpen
  \bibfield  {author} {\bibinfo {author} {\bibfnamefont {K.}~\bibnamefont
  {Yokomizo}}\ and\ \bibinfo {author} {\bibfnamefont {S.}~\bibnamefont
  {Murakami}},\ }\enquote {\bibinfo {title} {Non-{Bloch} band theory of
  non-{Hermitian} systems},}\ \href {\doibase 10.1103/PhysRevLett.123.066404}
  {\bibfield  {journal} {\bibinfo  {journal} {Phys. Rev. Lett.}\ }\textbf
  {\bibinfo {volume} {123}},\ \bibinfo {pages} {066404} (\bibinfo {year}
  {2019})}\BibitemShut {NoStop}%
\bibitem [{\citenamefont {Guo}\ \emph {et~al.}(2021)\citenamefont {Guo},
  \citenamefont {Liu}, \citenamefont {Zhao}, \citenamefont {Liu},\ and\
  \citenamefont {Chen}}]{Guo21}%
  \BibitemOpen
  \bibfield  {author} {\bibinfo {author} {\bibfnamefont {C.-X.}\ \bibnamefont
  {Guo}}, \bibinfo {author} {\bibfnamefont {C.-H.}\ \bibnamefont {Liu}},
  \bibinfo {author} {\bibfnamefont {X.-M.}\ \bibnamefont {Zhao}}, \bibinfo
  {author} {\bibfnamefont {Y.}~\bibnamefont {Liu}}, \ and\ \bibinfo {author}
  {\bibfnamefont {S.}~\bibnamefont {Chen}},\ }\enquote {\bibinfo {title} {Exact
  solution of non-{Hermitian} systems with generalized boundary conditions:
  Size-dependent boundary effect and fragility of the skin effect},}\ \href
  {\doibase 10.1103/PhysRevLett.127.116801} {\bibfield  {journal} {\bibinfo
  {journal} {Phys. Rev. Lett.}\ }\textbf {\bibinfo {volume} {127}},\ \bibinfo
  {pages} {116801} (\bibinfo {year} {2021})}\BibitemShut {NoStop}%
\bibitem [{\citenamefont {Hou}\ \emph {et~al.}(2023)\citenamefont {Hou},
  \citenamefont {Li}, \citenamefont {Wu}, \citenamefont {Ruan}, \citenamefont
  {Chen},\ and\ \citenamefont {Baronio}}]{Hou23}%
  \BibitemOpen
  \bibfield  {author} {\bibinfo {author} {\bibfnamefont {C.}~\bibnamefont
  {Hou}}, \bibinfo {author} {\bibfnamefont {L.}~\bibnamefont {Li}}, \bibinfo
  {author} {\bibfnamefont {G.}~\bibnamefont {Wu}}, \bibinfo {author}
  {\bibfnamefont {Y.}~\bibnamefont {Ruan}}, \bibinfo {author} {\bibfnamefont
  {S.}~\bibnamefont {Chen}}, \ and\ \bibinfo {author} {\bibfnamefont
  {F.}~\bibnamefont {Baronio}},\ }\enquote {\bibinfo {title} {Topological edge
  states in one-dimensional non-{Hermitian} {Su}-{Schrieffer}-{Heeger} systems
  of finite lattice size: {Analytical} solutions and exceptional points},}\
  \href {\doibase 10.1103/PhysRevB.108.085425} {\bibfield  {journal} {\bibinfo
  {journal} {Phys. Rev. B}\ }\textbf {\bibinfo {volume} {108}},\ \bibinfo
  {pages} {085425} (\bibinfo {year} {2023})}\BibitemShut {NoStop}%
\bibitem [{\citenamefont {Lee}(2022)}]{Lee22}%
  \BibitemOpen
  \bibfield  {author} {\bibinfo {author} {\bibfnamefont {C.~H.}\ \bibnamefont
  {Lee}},\ }\enquote {\bibinfo {title} {Exceptional bound states and negative
  entanglement entropy},}\ \href {\doibase 10.1103/PhysRevLett.128.010402}
  {\bibfield  {journal} {\bibinfo  {journal} {Phys. Rev. Lett.}\ }\textbf
  {\bibinfo {volume} {128}},\ \bibinfo {pages} {010402} (\bibinfo {year}
  {2022})}\BibitemShut {NoStop}%
\bibitem [{\citenamefont {Banerjee}\ \emph {et~al.}(2023)\citenamefont
  {Banerjee}, \citenamefont {Sarkar}, \citenamefont {Dey},\ and\ \citenamefont
  {Narayan}}]{Ban2023}%
  \BibitemOpen
  \bibfield  {author} {\bibinfo {author} {\bibfnamefont {A.}~\bibnamefont
  {Banerjee}}, \bibinfo {author} {\bibfnamefont {R.}~\bibnamefont {Sarkar}},
  \bibinfo {author} {\bibfnamefont {S.}~\bibnamefont {Dey}}, \ and\ \bibinfo
  {author} {\bibfnamefont {A.}~\bibnamefont {Narayan}},\ }\enquote {\bibinfo
  {title} {Non-{Hermitian} topological phases: principles and prospects},}\
  \href {\doibase 10.1088/1361-648X/acd1cb} {\bibfield  {journal} {\bibinfo
  {journal} {J. Phys.: Condens. Matter}\ }\textbf {\bibinfo {volume} {35}},\
  \bibinfo {pages} {333001} (\bibinfo {year} {2023})}\BibitemShut {NoStop}%
\bibitem [{\citenamefont {Dai}\ \emph {et~al.}(2024{\natexlab{a}})\citenamefont
  {Dai}, \citenamefont {Ao}, \citenamefont {Mao}, \citenamefont {Yang},
  \citenamefont {Zheng}, \citenamefont {Zhai}, \citenamefont {Li},
  \citenamefont {Yuan}, \citenamefont {Tang}, \citenamefont {Li}, \citenamefont
  {Luo}, \citenamefont {Wang}, \citenamefont {Hu}, \citenamefont {Gong},\ and\
  \citenamefont {Wang}}]{DaiT24}%
  \BibitemOpen
  \bibfield  {author} {\bibinfo {author} {\bibfnamefont {T.}~\bibnamefont
  {Dai}}, \bibinfo {author} {\bibfnamefont {Y.}~\bibnamefont {Ao}}, \bibinfo
  {author} {\bibfnamefont {J.}~\bibnamefont {Mao}}, \bibinfo {author}
  {\bibfnamefont {Y.}~\bibnamefont {Yang}}, \bibinfo {author} {\bibfnamefont
  {Y.}~\bibnamefont {Zheng}}, \bibinfo {author} {\bibfnamefont
  {C.}~\bibnamefont {Zhai}}, \bibinfo {author} {\bibfnamefont {Y.}~\bibnamefont
  {Li}}, \bibinfo {author} {\bibfnamefont {J.}~\bibnamefont {Yuan}}, \bibinfo
  {author} {\bibfnamefont {B.}~\bibnamefont {Tang}}, \bibinfo {author}
  {\bibfnamefont {Z.}~\bibnamefont {Li}}, \bibinfo {author} {\bibfnamefont
  {J.}~\bibnamefont {Luo}}, \bibinfo {author} {\bibfnamefont {W.}~\bibnamefont
  {Wang}}, \bibinfo {author} {\bibfnamefont {X.}~\bibnamefont {Hu}}, \bibinfo
  {author} {\bibfnamefont {Q.}~\bibnamefont {Gong}}, \ and\ \bibinfo {author}
  {\bibfnamefont {J.}~\bibnamefont {Wang}},\ }\enquote {\bibinfo {title}
  {Non-{Hermitian} topological phase transitions controlled by nonlinearity},}\
  \href {\doibase 10.1038/s41567-023-02244-8} {\bibfield  {journal} {\bibinfo
  {journal} {Nat. Phys.}\ }\textbf {\bibinfo {volume} {20}},\ \bibinfo {pages}
  {101} (\bibinfo {year} {2024}{\natexlab{a}})}\BibitemShut {NoStop}%
\bibitem [{\citenamefont {Wang}\ \emph {et~al.}(2024)\citenamefont {Wang},
  \citenamefont {Song},\ and\ \citenamefont {Wang}}]{Wang24}%
  \BibitemOpen
  \bibfield  {author} {\bibinfo {author} {\bibfnamefont {H.-Y.}\ \bibnamefont
  {Wang}}, \bibinfo {author} {\bibfnamefont {F.}~\bibnamefont {Song}}, \ and\
  \bibinfo {author} {\bibfnamefont {Z.}~\bibnamefont {Wang}},\ }\enquote
  {\bibinfo {title} {Amoeba formulation of non-{Bloch} band theory in arbitrary
  dimensions},}\ \href {\doibase 10.1103/PhysRevX.14.021011} {\bibfield
  {journal} {\bibinfo  {journal} {Phys. Rev. X}\ }\textbf {\bibinfo {volume}
  {14}},\ \bibinfo {pages} {021011} (\bibinfo {year} {2024})}\BibitemShut
  {NoStop}%
\bibitem [{\citenamefont {Zhen}\ \emph {et~al.}(2015)\citenamefont {Zhen},
  \citenamefont {Hsu}, \citenamefont {Igarashi}, \citenamefont {Lu},
  \citenamefont {Kaminer}, \citenamefont {Pick}, \citenamefont {Chua},
  \citenamefont {Joannopoulos},\ and\ \citenamefont
  {Solja{\v{c}}i{\'c}}}]{Zhen2015}%
  \BibitemOpen
  \bibfield  {author} {\bibinfo {author} {\bibfnamefont {B.}~\bibnamefont
  {Zhen}}, \bibinfo {author} {\bibfnamefont {C.~W.}\ \bibnamefont {Hsu}},
  \bibinfo {author} {\bibfnamefont {Y.}~\bibnamefont {Igarashi}}, \bibinfo
  {author} {\bibfnamefont {L.}~\bibnamefont {Lu}}, \bibinfo {author}
  {\bibfnamefont {I.}~\bibnamefont {Kaminer}}, \bibinfo {author} {\bibfnamefont
  {A.}~\bibnamefont {Pick}}, \bibinfo {author} {\bibfnamefont {S.-L.}\
  \bibnamefont {Chua}}, \bibinfo {author} {\bibfnamefont {J.~D.}\ \bibnamefont
  {Joannopoulos}}, \ and\ \bibinfo {author} {\bibfnamefont {M.}~\bibnamefont
  {Solja{\v{c}}i{\'c}}},\ }\enquote {\bibinfo {title} {Spawning rings of
  exceptional points out of {Dirac} cones},}\ \href
  {https://doi.org/10.1038/nature14889} {\bibfield  {journal} {\bibinfo
  {journal} {Nature}\ }\textbf {\bibinfo {volume} {525}},\ \bibinfo {pages}
  {354} (\bibinfo {year} {2015})}\BibitemShut {NoStop}%
\bibitem [{\citenamefont {Ding}\ \emph {et~al.}(2016)\citenamefont {Ding},
  \citenamefont {Ma}, \citenamefont {Xiao}, \citenamefont {Zhang},\ and\
  \citenamefont {Chan}}]{Ding16}%
  \BibitemOpen
  \bibfield  {author} {\bibinfo {author} {\bibfnamefont {K.}~\bibnamefont
  {Ding}}, \bibinfo {author} {\bibfnamefont {G.}~\bibnamefont {Ma}}, \bibinfo
  {author} {\bibfnamefont {M.}~\bibnamefont {Xiao}}, \bibinfo {author}
  {\bibfnamefont {Z.~Q.}\ \bibnamefont {Zhang}}, \ and\ \bibinfo {author}
  {\bibfnamefont {C.~T.}\ \bibnamefont {Chan}},\ }\enquote {\bibinfo {title}
  {Emergence, coalescence, and topological properties of multiple exceptional
  points and their experimental realization},}\ \href {\doibase
  10.1103/PhysRevX.6.021007} {\bibfield  {journal} {\bibinfo  {journal} {Phys.
  Rev. X}\ }\textbf {\bibinfo {volume} {6}},\ \bibinfo {pages} {021007}
  (\bibinfo {year} {2016})}\BibitemShut {NoStop}%
\bibitem [{\citenamefont {Peng}\ \emph {et~al.}(2016)\citenamefont {Peng},
  \citenamefont {\"{O}zdemir}, \citenamefont {Liertzer}, \citenamefont {Chen},
  \citenamefont {Kramer}, \citenamefont {Y{\i}lmaz}, \citenamefont {Wiersig},
  \citenamefont {Rotter},\ and\ \citenamefont {Yang}}]{Peng16}%
  \BibitemOpen
  \bibfield  {author} {\bibinfo {author} {\bibfnamefont {B.}~\bibnamefont
  {Peng}}, \bibinfo {author} {\bibfnamefont {{\c{S}}.~K.}\ \bibnamefont
  {\"{O}zdemir}}, \bibinfo {author} {\bibfnamefont {M.}~\bibnamefont
  {Liertzer}}, \bibinfo {author} {\bibfnamefont {W.}~\bibnamefont {Chen}},
  \bibinfo {author} {\bibfnamefont {J.}~\bibnamefont {Kramer}}, \bibinfo
  {author} {\bibfnamefont {H.}~\bibnamefont {Y{\i}lmaz}}, \bibinfo {author}
  {\bibfnamefont {J.}~\bibnamefont {Wiersig}}, \bibinfo {author} {\bibfnamefont
  {S.}~\bibnamefont {Rotter}}, \ and\ \bibinfo {author} {\bibfnamefont
  {L.}~\bibnamefont {Yang}},\ }\enquote {\bibinfo {title} {Chiral modes and
  directional lasing at exceptional points},}\ \href {\doibase
  10.1073/pnas.1603318113} {\bibfield  {journal} {\bibinfo  {journal} {Proc.
  Natl. Acad. Sci. USA}\ }\textbf {\bibinfo {volume} {113}},\ \bibinfo {pages}
  {6845} (\bibinfo {year} {2016})}\BibitemShut {NoStop}%
\bibitem [{\citenamefont {Chen}\ \emph {et~al.}(2017)\citenamefont {Chen},
  \citenamefont {\"{O}zdemir}, \citenamefont {Zhao}, \citenamefont {Wiersig},\
  and\ \citenamefont {Yang}}]{Chen17}%
  \BibitemOpen
  \bibfield  {author} {\bibinfo {author} {\bibfnamefont {W.}~\bibnamefont
  {Chen}}, \bibinfo {author} {\bibfnamefont {{\c{S}}.~K.}\ \bibnamefont
  {\"{O}zdemir}}, \bibinfo {author} {\bibfnamefont {G.}~\bibnamefont {Zhao}},
  \bibinfo {author} {\bibfnamefont {J.}~\bibnamefont {Wiersig}}, \ and\
  \bibinfo {author} {\bibfnamefont {L.}~\bibnamefont {Yang}},\ }\enquote
  {\bibinfo {title} {Exceptional points enhance sensing in an optical
  microcavity},}\ \href {\doibase 10.1038/nature23281} {\bibfield  {journal}
  {\bibinfo  {journal} {Nature}\ }\textbf {\bibinfo {volume} {548}},\ \bibinfo
  {pages} {192} (\bibinfo {year} {2017})}\BibitemShut {NoStop}%
\bibitem [{\citenamefont {{\"O}zdemir}\ \emph {et~al.}(2019)\citenamefont
  {{\"O}zdemir}, \citenamefont {Rotter}, \citenamefont {Nori},\ and\
  \citenamefont {Yang}}]{Ozd19}%
  \BibitemOpen
  \bibfield  {author} {\bibinfo {author} {\bibfnamefont {{\c{S}}.~K.}\
  \bibnamefont {{\"O}zdemir}}, \bibinfo {author} {\bibfnamefont
  {S.}~\bibnamefont {Rotter}}, \bibinfo {author} {\bibfnamefont
  {F.}~\bibnamefont {Nori}}, \ and\ \bibinfo {author} {\bibfnamefont
  {L.}~\bibnamefont {Yang}},\ }\enquote {\bibinfo {title} {Parity-time symmetry
  and exceptional points in photonics},}\ \href {\doibase
  10.1038/s41563-019-0304-9} {\bibfield  {journal} {\bibinfo  {journal} {Nat.
  Mater.}\ }\textbf {\bibinfo {volume} {18}},\ \bibinfo {pages} {783} (\bibinfo
  {year} {2019})}\BibitemShut {NoStop}%
\bibitem [{\citenamefont {Miri}\ and\ \citenamefont {Al\`u}(2019)}]{Ali2019}%
  \BibitemOpen
  \bibfield  {author} {\bibinfo {author} {\bibfnamefont {M.-A.}\ \bibnamefont
  {Miri}}\ and\ \bibinfo {author} {\bibfnamefont {A.}~\bibnamefont {Al\`u}},\
  }\enquote {\bibinfo {title} {Exceptional points in optics and photonics},}\
  \href {\doibase 10.1126/science.aar7709} {\bibfield  {journal} {\bibinfo
  {journal} {Science}\ }\textbf {\bibinfo {volume} {363}},\ \bibinfo {pages}
  {eaar7709} (\bibinfo {year} {2019})}\BibitemShut {NoStop}%
\bibitem [{\citenamefont {Kawabata}\ \emph
  {et~al.}(2019{\natexlab{b}})\citenamefont {Kawabata}, \citenamefont
  {Bessho},\ and\ \citenamefont {Sato}}]{Kaw19}%
  \BibitemOpen
  \bibfield  {author} {\bibinfo {author} {\bibfnamefont {K.}~\bibnamefont
  {Kawabata}}, \bibinfo {author} {\bibfnamefont {T.}~\bibnamefont {Bessho}}, \
  and\ \bibinfo {author} {\bibfnamefont {M.}~\bibnamefont {Sato}},\ }\enquote
  {\bibinfo {title} {Classification of exceptional points and non-{Hermitian}
  topological semimetals},}\ \href {\doibase 10.1103/PhysRevLett.123.066405}
  {\bibfield  {journal} {\bibinfo  {journal} {Phys. Rev. Lett.}\ }\textbf
  {\bibinfo {volume} {123}},\ \bibinfo {pages} {066405} (\bibinfo {year}
  {2019}{\natexlab{b}})}\BibitemShut {NoStop}%
\bibitem [{\citenamefont {Zhao}\ \emph {et~al.}(2025)\citenamefont {Zhao},
  \citenamefont {Wang}, \citenamefont {Liu}, \citenamefont {Shi},\ and\
  \citenamefont {Zi}}]{Zhao25}%
  \BibitemOpen
  \bibfield  {author} {\bibinfo {author} {\bibfnamefont {X.}~\bibnamefont
  {Zhao}}, \bibinfo {author} {\bibfnamefont {J.}~\bibnamefont {Wang}}, \bibinfo
  {author} {\bibfnamefont {W.}~\bibnamefont {Liu}}, \bibinfo {author}
  {\bibfnamefont {L.}~\bibnamefont {Shi}}, \ and\ \bibinfo {author}
  {\bibfnamefont {J.}~\bibnamefont {Zi}},\ }\enquote {\bibinfo {title}
  {Magnetically induced topological evolutions of exceptional points in
  photonic bands},}\ \href {\doibase 10.1103/wv2n-51qg} {\bibfield  {journal}
  {\bibinfo  {journal} {Phys. Rev. Lett.}\ }\textbf {\bibinfo {volume} {135}},\
  \bibinfo {pages} {046203} (\bibinfo {year} {2025})}\BibitemShut {NoStop}%
\bibitem [{\citenamefont {Kullig}\ \emph {et~al.}(2025)\citenamefont {Kullig},
  \citenamefont {Zhong}, \citenamefont {Wiersig},\ and\ \citenamefont
  {El-Ganainy}}]{Kul25}%
  \BibitemOpen
  \bibfield  {author} {\bibinfo {author} {\bibfnamefont {J.}~\bibnamefont
  {Kullig}}, \bibinfo {author} {\bibfnamefont {Q.}~\bibnamefont {Zhong}},
  \bibinfo {author} {\bibfnamefont {J.}~\bibnamefont {Wiersig}}, \ and\
  \bibinfo {author} {\bibfnamefont {R.}~\bibnamefont {El-Ganainy}},\ }\enquote
  {\bibinfo {title} {Exceptional points and lasing thresholds: When lower-{Q}
  modes win},}\ \href {\doibase 10.1103/zc6x-pfw2} {\bibfield  {journal}
  {\bibinfo  {journal} {Phys. Rev. Lett.}\ }\textbf {\bibinfo {volume} {135}},\
  \bibinfo {pages} {173802} (\bibinfo {year} {2025})}\BibitemShut {NoStop}%
\bibitem [{\citenamefont {Meng}\ \emph {et~al.}(2024)\citenamefont {Meng},
  \citenamefont {Ang},\ and\ \citenamefont {Lee}}]{Meng24}%
  \BibitemOpen
  \bibfield  {author} {\bibinfo {author} {\bibfnamefont {H.}~\bibnamefont
  {Meng}}, \bibinfo {author} {\bibfnamefont {Y.~S.}\ \bibnamefont {Ang}}, \
  and\ \bibinfo {author} {\bibfnamefont {C.~H.}\ \bibnamefont {Lee}},\
  }\enquote {\bibinfo {title} {Exceptional points in non-{Hermitian} systems:
  Applications and recent developments},}\ \href {\doibase 10.1063/5.0183826}
  {\bibfield  {journal} {\bibinfo  {journal} {Appl. Phys. Lett.}\ }\textbf
  {\bibinfo {volume} {124}},\ \bibinfo {pages} {060502} (\bibinfo {year}
  {2024})}\BibitemShut {NoStop}%
\bibitem [{\citenamefont {Arouca}\ \emph {et~al.}(2020)\citenamefont {Arouca},
  \citenamefont {Lee},\ and\ \citenamefont {Morais~Smith}}]{Arouca20}%
  \BibitemOpen
  \bibfield  {author} {\bibinfo {author} {\bibfnamefont {R.}~\bibnamefont
  {Arouca}}, \bibinfo {author} {\bibfnamefont {C.~H.}\ \bibnamefont {Lee}}, \
  and\ \bibinfo {author} {\bibfnamefont {C.}~\bibnamefont {Morais~Smith}},\
  }\enquote {\bibinfo {title} {Unconventional scaling at non-{Hermitian}
  critical points},}\ \href {\doibase 10.1103/PhysRevB.102.245145} {\bibfield
  {journal} {\bibinfo  {journal} {Phys. Rev. B}\ }\textbf {\bibinfo {volume}
  {102}},\ \bibinfo {pages} {245145} (\bibinfo {year} {2020})}\BibitemShut
  {NoStop}%
\bibitem [{\citenamefont {Xue}\ and\ \citenamefont {Lee}(2026)}]{Xue26}%
  \BibitemOpen
  \bibfield  {author} {\bibinfo {author} {\bibfnamefont {W.-T.}\ \bibnamefont
  {Xue}}\ and\ \bibinfo {author} {\bibfnamefont {C.~H.}\ \bibnamefont {Lee}},\
  }\enquote {\bibinfo {title} {Topologically protected negative
  entanglement},}\ \href {\doibase https://doi.org/10.1002/advs.202513868}
  {\bibfield  {journal} {\bibinfo  {journal} {Adv. Sci.}\ }\textbf {\bibinfo
  {volume} {13}},\ \bibinfo {pages} {e13868} (\bibinfo {year}
  {2026})}\BibitemShut {NoStop}%
\bibitem [{\citenamefont {Liu}\ \emph {et~al.}(2025)\citenamefont {Liu},
  \citenamefont {Jiang}, \citenamefont {Xue}, \citenamefont {Li}, \citenamefont
  {Gong}, \citenamefont {Liu},\ and\ \citenamefont {Lee}}]{Liu25}%
  \BibitemOpen
  \bibfield  {author} {\bibinfo {author} {\bibfnamefont {S.}~\bibnamefont
  {Liu}}, \bibinfo {author} {\bibfnamefont {H.}~\bibnamefont {Jiang}}, \bibinfo
  {author} {\bibfnamefont {W.-T.}\ \bibnamefont {Xue}}, \bibinfo {author}
  {\bibfnamefont {Q.}~\bibnamefont {Li}}, \bibinfo {author} {\bibfnamefont
  {J.}~\bibnamefont {Gong}}, \bibinfo {author} {\bibfnamefont {X.}~\bibnamefont
  {Liu}}, \ and\ \bibinfo {author} {\bibfnamefont {C.~H.}\ \bibnamefont
  {Lee}},\ }\enquote {\bibinfo {title} {Non-{Hermitian} entanglement dip from
  scaling-induced exceptional criticality},}\ \href {\doibase
  https://doi.org/10.1016/j.scib.2025.07.011} {\bibfield  {journal} {\bibinfo
  {journal} {Sci. Bull.}\ }\textbf {\bibinfo {volume} {70}},\ \bibinfo {pages}
  {2929} (\bibinfo {year} {2025})}\BibitemShut {NoStop}%
\bibitem [{\citenamefont {Kunst}\ \emph {et~al.}(2018)\citenamefont {Kunst},
  \citenamefont {Edvardsson}, \citenamefont {Budich},\ and\ \citenamefont
  {Bergholtz}}]{Kun18}%
  \BibitemOpen
  \bibfield  {author} {\bibinfo {author} {\bibfnamefont {F.~K.}\ \bibnamefont
  {Kunst}}, \bibinfo {author} {\bibfnamefont {E.}~\bibnamefont {Edvardsson}},
  \bibinfo {author} {\bibfnamefont {J.~C.}\ \bibnamefont {Budich}}, \ and\
  \bibinfo {author} {\bibfnamefont {E.~J.}\ \bibnamefont {Bergholtz}},\
  }\enquote {\bibinfo {title} {Biorthogonal bulk-boundary correspondence in
  non-{Hermitian} systems},}\ \href {\doibase 10.1103/PhysRevLett.121.026808}
  {\bibfield  {journal} {\bibinfo  {journal} {Phys. Rev. Lett.}\ }\textbf
  {\bibinfo {volume} {121}},\ \bibinfo {pages} {026808} (\bibinfo {year}
  {2018})}\BibitemShut {NoStop}%
\bibitem [{\citenamefont {Zirnstein}\ \emph {et~al.}(2021)\citenamefont
  {Zirnstein}, \citenamefont {Refael},\ and\ \citenamefont {Rosenow}}]{Zirn21}%
  \BibitemOpen
  \bibfield  {author} {\bibinfo {author} {\bibfnamefont {H.-G.}\ \bibnamefont
  {Zirnstein}}, \bibinfo {author} {\bibfnamefont {G.}~\bibnamefont {Refael}}, \
  and\ \bibinfo {author} {\bibfnamefont {B.}~\bibnamefont {Rosenow}},\
  }\enquote {\bibinfo {title} {Bulk-boundary correspondence for non-{Hermitian}
  {Hamiltonians} via {Green} functions},}\ \href {\doibase
  10.1103/PhysRevLett.126.216407} {\bibfield  {journal} {\bibinfo  {journal}
  {Phys. Rev. Lett.}\ }\textbf {\bibinfo {volume} {126}},\ \bibinfo {pages}
  {216407} (\bibinfo {year} {2021})}\BibitemShut {NoStop}%
\bibitem [{\citenamefont {Zhu}\ \emph {et~al.}(2020)\citenamefont {Zhu},
  \citenamefont {Wang}, \citenamefont {Gupta}, \citenamefont {Zhang},
  \citenamefont {Xie}, \citenamefont {Lu},\ and\ \citenamefont
  {Chen}}]{Zhu2020}%
  \BibitemOpen
  \bibfield  {author} {\bibinfo {author} {\bibfnamefont {X.}~\bibnamefont
  {Zhu}}, \bibinfo {author} {\bibfnamefont {H.}~\bibnamefont {Wang}}, \bibinfo
  {author} {\bibfnamefont {S.~K.}\ \bibnamefont {Gupta}}, \bibinfo {author}
  {\bibfnamefont {H.}~\bibnamefont {Zhang}}, \bibinfo {author} {\bibfnamefont
  {B.}~\bibnamefont {Xie}}, \bibinfo {author} {\bibfnamefont {M.}~\bibnamefont
  {Lu}}, \ and\ \bibinfo {author} {\bibfnamefont {Y.}~\bibnamefont {Chen}},\
  }\enquote {\bibinfo {title} {Photonic non-{Hermitian} skin effect and
  non-{Bloch} bulk-boundary correspondence},}\ \href {\doibase
  10.1103/PhysRevResearch.2.013280} {\bibfield  {journal} {\bibinfo  {journal}
  {Phys. Rev. Research}\ }\textbf {\bibinfo {volume} {2}},\ \bibinfo {pages}
  {013280} (\bibinfo {year} {2020})}\BibitemShut {NoStop}%
\bibitem [{\citenamefont {Yang}\ \emph {et~al.}(2020)\citenamefont {Yang},
  \citenamefont {Zhang}, \citenamefont {Fang},\ and\ \citenamefont
  {Hu}}]{Yang20}%
  \BibitemOpen
  \bibfield  {author} {\bibinfo {author} {\bibfnamefont {Z.}~\bibnamefont
  {Yang}}, \bibinfo {author} {\bibfnamefont {K.}~\bibnamefont {Zhang}},
  \bibinfo {author} {\bibfnamefont {C.}~\bibnamefont {Fang}}, \ and\ \bibinfo
  {author} {\bibfnamefont {J.}~\bibnamefont {Hu}},\ }\enquote {\bibinfo {title}
  {Non-{Hermitian} bulk-boundary correspondence and auxiliary generalized
  {Brillouin} zone theory},}\ \href {\doibase 10.1103/PhysRevLett.125.226402}
  {\bibfield  {journal} {\bibinfo  {journal} {Phys. Rev. Lett.}\ }\textbf
  {\bibinfo {volume} {125}},\ \bibinfo {pages} {226402} (\bibinfo {year}
  {2020})}\BibitemShut {NoStop}%
\bibitem [{\citenamefont {Helbig}\ \emph {et~al.}(2020)\citenamefont {Helbig},
  \citenamefont {Hofmann}, \citenamefont {Imhof}, \citenamefont {Abdelghany},
  \citenamefont {Kiessling}, \citenamefont {Molenkamp}, \citenamefont {Lee},
  \citenamefont {Szameit}, \citenamefont {Greiter},\ and\ \citenamefont
  {Thomale}}]{Hel20}%
  \BibitemOpen
  \bibfield  {author} {\bibinfo {author} {\bibfnamefont {T.}~\bibnamefont
  {Helbig}}, \bibinfo {author} {\bibfnamefont {T.}~\bibnamefont {Hofmann}},
  \bibinfo {author} {\bibfnamefont {S.}~\bibnamefont {Imhof}}, \bibinfo
  {author} {\bibfnamefont {M.}~\bibnamefont {Abdelghany}}, \bibinfo {author}
  {\bibfnamefont {T.}~\bibnamefont {Kiessling}}, \bibinfo {author}
  {\bibfnamefont {L.~W.}\ \bibnamefont {Molenkamp}}, \bibinfo {author}
  {\bibfnamefont {C.~H.}\ \bibnamefont {Lee}}, \bibinfo {author} {\bibfnamefont
  {A.}~\bibnamefont {Szameit}}, \bibinfo {author} {\bibfnamefont
  {M.}~\bibnamefont {Greiter}}, \ and\ \bibinfo {author} {\bibfnamefont
  {R.}~\bibnamefont {Thomale}},\ }\enquote {\bibinfo {title} {Generalized
  bulk-boundary correspondence in non-{Hermitian} topolectrical circuits},}\
  \href {https://doi.org/10.1038/s41567-020-0922-9} {\bibfield  {journal}
  {\bibinfo  {journal} {Nat. Phys.}\ }\textbf {\bibinfo {volume} {16}},\
  \bibinfo {pages} {747} (\bibinfo {year} {2020})}\BibitemShut {NoStop}%
\bibitem [{\citenamefont {Hou}\ \emph {et~al.}(2022)\citenamefont {Hou},
  \citenamefont {Li}, \citenamefont {Chen}, \citenamefont {Liu}, \citenamefont
  {Yuan}, \citenamefont {Zhang},\ and\ \citenamefont {Ni}}]{Hou22}%
  \BibitemOpen
  \bibfield  {author} {\bibinfo {author} {\bibfnamefont {C.}~\bibnamefont
  {Hou}}, \bibinfo {author} {\bibfnamefont {L.}~\bibnamefont {Li}}, \bibinfo
  {author} {\bibfnamefont {S.}~\bibnamefont {Chen}}, \bibinfo {author}
  {\bibfnamefont {Y.}~\bibnamefont {Liu}}, \bibinfo {author} {\bibfnamefont
  {L.}~\bibnamefont {Yuan}}, \bibinfo {author} {\bibfnamefont {Y.}~\bibnamefont
  {Zhang}}, \ and\ \bibinfo {author} {\bibfnamefont {Z.}~\bibnamefont {Ni}},\
  }\enquote {\bibinfo {title} {Deterministic bulk-boundary correspondences for
  skin and edge modes in a general two-band non-{Hermitian} system},}\ \href
  {\doibase 10.1103/PhysRevResearch.4.043222} {\bibfield  {journal} {\bibinfo
  {journal} {Phys. Rev. Research}\ }\textbf {\bibinfo {volume} {4}},\ \bibinfo
  {pages} {043222} (\bibinfo {year} {2022})}\BibitemShut {NoStop}%
\bibitem [{\citenamefont {Xiao}\ \emph {et~al.}(2020)\citenamefont {Xiao},
  \citenamefont {Deng}, \citenamefont {Wang}, \citenamefont {Zhu},
  \citenamefont {Wang}, \citenamefont {Yi},\ and\ \citenamefont
  {Xue}}]{Xiao20}%
  \BibitemOpen
  \bibfield  {author} {\bibinfo {author} {\bibfnamefont {L.}~\bibnamefont
  {Xiao}}, \bibinfo {author} {\bibfnamefont {T.}~\bibnamefont {Deng}}, \bibinfo
  {author} {\bibfnamefont {K.}~\bibnamefont {Wang}}, \bibinfo {author}
  {\bibfnamefont {G.}~\bibnamefont {Zhu}}, \bibinfo {author} {\bibfnamefont
  {Z.}~\bibnamefont {Wang}}, \bibinfo {author} {\bibfnamefont {W.}~\bibnamefont
  {Yi}}, \ and\ \bibinfo {author} {\bibfnamefont {P.}~\bibnamefont {Xue}},\
  }\enquote {\bibinfo {title} {Non-{Hermitian} bulk-boundary correspondence in
  quantum dynamics},}\ \href {https://doi.org/10.1038/s41567-020-0836-6}
  {\bibfield  {journal} {\bibinfo  {journal} {Nat. Phys.}\ }\textbf {\bibinfo
  {volume} {16}},\ \bibinfo {pages} {761} (\bibinfo {year} {2020})}\BibitemShut
  {NoStop}%
\bibitem [{\citenamefont {Nakamura}\ \emph {et~al.}(2024)\citenamefont
  {Nakamura}, \citenamefont {Bessho},\ and\ \citenamefont {Sato}}]{Naka24}%
  \BibitemOpen
  \bibfield  {author} {\bibinfo {author} {\bibfnamefont {D.}~\bibnamefont
  {Nakamura}}, \bibinfo {author} {\bibfnamefont {T.}~\bibnamefont {Bessho}}, \
  and\ \bibinfo {author} {\bibfnamefont {M.}~\bibnamefont {Sato}},\ }\enquote
  {\bibinfo {title} {Bulk-boundary correspondence in point-gap topological
  phases},}\ \href {\doibase 10.1103/PhysRevLett.132.136401} {\bibfield
  {journal} {\bibinfo  {journal} {Phys. Rev. Lett.}\ }\textbf {\bibinfo
  {volume} {132}},\ \bibinfo {pages} {136401} (\bibinfo {year}
  {2024})}\BibitemShut {NoStop}%
\bibitem [{\citenamefont {Verma}\ and\ \citenamefont {Park}(2024)}]{Verma24}%
  \BibitemOpen
  \bibfield  {author} {\bibinfo {author} {\bibfnamefont {S.}~\bibnamefont
  {Verma}}\ and\ \bibinfo {author} {\bibfnamefont {M.~J.}\ \bibnamefont
  {Park}},\ }\enquote {\bibinfo {title} {Topological phase transitions of
  generalized {Brillouin} zone},}\ \href {\doibase 10.1038/s42005-024-01519-7}
  {\bibfield  {journal} {\bibinfo  {journal} {Commun. Phys.}\ }\textbf
  {\bibinfo {volume} {7}},\ \bibinfo {pages} {21} (\bibinfo {year}
  {2024})}\BibitemShut {NoStop}%
\bibitem [{\citenamefont {Li}\ \emph {et~al.}(2019{\natexlab{a}})\citenamefont
  {Li}, \citenamefont {Harter}, \citenamefont {Liu}, \citenamefont {de~Melo},
  \citenamefont {Joglekar},\ and\ \citenamefont {Luo}}]{Li2019}%
  \BibitemOpen
  \bibfield  {author} {\bibinfo {author} {\bibfnamefont {J.}~\bibnamefont
  {Li}}, \bibinfo {author} {\bibfnamefont {A.~K.}\ \bibnamefont {Harter}},
  \bibinfo {author} {\bibfnamefont {J.}~\bibnamefont {Liu}}, \bibinfo {author}
  {\bibfnamefont {L.}~\bibnamefont {de~Melo}}, \bibinfo {author} {\bibfnamefont
  {Y.~N.}\ \bibnamefont {Joglekar}}, \ and\ \bibinfo {author} {\bibfnamefont
  {L.}~\bibnamefont {Luo}},\ }\enquote {\bibinfo {title} {Observation of
  parity-time symmetry breaking transitions in a dissipative {Floquet} system
  of ultracold atoms},}\ \href {\doibase 10.1038/s41467-019-08596-1} {\bibfield
   {journal} {\bibinfo  {journal} {Nat. Commun.}\ }\textbf {\bibinfo {volume}
  {10}},\ \bibinfo {pages} {855} (\bibinfo {year}
  {2019}{\natexlab{a}})}\BibitemShut {NoStop}%
\bibitem [{\citenamefont {Liang}\ \emph {et~al.}(2022)\citenamefont {Liang},
  \citenamefont {Xie}, \citenamefont {Dong}, \citenamefont {Li}, \citenamefont
  {Li}, \citenamefont {Gadway}, \citenamefont {Yi},\ and\ \citenamefont
  {Yan}}]{Liang22}%
  \BibitemOpen
  \bibfield  {author} {\bibinfo {author} {\bibfnamefont {Q.}~\bibnamefont
  {Liang}}, \bibinfo {author} {\bibfnamefont {D.}~\bibnamefont {Xie}}, \bibinfo
  {author} {\bibfnamefont {Z.}~\bibnamefont {Dong}}, \bibinfo {author}
  {\bibfnamefont {H.}~\bibnamefont {Li}}, \bibinfo {author} {\bibfnamefont
  {H.}~\bibnamefont {Li}}, \bibinfo {author} {\bibfnamefont {B.}~\bibnamefont
  {Gadway}}, \bibinfo {author} {\bibfnamefont {W.}~\bibnamefont {Yi}}, \ and\
  \bibinfo {author} {\bibfnamefont {B.}~\bibnamefont {Yan}},\ }\enquote
  {\bibinfo {title} {Dynamic signatures of non-{Hermitian} skin effect and
  topology in ultracold atoms},}\ \href {\doibase
  10.1103/PhysRevLett.129.070401} {\bibfield  {journal} {\bibinfo  {journal}
  {Phys. Rev. Lett.}\ }\textbf {\bibinfo {volume} {129}},\ \bibinfo {pages}
  {070401} (\bibinfo {year} {2022})}\BibitemShut {NoStop}%
\bibitem [{\citenamefont {Zhou}\ \emph {et~al.}(2022)\citenamefont {Zhou},
  \citenamefont {Li}, \citenamefont {Yi},\ and\ \citenamefont {Cui}}]{Zhou22}%
  \BibitemOpen
  \bibfield  {author} {\bibinfo {author} {\bibfnamefont {L.}~\bibnamefont
  {Zhou}}, \bibinfo {author} {\bibfnamefont {H.}~\bibnamefont {Li}}, \bibinfo
  {author} {\bibfnamefont {W.}~\bibnamefont {Yi}}, \ and\ \bibinfo {author}
  {\bibfnamefont {X.}~\bibnamefont {Cui}},\ }\enquote {\bibinfo {title}
  {Engineering non-{Hermitian} skin effect with band topology in ultracold
  gases},}\ \href {\doibase 10.1038/s42005-022-01021-y} {\bibfield  {journal}
  {\bibinfo  {journal} {Commun. Phys.}\ }\textbf {\bibinfo {volume} {5}},\
  \bibinfo {pages} {252} (\bibinfo {year} {2022})}\BibitemShut {NoStop}%
\bibitem [{\citenamefont {Shen}\ \emph {et~al.}(2023)\citenamefont {Shen},
  \citenamefont {Chen}, \citenamefont {Aliyu}, \citenamefont {Qin},
  \citenamefont {Zhong}, \citenamefont {Loh},\ and\ \citenamefont
  {Lee}}]{Shen23}%
  \BibitemOpen
  \bibfield  {author} {\bibinfo {author} {\bibfnamefont {R.}~\bibnamefont
  {Shen}}, \bibinfo {author} {\bibfnamefont {T.}~\bibnamefont {Chen}}, \bibinfo
  {author} {\bibfnamefont {M.~M.}\ \bibnamefont {Aliyu}}, \bibinfo {author}
  {\bibfnamefont {F.}~\bibnamefont {Qin}}, \bibinfo {author} {\bibfnamefont
  {Y.}~\bibnamefont {Zhong}}, \bibinfo {author} {\bibfnamefont
  {H.}~\bibnamefont {Loh}}, \ and\ \bibinfo {author} {\bibfnamefont {C.~H.}\
  \bibnamefont {Lee}},\ }\enquote {\bibinfo {title} {Proposal for observing
  {Yang}-{Lee} criticality in {Rydberg} atomic arrays},}\ \href {\doibase
  10.1103/PhysRevLett.131.080403} {\bibfield  {journal} {\bibinfo  {journal}
  {Phys. Rev. Lett.}\ }\textbf {\bibinfo {volume} {131}},\ \bibinfo {pages}
  {080403} (\bibinfo {year} {2023})}\BibitemShut {NoStop}%
\bibitem [{\citenamefont {Lee}\ \emph {et~al.}(2018)\citenamefont {Lee},
  \citenamefont {Imhof}, \citenamefont {Berger}, \citenamefont {Bayer},
  \citenamefont {Brehm}, \citenamefont {Molenkamp}, \citenamefont {Kiessling},\
  and\ \citenamefont {Thomale}}]{Lee18}%
  \BibitemOpen
  \bibfield  {author} {\bibinfo {author} {\bibfnamefont {C.~H.}\ \bibnamefont
  {Lee}}, \bibinfo {author} {\bibfnamefont {S.}~\bibnamefont {Imhof}}, \bibinfo
  {author} {\bibfnamefont {C.}~\bibnamefont {Berger}}, \bibinfo {author}
  {\bibfnamefont {F.}~\bibnamefont {Bayer}}, \bibinfo {author} {\bibfnamefont
  {J.}~\bibnamefont {Brehm}}, \bibinfo {author} {\bibfnamefont {L.~W.}\
  \bibnamefont {Molenkamp}}, \bibinfo {author} {\bibfnamefont {T.}~\bibnamefont
  {Kiessling}}, \ and\ \bibinfo {author} {\bibfnamefont {R.}~\bibnamefont
  {Thomale}},\ }\enquote {\bibinfo {title} {Topolectrical circuits},}\ \href
  {\doibase 10.1038/s42005-018-0035-2} {\bibfield  {journal} {\bibinfo
  {journal} {Commun. Phys.}\ }\textbf {\bibinfo {volume} {1}},\ \bibinfo
  {pages} {39} (\bibinfo {year} {2018})}\BibitemShut {NoStop}%
\bibitem [{\citenamefont {Stegmaier}\ \emph {et~al.}(2021)\citenamefont
  {Stegmaier}, \citenamefont {Imhof}, \citenamefont {Helbig}, \citenamefont
  {Hofmann}, \citenamefont {Lee}, \citenamefont {Kremer}, \citenamefont
  {Fritzsche}, \citenamefont {Feichtner}, \citenamefont {Klembt}, \citenamefont
  {H\"ofling}, \citenamefont {Boettcher}, \citenamefont {Fulga}, \citenamefont
  {Ma}, \citenamefont {Schmidt}, \citenamefont {Greiter}, \citenamefont
  {Kiessling}, \citenamefont {Szameit},\ and\ \citenamefont {Thomale}}]{Ste21}%
  \BibitemOpen
  \bibfield  {author} {\bibinfo {author} {\bibfnamefont {A.}~\bibnamefont
  {Stegmaier}}, \bibinfo {author} {\bibfnamefont {S.}~\bibnamefont {Imhof}},
  \bibinfo {author} {\bibfnamefont {T.}~\bibnamefont {Helbig}}, \bibinfo
  {author} {\bibfnamefont {T.}~\bibnamefont {Hofmann}}, \bibinfo {author}
  {\bibfnamefont {C.~H.}\ \bibnamefont {Lee}}, \bibinfo {author} {\bibfnamefont
  {M.}~\bibnamefont {Kremer}}, \bibinfo {author} {\bibfnamefont
  {A.}~\bibnamefont {Fritzsche}}, \bibinfo {author} {\bibfnamefont
  {T.}~\bibnamefont {Feichtner}}, \bibinfo {author} {\bibfnamefont
  {S.}~\bibnamefont {Klembt}}, \bibinfo {author} {\bibfnamefont
  {S.}~\bibnamefont {H\"ofling}}, \bibinfo {author} {\bibfnamefont
  {I.}~\bibnamefont {Boettcher}}, \bibinfo {author} {\bibfnamefont {I.~C.}\
  \bibnamefont {Fulga}}, \bibinfo {author} {\bibfnamefont {L.}~\bibnamefont
  {Ma}}, \bibinfo {author} {\bibfnamefont {O.~G.}\ \bibnamefont {Schmidt}},
  \bibinfo {author} {\bibfnamefont {M.}~\bibnamefont {Greiter}}, \bibinfo
  {author} {\bibfnamefont {T.}~\bibnamefont {Kiessling}}, \bibinfo {author}
  {\bibfnamefont {A.}~\bibnamefont {Szameit}}, \ and\ \bibinfo {author}
  {\bibfnamefont {R.}~\bibnamefont {Thomale}},\ }\enquote {\bibinfo {title}
  {Topological defect engineering and $\mathcal{P}\mathcal{T}$ symmetry in
  non-{Hermitian} electrical circuits},}\ \href {\doibase
  10.1103/PhysRevLett.126.215302} {\bibfield  {journal} {\bibinfo  {journal}
  {Phys. Rev. Lett.}\ }\textbf {\bibinfo {volume} {126}},\ \bibinfo {pages}
  {215302} (\bibinfo {year} {2021})}\BibitemShut {NoStop}%
\bibitem [{\citenamefont {Sahin}\ \emph {et~al.}(2025)\citenamefont {Sahin},
  \citenamefont {Jalil},\ and\ \citenamefont {Lee}}]{Sahin25}%
  \BibitemOpen
  \bibfield  {author} {\bibinfo {author} {\bibfnamefont {H.}~\bibnamefont
  {Sahin}}, \bibinfo {author} {\bibfnamefont {M.~B.~A.}\ \bibnamefont {Jalil}},
  \ and\ \bibinfo {author} {\bibfnamefont {C.~H.}\ \bibnamefont {Lee}},\
  }\enquote {\bibinfo {title} {Topolectrical circuits---{Recent} experimental
  advances and developments},}\ \href {\doibase 10.1063/5.0265293} {\bibfield
  {journal} {\bibinfo  {journal} {APL Electron. Dev.}\ }\textbf {\bibinfo
  {volume} {1}},\ \bibinfo {pages} {021503} (\bibinfo {year}
  {2025})}\BibitemShut {NoStop}%
\bibitem [{\citenamefont {Zhang}\ \emph {et~al.}(2024)\citenamefont {Zhang},
  \citenamefont {Zhang}, \citenamefont {Zhao},\ and\ \citenamefont
  {Lee}}]{Zhang24}%
  \BibitemOpen
  \bibfield  {author} {\bibinfo {author} {\bibfnamefont {X.}~\bibnamefont
  {Zhang}}, \bibinfo {author} {\bibfnamefont {B.}~\bibnamefont {Zhang}},
  \bibinfo {author} {\bibfnamefont {W.}~\bibnamefont {Zhao}}, \ and\ \bibinfo
  {author} {\bibfnamefont {C.~H.}\ \bibnamefont {Lee}},\ }\enquote {\bibinfo
  {title} {{Observation of non-local impedance response in a passive electrical
  circuit}},}\ \href {\doibase 10.21468/SciPostPhys.16.1.002} {\bibfield
  {journal} {\bibinfo  {journal} {SciPost Phys.}\ }\textbf {\bibinfo {volume}
  {16}},\ \bibinfo {pages} {002} (\bibinfo {year} {2024})}\BibitemShut
  {NoStop}%
\bibitem [{\citenamefont {Yang}\ \emph
  {et~al.}(2024{\natexlab{a}})\citenamefont {Yang}, \citenamefont {Song},
  \citenamefont {Cao},\ and\ \citenamefont {Yan}}]{YangH24}%
  \BibitemOpen
  \bibfield  {author} {\bibinfo {author} {\bibfnamefont {H.}~\bibnamefont
  {Yang}}, \bibinfo {author} {\bibfnamefont {L.}~\bibnamefont {Song}}, \bibinfo
  {author} {\bibfnamefont {Y.}~\bibnamefont {Cao}}, \ and\ \bibinfo {author}
  {\bibfnamefont {P.}~\bibnamefont {Yan}},\ }\enquote {\bibinfo {title}
  {Circuit realization of topological physics},}\ \href {\doibase
  10.1016/j.physrep.2024.09.007} {\bibfield  {journal} {\bibinfo  {journal}
  {Phys. Rep.}\ }\textbf {\bibinfo {volume} {1093}},\ \bibinfo {pages} {1}
  (\bibinfo {year} {2024}{\natexlab{a}})}\BibitemShut {NoStop}%
\bibitem [{\citenamefont {Zou}\ \emph {et~al.}(2024)\citenamefont {Zou},
  \citenamefont {Chen}, \citenamefont {Meng}, \citenamefont {Ang},
  \citenamefont {Zhang},\ and\ \citenamefont {Lee}}]{Zou24}%
  \BibitemOpen
  \bibfield  {author} {\bibinfo {author} {\bibfnamefont {D.}~\bibnamefont
  {Zou}}, \bibinfo {author} {\bibfnamefont {T.}~\bibnamefont {Chen}}, \bibinfo
  {author} {\bibfnamefont {H.}~\bibnamefont {Meng}}, \bibinfo {author}
  {\bibfnamefont {Y.~S.}\ \bibnamefont {Ang}}, \bibinfo {author} {\bibfnamefont
  {X.}~\bibnamefont {Zhang}}, \ and\ \bibinfo {author} {\bibfnamefont {C.~H.}\
  \bibnamefont {Lee}},\ }\enquote {\bibinfo {title} {Experimental observation
  of exceptional bound states in a classical circuit network},}\ \href
  {\doibase https://doi.org/10.1016/j.scib.2024.05.036} {\bibfield  {journal}
  {\bibinfo  {journal} {Sci. Bull.}\ }\textbf {\bibinfo {volume} {69}},\
  \bibinfo {pages} {2194} (\bibinfo {year} {2024})}\BibitemShut {NoStop}%
\bibitem [{\citenamefont {Bender}\ \emph {et~al.}(2013)\citenamefont {Bender},
  \citenamefont {Berntson}, \citenamefont {Parker},\ and\ \citenamefont
  {Samuel}}]{Bender13}%
  \BibitemOpen
  \bibfield  {author} {\bibinfo {author} {\bibfnamefont {C.~M.}\ \bibnamefont
  {Bender}}, \bibinfo {author} {\bibfnamefont {B.~K.}\ \bibnamefont
  {Berntson}}, \bibinfo {author} {\bibfnamefont {D.}~\bibnamefont {Parker}}, \
  and\ \bibinfo {author} {\bibfnamefont {E.}~\bibnamefont {Samuel}},\ }\enquote
  {\bibinfo {title} {Observation of $\mathcal{PT}$ phase transition in a simple
  mechanical system},}\ \href {\doibase 10.1119/1.4789549} {\bibfield
  {journal} {\bibinfo  {journal} {Am. J. Phys.}\ }\textbf {\bibinfo {volume}
  {81}},\ \bibinfo {pages} {173} (\bibinfo {year} {2013})}\BibitemShut
  {NoStop}%
\bibitem [{\citenamefont {Ghatak}\ \emph {et~al.}(2020)\citenamefont {Ghatak},
  \citenamefont {Brandenbourger}, \citenamefont {van Wezel},\ and\
  \citenamefont {Coulais}}]{Gha20}%
  \BibitemOpen
  \bibfield  {author} {\bibinfo {author} {\bibfnamefont {A.}~\bibnamefont
  {Ghatak}}, \bibinfo {author} {\bibfnamefont {M.}~\bibnamefont
  {Brandenbourger}}, \bibinfo {author} {\bibfnamefont {J.}~\bibnamefont {van
  Wezel}}, \ and\ \bibinfo {author} {\bibfnamefont {C.}~\bibnamefont
  {Coulais}},\ }\enquote {\bibinfo {title} {Observation of non-{Hermitia}n
  topology and its bulk-edge correspondence in an active mechanical
  metamaterial},}\ \href {\doibase 10.1073/pnas.2010580117} {\bibfield
  {journal} {\bibinfo  {journal} {Proc. Natl. Acad. Sci. U.S.A.}\ }\textbf
  {\bibinfo {volume} {117}},\ \bibinfo {pages} {29561} (\bibinfo {year}
  {2020})}\BibitemShut {NoStop}%
\bibitem [{\citenamefont {Xue}\ \emph {et~al.}(2022)\citenamefont {Xue},
  \citenamefont {Yang},\ and\ \citenamefont {Zhang}}]{Xue22}%
  \BibitemOpen
  \bibfield  {author} {\bibinfo {author} {\bibfnamefont {H.}~\bibnamefont
  {Xue}}, \bibinfo {author} {\bibfnamefont {Y.}~\bibnamefont {Yang}}, \ and\
  \bibinfo {author} {\bibfnamefont {B.}~\bibnamefont {Zhang}},\ }\enquote
  {\bibinfo {title} {Topological acoustics},}\ \href {\doibase
  10.1038/s41578-022-00465-6} {\bibfield  {journal} {\bibinfo  {journal} {Nat.
  Rev. Mater.}\ }\textbf {\bibinfo {volume} {7}},\ \bibinfo {pages} {974}
  (\bibinfo {year} {2022})}\BibitemShut {NoStop}%
\bibitem [{\citenamefont {Zhang}\ \emph {et~al.}(2026)\citenamefont {Zhang},
  \citenamefont {Xiong}, \citenamefont {Tong},\ and\ \citenamefont
  {Qiu}}]{Zhang26}%
  \BibitemOpen
  \bibfield  {author} {\bibinfo {author} {\bibfnamefont {Q.}~\bibnamefont
  {Zhang}}, \bibinfo {author} {\bibfnamefont {L.}~\bibnamefont {Xiong}},
  \bibinfo {author} {\bibfnamefont {S.}~\bibnamefont {Tong}}, \ and\ \bibinfo
  {author} {\bibfnamefont {C.}~\bibnamefont {Qiu}},\ }\enquote {\bibinfo
  {title} {Harmonic non-{Hermitian} skin effect},}\ \href {\doibase
  10.1038/s41467-026-69043-6} {\bibfield  {journal} {\bibinfo  {journal} {Nat.
  Commun.}\ } (\bibinfo {year} {2026}),\
  10.1038/s41467-026-69043-6}\BibitemShut {NoStop}%
\bibitem [{\citenamefont {Ozawa}\ \emph {et~al.}(2019)\citenamefont {Ozawa},
  \citenamefont {Price}, \citenamefont {Amo}, \citenamefont {Goldman},
  \citenamefont {Hafezi}, \citenamefont {Lu}, \citenamefont {Rechtsman},
  \citenamefont {Schuster}, \citenamefont {Simon}, \citenamefont {Zilberberg},\
  and\ \citenamefont {Carusotto}}]{Ozawa2019}%
  \BibitemOpen
  \bibfield  {author} {\bibinfo {author} {\bibfnamefont {T.}~\bibnamefont
  {Ozawa}}, \bibinfo {author} {\bibfnamefont {H.~M.}\ \bibnamefont {Price}},
  \bibinfo {author} {\bibfnamefont {A.}~\bibnamefont {Amo}}, \bibinfo {author}
  {\bibfnamefont {N.}~\bibnamefont {Goldman}}, \bibinfo {author} {\bibfnamefont
  {M.}~\bibnamefont {Hafezi}}, \bibinfo {author} {\bibfnamefont
  {L.}~\bibnamefont {Lu}}, \bibinfo {author} {\bibfnamefont {M.~C.}\
  \bibnamefont {Rechtsman}}, \bibinfo {author} {\bibfnamefont {D.}~\bibnamefont
  {Schuster}}, \bibinfo {author} {\bibfnamefont {J.}~\bibnamefont {Simon}},
  \bibinfo {author} {\bibfnamefont {O.}~\bibnamefont {Zilberberg}}, \ and\
  \bibinfo {author} {\bibfnamefont {I.}~\bibnamefont {Carusotto}},\ }\enquote
  {\bibinfo {title} {Topological photonics},}\ \href {\doibase
  10.1103/RevModPhys.91.015006} {\bibfield  {journal} {\bibinfo  {journal}
  {Rev. Mod. Phys.}\ }\textbf {\bibinfo {volume} {91}},\ \bibinfo {pages}
  {015006} (\bibinfo {year} {2019})}\BibitemShut {NoStop}%
\bibitem [{\citenamefont {Peng}\ \emph {et~al.}(2014)\citenamefont {Peng},
  \citenamefont {\"{O}zdemir}, \citenamefont {Lei}, \citenamefont {Monifi},
  \citenamefont {Gianfreda}, \citenamefont {Long}, \citenamefont {Fan},
  \citenamefont {Nori}, \citenamefont {Bender},\ and\ \citenamefont
  {Yang}}]{Peng14}%
  \BibitemOpen
  \bibfield  {author} {\bibinfo {author} {\bibfnamefont {B.}~\bibnamefont
  {Peng}}, \bibinfo {author} {\bibfnamefont {{\c{S}}.~K.}\ \bibnamefont
  {\"{O}zdemir}}, \bibinfo {author} {\bibfnamefont {F.}~\bibnamefont {Lei}},
  \bibinfo {author} {\bibfnamefont {F.}~\bibnamefont {Monifi}}, \bibinfo
  {author} {\bibfnamefont {M.}~\bibnamefont {Gianfreda}}, \bibinfo {author}
  {\bibfnamefont {G.~L.}\ \bibnamefont {Long}}, \bibinfo {author}
  {\bibfnamefont {S.}~\bibnamefont {Fan}}, \bibinfo {author} {\bibfnamefont
  {F.}~\bibnamefont {Nori}}, \bibinfo {author} {\bibfnamefont {C.~M.}\
  \bibnamefont {Bender}}, \ and\ \bibinfo {author} {\bibfnamefont
  {L.}~\bibnamefont {Yang}},\ }\enquote {\bibinfo {title}
  {Parity--time-symmetric whispering-gallery microcavities},}\ \href {\doibase
  10.1038/NPHYS2927} {\bibfield  {journal} {\bibinfo  {journal} {Nat. Phys.}\
  }\textbf {\bibinfo {volume} {10}},\ \bibinfo {pages} {394} (\bibinfo {year}
  {2014})}\BibitemShut {NoStop}%
\bibitem [{\citenamefont {Feng}\ \emph {et~al.}(2017)\citenamefont {Feng},
  \citenamefont {El-Ganainy},\ and\ \citenamefont {Ge}}]{Feng2017}%
  \BibitemOpen
  \bibfield  {author} {\bibinfo {author} {\bibfnamefont {L.}~\bibnamefont
  {Feng}}, \bibinfo {author} {\bibfnamefont {R.}~\bibnamefont {El-Ganainy}}, \
  and\ \bibinfo {author} {\bibfnamefont {L.}~\bibnamefont {Ge}},\ }\enquote
  {\bibinfo {title} {Non-{Hermitian} photonics based on parity-time
  symmetry},}\ \href {https://doi.org/10.1038/s41566-017-0031-1} {\bibfield
  {journal} {\bibinfo  {journal} {Nat. Photon.}\ }\textbf {\bibinfo {volume}
  {11}},\ \bibinfo {pages} {752} (\bibinfo {year} {2017})}\BibitemShut
  {NoStop}%
\bibitem [{\citenamefont {Parto}\ \emph {et~al.}(2018)\citenamefont {Parto},
  \citenamefont {Wittek}, \citenamefont {Hodaei}, \citenamefont {Harari},
  \citenamefont {Bandres}, \citenamefont {Ren}, \citenamefont {Rechtsman},
  \citenamefont {Segev}, \citenamefont {Christodoulides},\ and\ \citenamefont
  {Khajavikhan}}]{Par18}%
  \BibitemOpen
  \bibfield  {author} {\bibinfo {author} {\bibfnamefont {M.}~\bibnamefont
  {Parto}}, \bibinfo {author} {\bibfnamefont {S.}~\bibnamefont {Wittek}},
  \bibinfo {author} {\bibfnamefont {H.}~\bibnamefont {Hodaei}}, \bibinfo
  {author} {\bibfnamefont {G.}~\bibnamefont {Harari}}, \bibinfo {author}
  {\bibfnamefont {M.~A.}\ \bibnamefont {Bandres}}, \bibinfo {author}
  {\bibfnamefont {J.}~\bibnamefont {Ren}}, \bibinfo {author} {\bibfnamefont
  {M.~C.}\ \bibnamefont {Rechtsman}}, \bibinfo {author} {\bibfnamefont
  {M.}~\bibnamefont {Segev}}, \bibinfo {author} {\bibfnamefont {D.~N.}\
  \bibnamefont {Christodoulides}}, \ and\ \bibinfo {author} {\bibfnamefont
  {M.}~\bibnamefont {Khajavikhan}},\ }\enquote {\bibinfo {title} {Edge-mode
  lasing in {1D} topological active arrays},}\ \href {\doibase
  10.1103/PhysRevLett.120.113901} {\bibfield  {journal} {\bibinfo  {journal}
  {Phys. Rev. Lett.}\ }\textbf {\bibinfo {volume} {120}},\ \bibinfo {pages}
  {113901} (\bibinfo {year} {2018})}\BibitemShut {NoStop}%
\bibitem [{\citenamefont {Zhou}\ \emph {et~al.}(2018)\citenamefont {Zhou},
  \citenamefont {Peng}, \citenamefont {Yoon}, \citenamefont {Hsu},
  \citenamefont {Nelson}, \citenamefont {Fu}, \citenamefont {Joannopoulos},
  \citenamefont {Solja{\v{c}}i{\'c}},\ and\ \citenamefont {Zhen}}]{Zhou18}%
  \BibitemOpen
  \bibfield  {author} {\bibinfo {author} {\bibfnamefont {H.}~\bibnamefont
  {Zhou}}, \bibinfo {author} {\bibfnamefont {C.}~\bibnamefont {Peng}}, \bibinfo
  {author} {\bibfnamefont {Y.}~\bibnamefont {Yoon}}, \bibinfo {author}
  {\bibfnamefont {C.~W.}\ \bibnamefont {Hsu}}, \bibinfo {author} {\bibfnamefont
  {K.~A.}\ \bibnamefont {Nelson}}, \bibinfo {author} {\bibfnamefont
  {L.}~\bibnamefont {Fu}}, \bibinfo {author} {\bibfnamefont {J.~D.}\
  \bibnamefont {Joannopoulos}}, \bibinfo {author} {\bibfnamefont
  {M.}~\bibnamefont {Solja{\v{c}}i{\'c}}}, \ and\ \bibinfo {author}
  {\bibfnamefont {B.}~\bibnamefont {Zhen}},\ }\enquote {\bibinfo {title}
  {Observation of bulk {Fermi} arc and polarization half charge from paired
  exceptional points},}\ \href {\doibase 10.1126/science.aap9859} {\bibfield
  {journal} {\bibinfo  {journal} {Science}\ }\textbf {\bibinfo {volume}
  {359}},\ \bibinfo {pages} {1009} (\bibinfo {year} {2018})}\BibitemShut
  {NoStop}%
\bibitem [{\citenamefont {Slootman}\ \emph {et~al.}(2024)\citenamefont
  {Slootman}, \citenamefont {Cherifi}, \citenamefont {Eek}, \citenamefont
  {Arouca}, \citenamefont {Bergholtz}, \citenamefont {Bourennane},\ and\
  \citenamefont {Smith}}]{Slootman24}%
  \BibitemOpen
  \bibfield  {author} {\bibinfo {author} {\bibfnamefont {E.}~\bibnamefont
  {Slootman}}, \bibinfo {author} {\bibfnamefont {W.}~\bibnamefont {Cherifi}},
  \bibinfo {author} {\bibfnamefont {L.}~\bibnamefont {Eek}}, \bibinfo {author}
  {\bibfnamefont {R.}~\bibnamefont {Arouca}}, \bibinfo {author} {\bibfnamefont
  {E.~J.}\ \bibnamefont {Bergholtz}}, \bibinfo {author} {\bibfnamefont
  {M.}~\bibnamefont {Bourennane}}, \ and\ \bibinfo {author} {\bibfnamefont
  {C.~M.}\ \bibnamefont {Smith}},\ }\enquote {\bibinfo {title} {Breaking and
  resurgence of symmetry in the non-{Hermitian Su-Schrieffer-Heeger} model in
  photonic waveguides},}\ \href {\doibase 10.1103/PhysRevResearch.6.023140}
  {\bibfield  {journal} {\bibinfo  {journal} {Phys. Rev. Res.}\ }\textbf
  {\bibinfo {volume} {6}},\ \bibinfo {pages} {023140} (\bibinfo {year}
  {2024})}\BibitemShut {NoStop}%
\bibitem [{\citenamefont {Rudner}\ and\ \citenamefont {Levitov}(2009)}]{Rud09}%
  \BibitemOpen
  \bibfield  {author} {\bibinfo {author} {\bibfnamefont {M.~S.}\ \bibnamefont
  {Rudner}}\ and\ \bibinfo {author} {\bibfnamefont {L.~S.}\ \bibnamefont
  {Levitov}},\ }\enquote {\bibinfo {title} {Topological transition in a
  non-{Hermitian} quantum walk},}\ \href {\doibase
  10.1103/PhysRevLett.102.065703} {\bibfield  {journal} {\bibinfo  {journal}
  {Phys. Rev. Lett.}\ }\textbf {\bibinfo {volume} {102}},\ \bibinfo {pages}
  {065703} (\bibinfo {year} {2009})}\BibitemShut {NoStop}%
\bibitem [{\citenamefont {Xiao}\ \emph {et~al.}(2017)\citenamefont {Xiao},
  \citenamefont {Zhan}, \citenamefont {Bian}, \citenamefont {Wang},
  \citenamefont {Zhang}, \citenamefont {Wang}, \citenamefont {Li},
  \citenamefont {Mochizuki}, \citenamefont {Kim}, \citenamefont {Kawakami},
  \citenamefont {Yi}, \citenamefont {Obuse}, \citenamefont {Sanders},\ and\
  \citenamefont {Xue}}]{Xiao17}%
  \BibitemOpen
  \bibfield  {author} {\bibinfo {author} {\bibfnamefont {L.}~\bibnamefont
  {Xiao}}, \bibinfo {author} {\bibfnamefont {X.}~\bibnamefont {Zhan}}, \bibinfo
  {author} {\bibfnamefont {Z.~H.}\ \bibnamefont {Bian}}, \bibinfo {author}
  {\bibfnamefont {K.~K.}\ \bibnamefont {Wang}}, \bibinfo {author}
  {\bibfnamefont {X.}~\bibnamefont {Zhang}}, \bibinfo {author} {\bibfnamefont
  {X.~P.}\ \bibnamefont {Wang}}, \bibinfo {author} {\bibfnamefont
  {J.}~\bibnamefont {Li}}, \bibinfo {author} {\bibfnamefont {K.}~\bibnamefont
  {Mochizuki}}, \bibinfo {author} {\bibfnamefont {D.}~\bibnamefont {Kim}},
  \bibinfo {author} {\bibfnamefont {N.}~\bibnamefont {Kawakami}}, \bibinfo
  {author} {\bibfnamefont {W.}~\bibnamefont {Yi}}, \bibinfo {author}
  {\bibfnamefont {H.}~\bibnamefont {Obuse}}, \bibinfo {author} {\bibfnamefont
  {B.~C.}\ \bibnamefont {Sanders}}, \ and\ \bibinfo {author} {\bibfnamefont
  {P.}~\bibnamefont {Xue}},\ }\enquote {\bibinfo {title} {Observation of
  topological edge states in parity-time-symmetric quantum walks},}\ \href
  {https://doi.org/10.1038/nphys4204} {\bibfield  {journal} {\bibinfo
  {journal} {Nat. Phys.}\ }\textbf {\bibinfo {volume} {13}},\ \bibinfo {pages}
  {1117} (\bibinfo {year} {2017})}\BibitemShut {NoStop}%
\bibitem [{\citenamefont {Song}\ \emph {et~al.}(2019)\citenamefont {Song},
  \citenamefont {Yao},\ and\ \citenamefont {Wang}}]{FSong19}%
  \BibitemOpen
  \bibfield  {author} {\bibinfo {author} {\bibfnamefont {F.}~\bibnamefont
  {Song}}, \bibinfo {author} {\bibfnamefont {S.}~\bibnamefont {Yao}}, \ and\
  \bibinfo {author} {\bibfnamefont {Z.}~\bibnamefont {Wang}},\ }\enquote
  {\bibinfo {title} {Non-{Hermitian} skin effect and chiral damping in open
  quantum systems},}\ \href {\doibase 10.1103/PhysRevLett.123.170401}
  {\bibfield  {journal} {\bibinfo  {journal} {Phys. Rev. Lett.}\ }\textbf
  {\bibinfo {volume} {123}},\ \bibinfo {pages} {170401} (\bibinfo {year}
  {2019})}\BibitemShut {NoStop}%
\bibitem [{\citenamefont {Shen}\ \emph
  {et~al.}(2025{\natexlab{b}})\citenamefont {Shen}, \citenamefont {Chen},
  \citenamefont {Yang},\ and\ \citenamefont {Lee}}]{ShenR25}%
  \BibitemOpen
  \bibfield  {author} {\bibinfo {author} {\bibfnamefont {R.}~\bibnamefont
  {Shen}}, \bibinfo {author} {\bibfnamefont {T.}~\bibnamefont {Chen}}, \bibinfo
  {author} {\bibfnamefont {B.}~\bibnamefont {Yang}}, \ and\ \bibinfo {author}
  {\bibfnamefont {C.~H.}\ \bibnamefont {Lee}},\ }\enquote {\bibinfo {title}
  {Observation of the non-{Hermitian} skin effect and fermi skin on a digital
  quantum computer},}\ \href {\doibase 10.1038/s41467-025-55953-4} {\bibfield
  {journal} {\bibinfo  {journal} {Nat. Commun.}\ }\textbf {\bibinfo {volume}
  {16}},\ \bibinfo {pages} {1340} (\bibinfo {year}
  {2025}{\natexlab{b}})}\BibitemShut {NoStop}%
\bibitem [{\citenamefont {Shen}\ \emph {et~al.}(2024)\citenamefont {Shen},
  \citenamefont {Qin}, \citenamefont {Desaules}, \citenamefont
  {Papi\ifmmode~\acute{c}\else \'{c}\fi{}},\ and\ \citenamefont
  {Lee}}]{Shen24}%
  \BibitemOpen
  \bibfield  {author} {\bibinfo {author} {\bibfnamefont {R.}~\bibnamefont
  {Shen}}, \bibinfo {author} {\bibfnamefont {F.}~\bibnamefont {Qin}}, \bibinfo
  {author} {\bibfnamefont {J.-Y.}\ \bibnamefont {Desaules}}, \bibinfo {author}
  {\bibfnamefont {Z.}~\bibnamefont {Papi\ifmmode~\acute{c}\else \'{c}\fi{}}}, \
  and\ \bibinfo {author} {\bibfnamefont {C.~H.}\ \bibnamefont {Lee}},\
  }\enquote {\bibinfo {title} {Enhanced many-body quantum scars from the
  non-{Hermitian} {Fock} skin effect},}\ \href {\doibase
  10.1103/PhysRevLett.133.216601} {\bibfield  {journal} {\bibinfo  {journal}
  {Phys. Rev. Lett.}\ }\textbf {\bibinfo {volume} {133}},\ \bibinfo {pages}
  {216601} (\bibinfo {year} {2024})}\BibitemShut {NoStop}%
\bibitem [{\citenamefont {Zhang}\ \emph {et~al.}(2021)\citenamefont {Zhang},
  \citenamefont {Ouyang}, \citenamefont {Huang}, \citenamefont {Wang},
  \citenamefont {Zhang}, \citenamefont {Yu}, \citenamefont {Chang},
  \citenamefont {Liu}, \citenamefont {Deng},\ and\ \citenamefont
  {Duan}}]{Zhan21}%
  \BibitemOpen
  \bibfield  {author} {\bibinfo {author} {\bibfnamefont {W.}~\bibnamefont
  {Zhang}}, \bibinfo {author} {\bibfnamefont {X.}~\bibnamefont {Ouyang}},
  \bibinfo {author} {\bibfnamefont {X.}~\bibnamefont {Huang}}, \bibinfo
  {author} {\bibfnamefont {X.}~\bibnamefont {Wang}}, \bibinfo {author}
  {\bibfnamefont {H.}~\bibnamefont {Zhang}}, \bibinfo {author} {\bibfnamefont
  {Y.}~\bibnamefont {Yu}}, \bibinfo {author} {\bibfnamefont {X.}~\bibnamefont
  {Chang}}, \bibinfo {author} {\bibfnamefont {Y.}~\bibnamefont {Liu}}, \bibinfo
  {author} {\bibfnamefont {D.-L.}\ \bibnamefont {Deng}}, \ and\ \bibinfo
  {author} {\bibfnamefont {L.-M.}\ \bibnamefont {Duan}},\ }\enquote {\bibinfo
  {title} {Observation of non-{Hermitian} topology with nonunitary dynamics of
  solid-state spins},}\ \href {\doibase 10.1103/PhysRevLett.127.090501}
  {\bibfield  {journal} {\bibinfo  {journal} {Phys. Rev. Lett.}\ }\textbf
  {\bibinfo {volume} {127}},\ \bibinfo {pages} {090501} (\bibinfo {year}
  {2021})}\BibitemShut {NoStop}%
\bibitem [{\citenamefont {Yang}\ \emph
  {et~al.}(2024{\natexlab{b}})\citenamefont {Yang}, \citenamefont {Li},
  \citenamefont {K\"{o}nig}, \citenamefont {R\o{}dland}, \citenamefont
  {St\r{a}lhammar},\ and\ \citenamefont {Bergholtz}}]{YangK24}%
  \BibitemOpen
  \bibfield  {author} {\bibinfo {author} {\bibfnamefont {K.}~\bibnamefont
  {Yang}}, \bibinfo {author} {\bibfnamefont {Z.}~\bibnamefont {Li}}, \bibinfo
  {author} {\bibfnamefont {J.~L.~K.}\ \bibnamefont {K\"{o}nig}}, \bibinfo
  {author} {\bibfnamefont {L.}~\bibnamefont {R\o{}dland}}, \bibinfo {author}
  {\bibfnamefont {M.}~\bibnamefont {St\r{a}lhammar}}, \ and\ \bibinfo {author}
  {\bibfnamefont {E.~J.}\ \bibnamefont {Bergholtz}},\ }\enquote {\bibinfo
  {title} {Homotopy, symmetry, and non-{Hermitian} band topology},}\ \href
  {\doibase 10.1088/1361-6633/ad4e64} {\bibfield  {journal} {\bibinfo
  {journal} {Rep. Prog. Phys.}\ }\textbf {\bibinfo {volume} {87}},\ \bibinfo
  {pages} {078002} (\bibinfo {year} {2024}{\natexlab{b}})}\BibitemShut
  {NoStop}%
\bibitem [{\citenamefont {Jia}\ \emph {et~al.}(2025)\citenamefont {Jia},
  \citenamefont {Hu}, \citenamefont {Zhang}, \citenamefont {Xiao},
  \citenamefont {Wang}, \citenamefont {Wang}, \citenamefont {Ma}, \citenamefont
  {Ouyang}, \citenamefont {Zhu},\ and\ \citenamefont {Chan}}]{Jia25}%
  \BibitemOpen
  \bibfield  {author} {\bibinfo {author} {\bibfnamefont {H.}~\bibnamefont
  {Jia}}, \bibinfo {author} {\bibfnamefont {J.}~\bibnamefont {Hu}}, \bibinfo
  {author} {\bibfnamefont {R.-Y.}\ \bibnamefont {Zhang}}, \bibinfo {author}
  {\bibfnamefont {Y.}~\bibnamefont {Xiao}}, \bibinfo {author} {\bibfnamefont
  {D.}~\bibnamefont {Wang}}, \bibinfo {author} {\bibfnamefont {M.}~\bibnamefont
  {Wang}}, \bibinfo {author} {\bibfnamefont {S.}~\bibnamefont {Ma}}, \bibinfo
  {author} {\bibfnamefont {X.}~\bibnamefont {Ouyang}}, \bibinfo {author}
  {\bibfnamefont {Y.}~\bibnamefont {Zhu}}, \ and\ \bibinfo {author}
  {\bibfnamefont {C.}~\bibnamefont {Chan}},\ }\enquote {\bibinfo {title}
  {Unconventional topological edge states in one-dimensional non-{Hermitian}
  gapless systems stemming from nonisolated hypersurface singularities},}\
  \href {\doibase 10.1103/PhysRevLett.134.206603} {\bibfield  {journal}
  {\bibinfo  {journal} {Phys. Rev. Lett.}\ }\textbf {\bibinfo {volume} {134}},\
  \bibinfo {pages} {206603} (\bibinfo {year} {2025})}\BibitemShut {NoStop}%
\bibitem [{\citenamefont {Jiang}\ and\ \citenamefont {Lee}(2023)}]{Jiang23}%
  \BibitemOpen
  \bibfield  {author} {\bibinfo {author} {\bibfnamefont {H.}~\bibnamefont
  {Jiang}}\ and\ \bibinfo {author} {\bibfnamefont {C.~H.}\ \bibnamefont
  {Lee}},\ }\enquote {\bibinfo {title} {Dimensional transmutation from
  non-{Hermiticity}},}\ \href {\doibase 10.1103/PhysRevLett.131.076401}
  {\bibfield  {journal} {\bibinfo  {journal} {Phys. Rev. Lett.}\ }\textbf
  {\bibinfo {volume} {131}},\ \bibinfo {pages} {076401} (\bibinfo {year}
  {2023})}\BibitemShut {NoStop}%
\bibitem [{\citenamefont {Jiang}\ \emph {et~al.}(2018)\citenamefont {Jiang},
  \citenamefont {Yang},\ and\ \citenamefont {Chen}}]{Jiang2018}%
  \BibitemOpen
  \bibfield  {author} {\bibinfo {author} {\bibfnamefont {H.}~\bibnamefont
  {Jiang}}, \bibinfo {author} {\bibfnamefont {C.}~\bibnamefont {Yang}}, \ and\
  \bibinfo {author} {\bibfnamefont {S.}~\bibnamefont {Chen}},\ }\enquote
  {\bibinfo {title} {Topological invariants and phase diagrams for
  one-dimensional two-band non-{Hermitian} systems without chiral symmetry},}\
  \href {\doibase 10.1103/PhysRevA.98.052116} {\bibfield  {journal} {\bibinfo
  {journal} {Phys. Rev. A}\ }\textbf {\bibinfo {volume} {98}},\ \bibinfo
  {pages} {052116} (\bibinfo {year} {2018})}\BibitemShut {NoStop}%
\bibitem [{\citenamefont {Li}\ \emph {et~al.}(2024{\natexlab{a}})\citenamefont
  {Li}, \citenamefont {Hou}, \citenamefont {Wu}, \citenamefont {Ruan},
  \citenamefont {Chen}, \citenamefont {Yuan},\ and\ \citenamefont {Ni}}]{Li24}%
  \BibitemOpen
  \bibfield  {author} {\bibinfo {author} {\bibfnamefont {L.}~\bibnamefont
  {Li}}, \bibinfo {author} {\bibfnamefont {C.}~\bibnamefont {Hou}}, \bibinfo
  {author} {\bibfnamefont {G.}~\bibnamefont {Wu}}, \bibinfo {author}
  {\bibfnamefont {Y.}~\bibnamefont {Ruan}}, \bibinfo {author} {\bibfnamefont
  {S.}~\bibnamefont {Chen}}, \bibinfo {author} {\bibfnamefont {L.}~\bibnamefont
  {Yuan}}, \ and\ \bibinfo {author} {\bibfnamefont {Z.}~\bibnamefont {Ni}},\
  }\enquote {\bibinfo {title} {Dual bulk-boundary correspondence in a
  nonreciprocal spin-orbit coupled zigzag lattice},}\ \href {\doibase
  10.1103/PhysRevB.110.L041103} {\bibfield  {journal} {\bibinfo  {journal}
  {Phys. Rev. B}\ }\textbf {\bibinfo {volume} {110}},\ \bibinfo {pages}
  {L041103} (\bibinfo {year} {2024}{\natexlab{a}})}\BibitemShut {NoStop}%
\bibitem [{\citenamefont {Yang}\ and\ \citenamefont {Lee}(2024)}]{Yang24}%
  \BibitemOpen
  \bibfield  {author} {\bibinfo {author} {\bibfnamefont {M.}~\bibnamefont
  {Yang}}\ and\ \bibinfo {author} {\bibfnamefont {C.~H.}\ \bibnamefont {Lee}},\
  }\enquote {\bibinfo {title} {Percolation-induced $\mathcal{P}\mathcal{T}$
  symmetry breaking},}\ \href {\doibase 10.1103/PhysRevLett.133.136602}
  {\bibfield  {journal} {\bibinfo  {journal} {Phys. Rev. Lett.}\ }\textbf
  {\bibinfo {volume} {133}},\ \bibinfo {pages} {136602} (\bibinfo {year}
  {2024})}\BibitemShut {NoStop}%
\bibitem [{\citenamefont {Zhang}\ \emph
  {et~al.}(2025{\natexlab{a}})\citenamefont {Zhang}, \citenamefont {Zhang},
  \citenamefont {Li}, \citenamefont {an~Li}, \citenamefont {Wei}, \citenamefont
  {Tian}, \citenamefont {Wu}, \citenamefont {Shi}, \citenamefont {Gao},
  \citenamefont {Li},\ and\ \citenamefont {Liu}}]{Zhang25}%
  \BibitemOpen
  \bibfield  {author} {\bibinfo {author} {\bibfnamefont {M.}~\bibnamefont
  {Zhang}}, \bibinfo {author} {\bibfnamefont {Y.}~\bibnamefont {Zhang}},
  \bibinfo {author} {\bibfnamefont {S.}~\bibnamefont {Li}}, \bibinfo {author}
  {\bibfnamefont {Y.}~\bibnamefont {an~Li}}, \bibinfo {author} {\bibfnamefont
  {Y.}~\bibnamefont {Wei}}, \bibinfo {author} {\bibfnamefont {R.}~\bibnamefont
  {Tian}}, \bibinfo {author} {\bibfnamefont {T.}~\bibnamefont {Wu}}, \bibinfo
  {author} {\bibfnamefont {H.}~\bibnamefont {Shi}}, \bibinfo {author}
  {\bibfnamefont {H.}~\bibnamefont {Gao}}, \bibinfo {author} {\bibfnamefont
  {F.}~\bibnamefont {Li}}, \ and\ \bibinfo {author} {\bibfnamefont
  {B.}~\bibnamefont {Liu}},\ }\enquote {\bibinfo {title} {Observation of
  non-{Hermitian} bulk-boundary correspondence in nonchiral nonunitary quantum
  dynamics of single photons},}\ \href {\doibase 10.1103/2zx3-rs57} {\bibfield
  {journal} {\bibinfo  {journal} {Phys. Rev. Lett.}\ }\textbf {\bibinfo
  {volume} {135}},\ \bibinfo {pages} {213601} (\bibinfo {year}
  {2025}{\natexlab{a}})}\BibitemShut {NoStop}%
\bibitem [{\citenamefont {Zhang}\ \emph
  {et~al.}(2025{\natexlab{b}})\citenamefont {Zhang}, \citenamefont {Li},
  \citenamefont {Xu}, \citenamefont {Tian}, \citenamefont {Zhang},
  \citenamefont {Li}, \citenamefont {Gao}, \citenamefont {Zubairy},
  \citenamefont {Li},\ and\ \citenamefont {Liu}}]{ZhangY25}%
  \BibitemOpen
  \bibfield  {author} {\bibinfo {author} {\bibfnamefont {Y.}~\bibnamefont
  {Zhang}}, \bibinfo {author} {\bibfnamefont {S.}~\bibnamefont {Li}}, \bibinfo
  {author} {\bibfnamefont {Y.}~\bibnamefont {Xu}}, \bibinfo {author}
  {\bibfnamefont {R.}~\bibnamefont {Tian}}, \bibinfo {author} {\bibfnamefont
  {M.}~\bibnamefont {Zhang}}, \bibinfo {author} {\bibfnamefont
  {H.}~\bibnamefont {Li}}, \bibinfo {author} {\bibfnamefont {H.}~\bibnamefont
  {Gao}}, \bibinfo {author} {\bibfnamefont {M.~S.}\ \bibnamefont {Zubairy}},
  \bibinfo {author} {\bibfnamefont {F.}~\bibnamefont {Li}}, \ and\ \bibinfo
  {author} {\bibfnamefont {B.}~\bibnamefont {Liu}},\ }\enquote {\bibinfo
  {title} {Nonchiral non-{Bloch} invariants and topological phase diagram in
  nonunitary quantum dynamics without chiral symmetry},}\ \href {\doibase
  10.1103/PhysRevLett.134.113603} {\bibfield  {journal} {\bibinfo  {journal}
  {Phys. Rev. Lett.}\ }\textbf {\bibinfo {volume} {134}},\ \bibinfo {pages}
  {113603} (\bibinfo {year} {2025}{\natexlab{b}})}\BibitemShut {NoStop}%
\bibitem [{\citenamefont {Jiao}\ \emph {et~al.}(2021)\citenamefont {Jiao},
  \citenamefont {Longhi}, \citenamefont {Wang}, \citenamefont {Gao},
  \citenamefont {Zhou}, \citenamefont {Wang}, \citenamefont {Fu}, \citenamefont
  {Wang}, \citenamefont {Ren}, \citenamefont {Qiao},\ and\ \citenamefont
  {Jin}}]{Jiao21}%
  \BibitemOpen
  \bibfield  {author} {\bibinfo {author} {\bibfnamefont {Z.-Q.}\ \bibnamefont
  {Jiao}}, \bibinfo {author} {\bibfnamefont {S.}~\bibnamefont {Longhi}},
  \bibinfo {author} {\bibfnamefont {X.-W.}\ \bibnamefont {Wang}}, \bibinfo
  {author} {\bibfnamefont {J.}~\bibnamefont {Gao}}, \bibinfo {author}
  {\bibfnamefont {W.-H.}\ \bibnamefont {Zhou}}, \bibinfo {author}
  {\bibfnamefont {Y.}~\bibnamefont {Wang}}, \bibinfo {author} {\bibfnamefont
  {Y.-X.}\ \bibnamefont {Fu}}, \bibinfo {author} {\bibfnamefont
  {L.}~\bibnamefont {Wang}}, \bibinfo {author} {\bibfnamefont {R.-J.}\
  \bibnamefont {Ren}}, \bibinfo {author} {\bibfnamefont {L.-F.}\ \bibnamefont
  {Qiao}}, \ and\ \bibinfo {author} {\bibfnamefont {X.-M.}\ \bibnamefont
  {Jin}},\ }\enquote {\bibinfo {title} {Experimentally detecting quantized
  {Zak} phases without chiral symmetry in photonic lattices},}\ \href {\doibase
  10.1103/PhysRevLett.127.147401} {\bibfield  {journal} {\bibinfo  {journal}
  {Phys. Rev. Lett.}\ }\textbf {\bibinfo {volume} {127}},\ \bibinfo {pages}
  {147401} (\bibinfo {year} {2021})}\BibitemShut {NoStop}%
\bibitem [{\citenamefont {Wu}\ \emph {et~al.}(2023)\citenamefont {Wu},
  \citenamefont {Wang}, \citenamefont {Chen}, \citenamefont {Chen},\ and\
  \citenamefont {Yuan}}]{Wu2023}%
  \BibitemOpen
  \bibfield  {author} {\bibinfo {author} {\bibfnamefont {X.}~\bibnamefont
  {Wu}}, \bibinfo {author} {\bibfnamefont {L.}~\bibnamefont {Wang}}, \bibinfo
  {author} {\bibfnamefont {S.}~\bibnamefont {Chen}}, \bibinfo {author}
  {\bibfnamefont {X.}~\bibnamefont {Chen}}, \ and\ \bibinfo {author}
  {\bibfnamefont {L.}~\bibnamefont {Yuan}},\ }\enquote {\bibinfo {title}
  {Transition characteristics of non-{Hermitian} skin effects in a zigzag
  lattice without chiral symmetry},}\ \href {\doibase 10.1002/apxr.202300007}
  {\bibfield  {journal} {\bibinfo  {journal} {Adv. Phys. Res.}\ }\textbf
  {\bibinfo {volume} {2}},\ \bibinfo {pages} {2300007} (\bibinfo {year}
  {2023})}\BibitemShut {NoStop}%
\bibitem [{\citenamefont {Zhong}\ \emph {et~al.}(2025)\citenamefont {Zhong},
  \citenamefont {Wang}, \citenamefont {Poddubny},\ and\ \citenamefont
  {Fan}}]{Zhong25}%
  \BibitemOpen
  \bibfield  {author} {\bibinfo {author} {\bibfnamefont {J.}~\bibnamefont
  {Zhong}}, \bibinfo {author} {\bibfnamefont {H.}~\bibnamefont {Wang}},
  \bibinfo {author} {\bibfnamefont {A.~N.}\ \bibnamefont {Poddubny}}, \ and\
  \bibinfo {author} {\bibfnamefont {S.}~\bibnamefont {Fan}},\ }\enquote
  {\bibinfo {title} {Topological nature of edge states for one-dimensional
  systems without symmetry protection},}\ \href {\doibase 10.1103/k77w-ft26}
  {\bibfield  {journal} {\bibinfo  {journal} {Phys. Rev. Lett.}\ }\textbf
  {\bibinfo {volume} {135}},\ \bibinfo {pages} {016601} (\bibinfo {year}
  {2025})}\BibitemShut {NoStop}%
\bibitem [{\citenamefont {Wei}\ \emph {et~al.}(2026)\citenamefont {Wei},
  \citenamefont {Ruan}, \citenamefont {Wu}, \citenamefont {Li}, \citenamefont
  {Chen}, \citenamefont {Lin},\ and\ \citenamefont {Ni}}]{Wei26}%
  \BibitemOpen
  \bibfield  {author} {\bibinfo {author} {\bibfnamefont {Y.}~\bibnamefont
  {Wei}}, \bibinfo {author} {\bibfnamefont {Y.}~\bibnamefont {Ruan}}, \bibinfo
  {author} {\bibfnamefont {G.}~\bibnamefont {Wu}}, \bibinfo {author}
  {\bibfnamefont {L.}~\bibnamefont {Li}}, \bibinfo {author} {\bibfnamefont
  {S.}~\bibnamefont {Chen}}, \bibinfo {author} {\bibfnamefont {T.}~\bibnamefont
  {Lin}}, \ and\ \bibinfo {author} {\bibfnamefont {Z.}~\bibnamefont {Ni}},\
  }\enquote {\bibinfo {title} {Beyond predicting existence: {Unified} winding
  numbers for topological edge states in non-{Hermitian} quadripartite
  lattices},}\ \href {\doibase 10.1103/h5wk-8cdl} {\bibfield  {journal}
  {\bibinfo  {journal} {Phys. Rev. B}\ }\textbf {\bibinfo {volume} {113}},\
  \bibinfo {pages} {075136} (\bibinfo {year} {2026})}\BibitemShut {NoStop}%
\bibitem [{\citenamefont {Li}\ \emph {et~al.}(2019{\natexlab{b}})\citenamefont
  {Li}, \citenamefont {Lee},\ and\ \citenamefont {Gong}}]{Li19}%
  \BibitemOpen
  \bibfield  {author} {\bibinfo {author} {\bibfnamefont {L.}~\bibnamefont
  {Li}}, \bibinfo {author} {\bibfnamefont {C.~H.}\ \bibnamefont {Lee}}, \ and\
  \bibinfo {author} {\bibfnamefont {J.}~\bibnamefont {Gong}},\ }\enquote
  {\bibinfo {title} {Geometric characterization of non-{Hermitian} topological
  systems through the singularity ring in pseudospin vector space},}\ \href
  {\doibase 10.1103/PhysRevB.100.075403} {\bibfield  {journal} {\bibinfo
  {journal} {Phys. Rev. B}\ }\textbf {\bibinfo {volume} {100}},\ \bibinfo
  {pages} {075403} (\bibinfo {year} {2019}{\natexlab{b}})}\BibitemShut
  {NoStop}%
\bibitem [{\citenamefont {Hu}\ \emph {et~al.}(2024)\citenamefont {Hu},
  \citenamefont {Wang}, \citenamefont {Wang},\ and\ \citenamefont
  {Song}}]{Hu24}%
  \BibitemOpen
  \bibfield  {author} {\bibinfo {author} {\bibfnamefont {Y.-M.}\ \bibnamefont
  {Hu}}, \bibinfo {author} {\bibfnamefont {H.-Y.}\ \bibnamefont {Wang}},
  \bibinfo {author} {\bibfnamefont {Z.}~\bibnamefont {Wang}}, \ and\ \bibinfo
  {author} {\bibfnamefont {F.}~\bibnamefont {Song}},\ }\enquote {\bibinfo
  {title} {Geometric origin of non-{Bloch}- $\mathcal{P}\mathcal{T}$ symmetry
  breaking},}\ \href {\doibase 10.1103/PhysRevLett.132.050402} {\bibfield
  {journal} {\bibinfo  {journal} {Phys. Rev. Lett.}\ }\textbf {\bibinfo
  {volume} {132}},\ \bibinfo {pages} {050402} (\bibinfo {year}
  {2024})}\BibitemShut {NoStop}%
\bibitem [{\citenamefont {Meng}\ \emph {et~al.}(2025)\citenamefont {Meng},
  \citenamefont {Ang},\ and\ \citenamefont {Lee}}]{Meng25}%
  \BibitemOpen
  \bibfield  {author} {\bibinfo {author} {\bibfnamefont {H.}~\bibnamefont
  {Meng}}, \bibinfo {author} {\bibfnamefont {Y.~S.}\ \bibnamefont {Ang}}, \
  and\ \bibinfo {author} {\bibfnamefont {C.~H.}\ \bibnamefont {Lee}},\
  }\enquote {\bibinfo {title} {Generalized {Brillouin} zone fragmentation},}\
  \href@noop {} {\  (\bibinfo {year} {2025})},\ \Eprint
  {http://arxiv.org/abs/2508.13275} {arXiv:2508.13275} \BibitemShut {NoStop}%
\bibitem [{\citenamefont {Li}\ \emph {et~al.}(2020)\citenamefont {Li},
  \citenamefont {Lee}, \citenamefont {Mu},\ and\ \citenamefont
  {Gong}}]{Li2020}%
  \BibitemOpen
  \bibfield  {author} {\bibinfo {author} {\bibfnamefont {L.}~\bibnamefont
  {Li}}, \bibinfo {author} {\bibfnamefont {C.~H.}\ \bibnamefont {Lee}},
  \bibinfo {author} {\bibfnamefont {S.}~\bibnamefont {Mu}}, \ and\ \bibinfo
  {author} {\bibfnamefont {J.}~\bibnamefont {Gong}},\ }\enquote {\bibinfo
  {title} {Critical non-{Hermitian} skin effect},}\ \href {\doibase
  10.1038/s41467-020-18917-4} {\bibfield  {journal} {\bibinfo  {journal} {Nat.
  Commun.}\ }\textbf {\bibinfo {volume} {11}},\ \bibinfo {pages} {5491}
  (\bibinfo {year} {2020})}\BibitemShut {NoStop}%
\bibitem [{\citenamefont {Qin}\ \emph {et~al.}(2023)\citenamefont {Qin},
  \citenamefont {Ma}, \citenamefont {Shen},\ and\ \citenamefont {Lee}}]{Qin23}%
  \BibitemOpen
  \bibfield  {author} {\bibinfo {author} {\bibfnamefont {F.}~\bibnamefont
  {Qin}}, \bibinfo {author} {\bibfnamefont {Y.}~\bibnamefont {Ma}}, \bibinfo
  {author} {\bibfnamefont {R.}~\bibnamefont {Shen}}, \ and\ \bibinfo {author}
  {\bibfnamefont {C.~H.}\ \bibnamefont {Lee}},\ }\enquote {\bibinfo {title}
  {Universal competitive spectral scaling from the critical non-{Hermitian}
  skin effect},}\ \href {\doibase 10.1103/PhysRevB.107.155430} {\bibfield
  {journal} {\bibinfo  {journal} {Phys. Rev. B}\ }\textbf {\bibinfo {volume}
  {107}},\ \bibinfo {pages} {155430} (\bibinfo {year} {2023})}\BibitemShut
  {NoStop}%
\bibitem [{\citenamefont {Yokomizo}\ and\ \citenamefont
  {Murakami}(2021)}]{Kazuki21}%
  \BibitemOpen
  \bibfield  {author} {\bibinfo {author} {\bibfnamefont {K.}~\bibnamefont
  {Yokomizo}}\ and\ \bibinfo {author} {\bibfnamefont {S.}~\bibnamefont
  {Murakami}},\ }\enquote {\bibinfo {title} {Scaling rule for the critical
  non-{Hermitian} skin effect},}\ \href {\doibase 10.1103/PhysRevB.104.165117}
  {\bibfield  {journal} {\bibinfo  {journal} {Phys. Rev. B}\ }\textbf {\bibinfo
  {volume} {104}},\ \bibinfo {pages} {165117} (\bibinfo {year}
  {2021})}\BibitemShut {NoStop}%
\bibitem [{\citenamefont {Qin}\ \emph {et~al.}(2026)\citenamefont {Qin},
  \citenamefont {Ang}, \citenamefont {Lee},\ and\ \citenamefont {Li}}]{Qin26}%
  \BibitemOpen
  \bibfield  {author} {\bibinfo {author} {\bibfnamefont {Y.}~\bibnamefont
  {Qin}}, \bibinfo {author} {\bibfnamefont {Y.~S.}\ \bibnamefont {Ang}},
  \bibinfo {author} {\bibfnamefont {C.~H.}\ \bibnamefont {Lee}}, \ and\
  \bibinfo {author} {\bibfnamefont {L.}~\bibnamefont {Li}},\ }\enquote
  {\bibinfo {title} {Many-body critical non-{Hermitian} skin effect},}\ \href
  {\doibase 10.1038/s42005-025-02448-9} {\bibfield  {journal} {\bibinfo
  {journal} {Commun. Phys.}\ }\textbf {\bibinfo {volume} {9}},\ \bibinfo
  {pages} {16} (\bibinfo {year} {2026})}\BibitemShut {NoStop}%
\bibitem [{\citenamefont {Murakami}(2006)}]{Mura06}%
  \BibitemOpen
  \bibfield  {author} {\bibinfo {author} {\bibfnamefont {S.}~\bibnamefont
  {Murakami}},\ }\enquote {\bibinfo {title} {Quantum spin {Hall} effect and
  enhanced magnetic response by spin-orbit coupling},}\ \href {\doibase
  10.1103/PhysRevLett.97.236805} {\bibfield  {journal} {\bibinfo  {journal}
  {Phys. Rev. Lett.}\ }\textbf {\bibinfo {volume} {97}},\ \bibinfo {pages}
  {236805} (\bibinfo {year} {2006})}\BibitemShut {NoStop}%
\bibitem [{\citenamefont {Soumyanarayanan}\ \emph {et~al.}(2016)\citenamefont
  {Soumyanarayanan}, \citenamefont {reyren}, \citenamefont {Fert},\ and\
  \citenamefont {Panagopoulos}}]{Sou16}%
  \BibitemOpen
  \bibfield  {author} {\bibinfo {author} {\bibfnamefont {A.}~\bibnamefont
  {Soumyanarayanan}}, \bibinfo {author} {\bibfnamefont {N.}~\bibnamefont
  {reyren}}, \bibinfo {author} {\bibfnamefont {A.}~\bibnamefont {Fert}}, \ and\
  \bibinfo {author} {\bibfnamefont {C.}~\bibnamefont {Panagopoulos}},\
  }\enquote {\bibinfo {title} {Emergent phenomena induced by spin-orbit
  coupling at surfaces and interfaces},}\ \href {\doibase 10.1038/nature19820}
  {\bibfield  {journal} {\bibinfo  {journal} {Nature}\ }\textbf {\bibinfo
  {volume} {539}},\ \bibinfo {pages} {509} (\bibinfo {year}
  {2016})}\BibitemShut {NoStop}%
\bibitem [{\citenamefont {Schaffer}\ \emph {et~al.}(2016)\citenamefont
  {Schaffer}, \citenamefont {Lee}, \citenamefont {Yang},\ and\ \citenamefont
  {Kim}}]{Sch16}%
  \BibitemOpen
  \bibfield  {author} {\bibinfo {author} {\bibfnamefont {R.}~\bibnamefont
  {Schaffer}}, \bibinfo {author} {\bibfnamefont {E.~K.-H.}\ \bibnamefont
  {Lee}}, \bibinfo {author} {\bibfnamefont {B.-J.}\ \bibnamefont {Yang}}, \
  and\ \bibinfo {author} {\bibfnamefont {Y.~B.}\ \bibnamefont {Kim}},\
  }\enquote {\bibinfo {title} {Recent progress on correlated electron systems
  with strong spin-orbit coupling},}\ \href {\doibase
  10.1088/0034-4885/79/9/094504} {\bibfield  {journal} {\bibinfo  {journal}
  {Rep. Prog. Phys.}\ }\textbf {\bibinfo {volume} {79}},\ \bibinfo {pages}
  {094504} (\bibinfo {year} {2016})}\BibitemShut {NoStop}%
\bibitem [{\citenamefont {Manchon}\ \emph {et~al.}(2015)\citenamefont
  {Manchon}, \citenamefont {Koo}, \citenamefont {Nitta}, \citenamefont
  {Frolov},\ and\ \citenamefont {Duine}}]{Man15}%
  \BibitemOpen
  \bibfield  {author} {\bibinfo {author} {\bibfnamefont {A.}~\bibnamefont
  {Manchon}}, \bibinfo {author} {\bibfnamefont {H.~C.}\ \bibnamefont {Koo}},
  \bibinfo {author} {\bibfnamefont {J.}~\bibnamefont {Nitta}}, \bibinfo
  {author} {\bibfnamefont {S.~M.}\ \bibnamefont {Frolov}}, \ and\ \bibinfo
  {author} {\bibfnamefont {R.~A.}\ \bibnamefont {Duine}},\ }\enquote {\bibinfo
  {title} {New perspectives for {Rashba} spin-orbit coupling},}\ \href
  {\doibase 10.1038/NMAT4360} {\bibfield  {journal} {\bibinfo  {journal} {Nat.
  Mater.}\ }\textbf {\bibinfo {volume} {14}},\ \bibinfo {pages} {871} (\bibinfo
  {year} {2015})}\BibitemShut {NoStop}%
\bibitem [{\citenamefont {Hatano}\ and\ \citenamefont
  {Nelson}(1996)}]{Hatano96}%
  \BibitemOpen
  \bibfield  {author} {\bibinfo {author} {\bibfnamefont {N.}~\bibnamefont
  {Hatano}}\ and\ \bibinfo {author} {\bibfnamefont {D.~R.}\ \bibnamefont
  {Nelson}},\ }\enquote {\bibinfo {title} {Localization transitions in
  non-{Hermitian} quantum mechanics},}\ \href {\doibase
  10.1103/PhysRevLett.77.570} {\bibfield  {journal} {\bibinfo  {journal} {Phys.
  Rev. Lett.}\ }\textbf {\bibinfo {volume} {77}},\ \bibinfo {pages} {570}
  (\bibinfo {year} {1996})}\BibitemShut {NoStop}%
\bibitem [{\citenamefont {Liu}\ \emph {et~al.}(2022)\citenamefont {Liu},
  \citenamefont {Wei}, \citenamefont {Hemmatyar}, \citenamefont {Pyrialakos},
  \citenamefont {Jung}, \citenamefont {Christodoulides},\ and\ \citenamefont
  {Khajavikhan}}]{Liu2022}%
  \BibitemOpen
  \bibfield  {author} {\bibinfo {author} {\bibfnamefont {Y.~G.~N.}\
  \bibnamefont {Liu}}, \bibinfo {author} {\bibfnamefont {Y.}~\bibnamefont
  {Wei}}, \bibinfo {author} {\bibfnamefont {O.}~\bibnamefont {Hemmatyar}},
  \bibinfo {author} {\bibfnamefont {G.~G.}\ \bibnamefont {Pyrialakos}},
  \bibinfo {author} {\bibfnamefont {P.~S.}\ \bibnamefont {Jung}}, \bibinfo
  {author} {\bibfnamefont {D.~N.}\ \bibnamefont {Christodoulides}}, \ and\
  \bibinfo {author} {\bibfnamefont {M.}~\bibnamefont {Khajavikhan}},\ }\enquote
  {\bibinfo {title} {Complex skin modes in non-{Hermitian} coupled laser
  arrays},}\ \href {\doibase 10.1038/s41377-022-01030-0} {\bibfield  {journal}
  {\bibinfo  {journal} {Light Sci. Appl.}\ }\textbf {\bibinfo {volume} {11}},\
  \bibinfo {pages} {336} (\bibinfo {year} {2022})}\BibitemShut {NoStop}%
\bibitem [{\citenamefont {Li}\ \emph {et~al.}(2024{\natexlab{b}})\citenamefont
  {Li}, \citenamefont {Wang}, \citenamefont {Wang}, \citenamefont {Lin},
  \citenamefont {Ma},\ and\ \citenamefont {Jiang}}]{LiZ2024}%
  \BibitemOpen
  \bibfield  {author} {\bibinfo {author} {\bibfnamefont {Z.}~\bibnamefont
  {Li}}, \bibinfo {author} {\bibfnamefont {L.-W.}\ \bibnamefont {Wang}},
  \bibinfo {author} {\bibfnamefont {X.}~\bibnamefont {Wang}}, \bibinfo {author}
  {\bibfnamefont {Z.-K.}\ \bibnamefont {Lin}}, \bibinfo {author} {\bibfnamefont
  {G.}~\bibnamefont {Ma}}, \ and\ \bibinfo {author} {\bibfnamefont {J.-H.}\
  \bibnamefont {Jiang}},\ }\enquote {\bibinfo {title} {Observation of dynamic
  non-{Hermitian} skin effects},}\ \href {\doibase 10.1038/s41467-024-50776-1}
  {\bibfield  {journal} {\bibinfo  {journal} {Nat. Commun.}\ }\textbf {\bibinfo
  {volume} {15}},\ \bibinfo {pages} {6544} (\bibinfo {year}
  {2024}{\natexlab{b}})}\BibitemShut {NoStop}%
\bibitem [{\citenamefont {Longhi}(2017)}]{Longhi17}%
  \BibitemOpen
  \bibfield  {author} {\bibinfo {author} {\bibfnamefont {S.}~\bibnamefont
  {Longhi}},\ }\enquote {\bibinfo {title} {Non-{Hermitian} bidirectional robust
  transport},}\ \href {\doibase 10.1103/PhysRevB.95.014201} {\bibfield
  {journal} {\bibinfo  {journal} {Phys. Rev. B}\ }\textbf {\bibinfo {volume}
  {95}},\ \bibinfo {pages} {014201} (\bibinfo {year} {2017})}\BibitemShut
  {NoStop}%
\bibitem [{\citenamefont {Zhang}\ \emph {et~al.}(2022)\citenamefont {Zhang},
  \citenamefont {Denner}, \citenamefont {Bzdu\v{s}ek}, \citenamefont {Sentef},\
  and\ \citenamefont {Neupert}}]{Zhang22}%
  \BibitemOpen
  \bibfield  {author} {\bibinfo {author} {\bibfnamefont {S.-B.}\ \bibnamefont
  {Zhang}}, \bibinfo {author} {\bibfnamefont {M.~M.}\ \bibnamefont {Denner}},
  \bibinfo {author} {\bibfnamefont {T.}~\bibnamefont {Bzdu\v{s}ek}}, \bibinfo
  {author} {\bibfnamefont {M.~A.}\ \bibnamefont {Sentef}}, \ and\ \bibinfo
  {author} {\bibfnamefont {T.}~\bibnamefont {Neupert}},\ }\enquote {\bibinfo
  {title} {Symmetry breaking and spectral structure of the interacting
  {Hatano-Nelson} model},}\ \href {\doibase 10.1103/PhysRevB.106.L121102}
  {\bibfield  {journal} {\bibinfo  {journal} {Phys. Rev. B}\ }\textbf {\bibinfo
  {volume} {106}},\ \bibinfo {pages} {L121102} (\bibinfo {year}
  {2022})}\BibitemShut {NoStop}%
\bibitem [{\citenamefont {Chen}\ \emph {et~al.}(2025)\citenamefont {Chen},
  \citenamefont {Basit}, \citenamefont {Li}, \citenamefont {Hou}, \citenamefont
  {Ruan}, \citenamefont {Wei},\ and\ \citenamefont {Ni}}]{Chen25}%
  \BibitemOpen
  \bibfield  {author} {\bibinfo {author} {\bibfnamefont {S.}~\bibnamefont
  {Chen}}, \bibinfo {author} {\bibfnamefont {A.}~\bibnamefont {Basit}},
  \bibinfo {author} {\bibfnamefont {L.}~\bibnamefont {Li}}, \bibinfo {author}
  {\bibfnamefont {C.}~\bibnamefont {Hou}}, \bibinfo {author} {\bibfnamefont
  {Y.}~\bibnamefont {Ruan}}, \bibinfo {author} {\bibfnamefont {Y.}~\bibnamefont
  {Wei}}, \ and\ \bibinfo {author} {\bibfnamefont {Z.}~\bibnamefont {Ni}},\
  }\enquote {\bibinfo {title} {Non-{Hermitian} topological lattice photonics:
  An analytic perspective},}\ \href {\doibase 10.1002/adpr.202500083}
  {\bibfield  {journal} {\bibinfo  {journal} {Adv. Photon. Res.}\ }\textbf
  {\bibinfo {volume} {6}},\ \bibinfo {pages} {2500083} (\bibinfo {year}
  {2025})}\BibitemShut {NoStop}%
\bibitem [{\citenamefont {\"{O}rsel}\ \emph {et~al.}(2025)\citenamefont
  {\"{O}rsel}, \citenamefont {Noh}, \citenamefont {Zhu}, \citenamefont {Yim},
  \citenamefont {Hughes}, \citenamefont {Thomale},\ and\ \citenamefont
  {Bahl}}]{Ors25}%
  \BibitemOpen
  \bibfield  {author} {\bibinfo {author} {\bibfnamefont {O.~E.}\ \bibnamefont
  {\"{O}rsel}}, \bibinfo {author} {\bibfnamefont {J.}~\bibnamefont {Noh}},
  \bibinfo {author} {\bibfnamefont {P.}~\bibnamefont {Zhu}}, \bibinfo {author}
  {\bibfnamefont {J.}~\bibnamefont {Yim}}, \bibinfo {author} {\bibfnamefont
  {T.~L.}\ \bibnamefont {Hughes}}, \bibinfo {author} {\bibfnamefont
  {R.}~\bibnamefont {Thomale}}, \ and\ \bibinfo {author} {\bibfnamefont
  {G.}~\bibnamefont {Bahl}},\ }\enquote {\bibinfo {title} {Giant nonreciprocity
  and gyration through modulation-induced {Hatano-Nelson} coupling in
  integrated photonics},}\ \href {\doibase 10.1103/PhysRevLett.134.153801}
  {\bibfield  {journal} {\bibinfo  {journal} {Phys. Rev. Lett.}\ }\textbf
  {\bibinfo {volume} {134}},\ \bibinfo {pages} {153801} (\bibinfo {year}
  {2025})}\BibitemShut {NoStop}%
\bibitem [{Sup()}]{SuppMat}%
  \BibitemOpen
  \bibinfo {note} {See Supplemental Material at http://link.aps.org/
  supplemental/10.1103/PhysRevLett.xxx for details of solving the non-Hermitian
  SO-coupled HN model under OBC, deriving the edge-state parameter conditions,
  constructing the exact GBZ analytically, and extending the framework to the
  most general symmetry-free model, along with a discussion of the NHSE,
  topological islands, as well as the edge-state robustness against generic
  disorder.}\BibitemShut {Stop}%
\bibitem [{\citenamefont {Li}\ \emph {et~al.}(2025{\natexlab{b}})\citenamefont
  {Li}, \citenamefont {Wei}, \citenamefont {Wu}, \citenamefont {Ruan},
  \citenamefont {Chen}, \citenamefont {Lee},\ and\ \citenamefont {Ni}}]{Li25}%
  \BibitemOpen
  \bibfield  {author} {\bibinfo {author} {\bibfnamefont {L.}~\bibnamefont
  {Li}}, \bibinfo {author} {\bibfnamefont {Y.}~\bibnamefont {Wei}}, \bibinfo
  {author} {\bibfnamefont {G.}~\bibnamefont {Wu}}, \bibinfo {author}
  {\bibfnamefont {Y.}~\bibnamefont {Ruan}}, \bibinfo {author} {\bibfnamefont
  {S.}~\bibnamefont {Chen}}, \bibinfo {author} {\bibfnamefont {C.~H.}\
  \bibnamefont {Lee}}, \ and\ \bibinfo {author} {\bibfnamefont
  {Z.}~\bibnamefont {Ni}},\ }\enquote {\bibinfo {title} {Exact solutions
  disentangle higher-order topology in two-dimensional non-{Hermitian}
  lattices},}\ \href {\doibase 10.1103/PhysRevB.111.075132} {\bibfield
  {journal} {\bibinfo  {journal} {Phys. Rev. B}\ }\textbf {\bibinfo {volume}
  {111}},\ \bibinfo {pages} {075132} (\bibinfo {year}
  {2025}{\natexlab{b}})}\BibitemShut {NoStop}%
\bibitem [{\citenamefont {Longhi}\ \emph {et~al.}(2015)\citenamefont {Longhi},
  \citenamefont {Gatti},\ and\ \citenamefont {Valle}}]{Long15}%
  \BibitemOpen
  \bibfield  {author} {\bibinfo {author} {\bibfnamefont {S.}~\bibnamefont
  {Longhi}}, \bibinfo {author} {\bibfnamefont {D.}~\bibnamefont {Gatti}}, \
  and\ \bibinfo {author} {\bibfnamefont {G.~D.}\ \bibnamefont {Valle}},\
  }\enquote {\bibinfo {title} {Non-{Hermitian} transparency and one-way
  transport in low-dimensional lattices by an imaginary gauge field},}\ \href
  {\doibase 10.1103/PhysRevB.92.094204} {\bibfield  {journal} {\bibinfo
  {journal} {Phys. Rev. B}\ }\textbf {\bibinfo {volume} {92}},\ \bibinfo
  {pages} {094204} (\bibinfo {year} {2015})}\BibitemShut {NoStop}%
\bibitem [{\citenamefont {Sanahal}\ \emph {et~al.}(2025)\citenamefont
  {Sanahal}, \citenamefont {Panda},\ and\ \citenamefont {Nandy}}]{San25}%
  \BibitemOpen
  \bibfield  {author} {\bibinfo {author} {\bibfnamefont {M.}~\bibnamefont
  {Sanahal}}, \bibinfo {author} {\bibfnamefont {S.}~\bibnamefont {Panda}}, \
  and\ \bibinfo {author} {\bibfnamefont {S.}~\bibnamefont {Nandy}},\ }\enquote
  {\bibinfo {title} {Gauge field induced skin effect in spinful non-{Hermitian}
  systems},}\ \href {\doibase 10.1103/vhz9-xwf4} {\bibfield  {journal}
  {\bibinfo  {journal} {Phys. Rev. B}\ }\textbf {\bibinfo {volume} {112}},\
  \bibinfo {pages} {125149} (\bibinfo {year} {2025})}\BibitemShut {NoStop}%
\bibitem [{\citenamefont {Chai}\ \emph {et~al.}(2020)\citenamefont {Chai},
  \citenamefont {Zhao}, \citenamefont {Tang}, \citenamefont {Guo},
  \citenamefont {Zou},\ and\ \citenamefont {Dong}}]{Chai20}%
  \BibitemOpen
  \bibfield  {author} {\bibinfo {author} {\bibfnamefont {C.-Z.}\ \bibnamefont
  {Chai}}, \bibinfo {author} {\bibfnamefont {H.-Q.}\ \bibnamefont {Zhao}},
  \bibinfo {author} {\bibfnamefont {H.~X.}\ \bibnamefont {Tang}}, \bibinfo
  {author} {\bibfnamefont {G.-C.}\ \bibnamefont {Guo}}, \bibinfo {author}
  {\bibfnamefont {C.-L.}\ \bibnamefont {Zou}}, \ and\ \bibinfo {author}
  {\bibfnamefont {C.-H.}\ \bibnamefont {Dong}},\ }\enquote {\bibinfo {title}
  {Non-reciprocity in high-{Q} ferromagnetic microspheres via photonic
  spin-orbit coupling},}\ \href {\doibase 10.1002/lpor.201900252} {\bibfield
  {journal} {\bibinfo  {journal} {Laser Photon. Rev.}\ }\textbf {\bibinfo
  {volume} {14}},\ \bibinfo {pages} {1900252} (\bibinfo {year}
  {2020})}\BibitemShut {NoStop}%
\bibitem [{\citenamefont {Mittal}\ \emph {et~al.}(2021)\citenamefont {Mittal},
  \citenamefont {Moille}, \citenamefont {Srinivasan}, \citenamefont {Chembo},\
  and\ \citenamefont {Hafezi}}]{Mittal21}%
  \BibitemOpen
  \bibfield  {author} {\bibinfo {author} {\bibfnamefont {S.}~\bibnamefont
  {Mittal}}, \bibinfo {author} {\bibfnamefont {G.}~\bibnamefont {Moille}},
  \bibinfo {author} {\bibfnamefont {K.}~\bibnamefont {Srinivasan}}, \bibinfo
  {author} {\bibfnamefont {Y.~K.}\ \bibnamefont {Chembo}}, \ and\ \bibinfo
  {author} {\bibfnamefont {M.}~\bibnamefont {Hafezi}},\ }\enquote {\bibinfo
  {title} {Topological frequency combs and nested temporal solitons},}\ \href
  {\doibase 10.1038/s41567-021-01302-3} {\bibfield  {journal} {\bibinfo
  {journal} {Nat. Phys.}\ }\textbf {\bibinfo {volume} {17}},\ \bibinfo {pages}
  {1169} (\bibinfo {year} {2021})}\BibitemShut {NoStop}%
\bibitem [{\citenamefont {Xin}\ \emph {et~al.}(2023)\citenamefont {Xin},
  \citenamefont {Song}, \citenamefont {Wu}, \citenamefont {Lin}, \citenamefont
  {Zhu},\ and\ \citenamefont {Li}}]{Xin23}%
  \BibitemOpen
  \bibfield  {author} {\bibinfo {author} {\bibfnamefont {H.}~\bibnamefont
  {Xin}}, \bibinfo {author} {\bibfnamefont {W.}~\bibnamefont {Song}}, \bibinfo
  {author} {\bibfnamefont {S.}~\bibnamefont {Wu}}, \bibinfo {author}
  {\bibfnamefont {Z.}~\bibnamefont {Lin}}, \bibinfo {author} {\bibfnamefont
  {S.}~\bibnamefont {Zhu}}, \ and\ \bibinfo {author} {\bibfnamefont
  {T.}~\bibnamefont {Li}},\ }\enquote {\bibinfo {title} {Manipulating the
  non-{Hermitian} skin effect in optical ring resonators},}\ \href {\doibase
  10.1103/PhysRevB.107.165401} {\bibfield  {journal} {\bibinfo  {journal}
  {Phys. Rev. B}\ }\textbf {\bibinfo {volume} {107}},\ \bibinfo {pages}
  {165401} (\bibinfo {year} {2023})}\BibitemShut {NoStop}%
\bibitem [{\citenamefont {Wong}\ \emph {et~al.}(2025)\citenamefont {Wong},
  \citenamefont {Yang}, \citenamefont {Pang},\ and\ \citenamefont
  {Yang}}]{Wong25}%
  \BibitemOpen
  \bibfield  {author} {\bibinfo {author} {\bibfnamefont {B.~T.~T.}\
  \bibnamefont {Wong}}, \bibinfo {author} {\bibfnamefont {S.}~\bibnamefont
  {Yang}}, \bibinfo {author} {\bibfnamefont {Z.}~\bibnamefont {Pang}}, \ and\
  \bibinfo {author} {\bibfnamefont {Y.}~\bibnamefont {Yang}},\ }\enquote
  {\bibinfo {title} {Synthetic non-{Abelian} electric fields and spin-orbit
  coupling in photonic synthetic dimensions},}\ \href {\doibase
  10.1103/PhysRevLett.134.163803} {\bibfield  {journal} {\bibinfo  {journal}
  {Phys. Rev. Lett.}\ }\textbf {\bibinfo {volume} {134}},\ \bibinfo {pages}
  {163803} (\bibinfo {year} {2025})}\BibitemShut {NoStop}%
\bibitem [{\citenamefont {Flower}\ \emph {et~al.}(2024)\citenamefont {Flower},
  \citenamefont {Mehrabad}, \citenamefont {Xu}, \citenamefont {Moille},
  \citenamefont {Suarez-Forero}, \citenamefont {\"{O}rsel}, \citenamefont
  {Bahl}, \citenamefont {Chembo}, \citenamefont {Srinivasan}, \citenamefont
  {Mittal},\ and\ \citenamefont {Hafezi}}]{Flower24}%
  \BibitemOpen
  \bibfield  {author} {\bibinfo {author} {\bibfnamefont {C.~J.}\ \bibnamefont
  {Flower}}, \bibinfo {author} {\bibfnamefont {M.~J.}\ \bibnamefont
  {Mehrabad}}, \bibinfo {author} {\bibfnamefont {L.}~\bibnamefont {Xu}},
  \bibinfo {author} {\bibfnamefont {G.}~\bibnamefont {Moille}}, \bibinfo
  {author} {\bibfnamefont {D.~G.}\ \bibnamefont {Suarez-Forero}}, \bibinfo
  {author} {\bibfnamefont {O.}~\bibnamefont {\"{O}rsel}}, \bibinfo {author}
  {\bibfnamefont {G.}~\bibnamefont {Bahl}}, \bibinfo {author} {\bibfnamefont
  {Y.}~\bibnamefont {Chembo}}, \bibinfo {author} {\bibfnamefont
  {K.}~\bibnamefont {Srinivasan}}, \bibinfo {author} {\bibfnamefont
  {S.}~\bibnamefont {Mittal}}, \ and\ \bibinfo {author} {\bibfnamefont
  {M.}~\bibnamefont {Hafezi}},\ }\enquote {\bibinfo {title} {Observation of
  topological frequency combs},}\ \href {\doibase 10.1126/science.ado0053}
  {\bibfield  {journal} {\bibinfo  {journal} {Science}\ }\textbf {\bibinfo
  {volume} {384}},\ \bibinfo {pages} {1356} (\bibinfo {year}
  {2024})}\BibitemShut {NoStop}%
\bibitem [{\citenamefont {Wu}\ \emph {et~al.}(2025)\citenamefont {Wu},
  \citenamefont {Wei}, \citenamefont {Li}, \citenamefont {Chen}, \citenamefont
  {Bu}, \citenamefont {Baronio}, \citenamefont {Lin}, \citenamefont {Zhu},
  \citenamefont {Trillo},\ and\ \citenamefont {Ni}}]{Wu25}%
  \BibitemOpen
  \bibfield  {author} {\bibinfo {author} {\bibfnamefont {G.}~\bibnamefont
  {Wu}}, \bibinfo {author} {\bibfnamefont {Y.}~\bibnamefont {Wei}}, \bibinfo
  {author} {\bibfnamefont {L.}~\bibnamefont {Li}}, \bibinfo {author}
  {\bibfnamefont {S.}~\bibnamefont {Chen}}, \bibinfo {author} {\bibfnamefont
  {L.}~\bibnamefont {Bu}}, \bibinfo {author} {\bibfnamefont {F.}~\bibnamefont
  {Baronio}}, \bibinfo {author} {\bibfnamefont {T.}~\bibnamefont {Lin}},
  \bibinfo {author} {\bibfnamefont {M.}~\bibnamefont {Zhu}}, \bibinfo {author}
  {\bibfnamefont {S.}~\bibnamefont {Trillo}}, \ and\ \bibinfo {author}
  {\bibfnamefont {Z.}~\bibnamefont {Ni}},\ }\enquote {\bibinfo {title}
  {Ultraflat soliton microcombs in driven quadratic-{Kerr} nonlinear
  microresonators},}\ \href {\doibase 10.1103/cf1p-k6v6} {\bibfield  {journal}
  {\bibinfo  {journal} {Phys. Rev. Lett.}\ }\textbf {\bibinfo {volume} {135}},\
  \bibinfo {pages} {113801} (\bibinfo {year} {2025})}\BibitemShut {NoStop}%
\bibitem [{\citenamefont {Dai}\ \emph {et~al.}(2022)\citenamefont {Dai},
  \citenamefont {Ao}, \citenamefont {Bao}, \citenamefont {Mao}, \citenamefont
  {Chi}, \citenamefont {Fu}, \citenamefont {You}, \citenamefont {Chen},
  \citenamefont {Zhai}, \citenamefont {Tang}, \citenamefont {Yang},
  \citenamefont {Li}, \citenamefont {Yuan}, \citenamefont {Gao}, \citenamefont
  {Lin}, \citenamefont {Thompson}, \citenamefont {O'Brien}, \citenamefont {Li},
  \citenamefont {Hu}, \citenamefont {Gong},\ and\ \citenamefont
  {Wang}}]{Dai22}%
  \BibitemOpen
  \bibfield  {author} {\bibinfo {author} {\bibfnamefont {T.}~\bibnamefont
  {Dai}}, \bibinfo {author} {\bibfnamefont {Y.}~\bibnamefont {Ao}}, \bibinfo
  {author} {\bibfnamefont {J.}~\bibnamefont {Bao}}, \bibinfo {author}
  {\bibfnamefont {J.}~\bibnamefont {Mao}}, \bibinfo {author} {\bibfnamefont
  {Y.}~\bibnamefont {Chi}}, \bibinfo {author} {\bibfnamefont {Z.}~\bibnamefont
  {Fu}}, \bibinfo {author} {\bibfnamefont {Y.}~\bibnamefont {You}}, \bibinfo
  {author} {\bibfnamefont {X.}~\bibnamefont {Chen}}, \bibinfo {author}
  {\bibfnamefont {C.}~\bibnamefont {Zhai}}, \bibinfo {author} {\bibfnamefont
  {B.}~\bibnamefont {Tang}}, \bibinfo {author} {\bibfnamefont {Y.}~\bibnamefont
  {Yang}}, \bibinfo {author} {\bibfnamefont {Z.}~\bibnamefont {Li}}, \bibinfo
  {author} {\bibfnamefont {L.}~\bibnamefont {Yuan}}, \bibinfo {author}
  {\bibfnamefont {F.}~\bibnamefont {Gao}}, \bibinfo {author} {\bibfnamefont
  {X.}~\bibnamefont {Lin}}, \bibinfo {author} {\bibfnamefont {M.~G.}\
  \bibnamefont {Thompson}}, \bibinfo {author} {\bibfnamefont {J.~L.}\
  \bibnamefont {O'Brien}}, \bibinfo {author} {\bibfnamefont {Y.}~\bibnamefont
  {Li}}, \bibinfo {author} {\bibfnamefont {X.}~\bibnamefont {Hu}}, \bibinfo
  {author} {\bibfnamefont {Q.}~\bibnamefont {Gong}}, \ and\ \bibinfo {author}
  {\bibfnamefont {J.}~\bibnamefont {Wang}},\ }\enquote {\bibinfo {title}
  {Topologically protected quantum entanglement emitters},}\ \href {\doibase
  10.1038/s41566-021-00944-2} {\bibfield  {journal} {\bibinfo  {journal} {Nat.
  Photon.}\ }\textbf {\bibinfo {volume} {16}},\ \bibinfo {pages} {248}
  (\bibinfo {year} {2022})}\BibitemShut {NoStop}%
\bibitem [{\citenamefont {Mei}\ \emph {et~al.}(2020)\citenamefont {Mei},
  \citenamefont {Guo}, \citenamefont {Yu}, \citenamefont {Xiao}, \citenamefont
  {Zhu},\ and\ \citenamefont {Jia}}]{Mei20}%
  \BibitemOpen
  \bibfield  {author} {\bibinfo {author} {\bibfnamefont {F.}~\bibnamefont
  {Mei}}, \bibinfo {author} {\bibfnamefont {Q.}~\bibnamefont {Guo}}, \bibinfo
  {author} {\bibfnamefont {Y.-F.}\ \bibnamefont {Yu}}, \bibinfo {author}
  {\bibfnamefont {L.}~\bibnamefont {Xiao}}, \bibinfo {author} {\bibfnamefont
  {S.-L.}\ \bibnamefont {Zhu}}, \ and\ \bibinfo {author} {\bibfnamefont
  {S.}~\bibnamefont {Jia}},\ }\enquote {\bibinfo {title} {Digital simulation of
  topological matter on programmable quantum processors},}\ \href {\doibase
  10.1103/PhysRevLett.125.160503} {\bibfield  {journal} {\bibinfo  {journal}
  {Phys. Rev. Lett.}\ }\textbf {\bibinfo {volume} {125}},\ \bibinfo {pages}
  {160503} (\bibinfo {year} {2020})}\BibitemShut {NoStop}%
\bibitem [{\citenamefont {Dai}\ \emph {et~al.}(2024{\natexlab{b}})\citenamefont
  {Dai}, \citenamefont {Ma}, \citenamefont {Mao}, \citenamefont {Ao},
  \citenamefont {Jia}, \citenamefont {Zheng}, \citenamefont {Zhai},
  \citenamefont {Yang}, \citenamefont {Li}, \citenamefont {Tang}, \citenamefont
  {Luo}, \citenamefont {Zhang}, \citenamefont {Hu}, \citenamefont {Gong},\ and\
  \citenamefont {Wang}}]{Dai24}%
  \BibitemOpen
  \bibfield  {author} {\bibinfo {author} {\bibfnamefont {T.}~\bibnamefont
  {Dai}}, \bibinfo {author} {\bibfnamefont {A.}~\bibnamefont {Ma}}, \bibinfo
  {author} {\bibfnamefont {J.}~\bibnamefont {Mao}}, \bibinfo {author}
  {\bibfnamefont {Y.}~\bibnamefont {Ao}}, \bibinfo {author} {\bibfnamefont
  {X.}~\bibnamefont {Jia}}, \bibinfo {author} {\bibfnamefont {Y.}~\bibnamefont
  {Zheng}}, \bibinfo {author} {\bibfnamefont {C.}~\bibnamefont {Zhai}},
  \bibinfo {author} {\bibfnamefont {Y.}~\bibnamefont {Yang}}, \bibinfo {author}
  {\bibfnamefont {Z.}~\bibnamefont {Li}}, \bibinfo {author} {\bibfnamefont
  {B.}~\bibnamefont {Tang}}, \bibinfo {author} {\bibfnamefont {J.}~\bibnamefont
  {Luo}}, \bibinfo {author} {\bibfnamefont {B.}~\bibnamefont {Zhang}}, \bibinfo
  {author} {\bibfnamefont {X.}~\bibnamefont {Hu}}, \bibinfo {author}
  {\bibfnamefont {Q.}~\bibnamefont {Gong}}, \ and\ \bibinfo {author}
  {\bibfnamefont {J.}~\bibnamefont {Wang}},\ }\enquote {\bibinfo {title} {A
  programmable topological photonic chip},}\ \href {\doibase
  10.1038/s41563-024-01904-1} {\bibfield  {journal} {\bibinfo  {journal} {Nat.
  Mater.}\ }\textbf {\bibinfo {volume} {23}},\ \bibinfo {pages} {928} (\bibinfo
  {year} {2024}{\natexlab{b}})}\BibitemShut {NoStop}%
\bibitem [{\citenamefont {Li}\ \emph {et~al.}(2026)\citenamefont {Li},
  \citenamefont {Wei}, \citenamefont {Ruan}, \citenamefont {Wu}, \citenamefont
  {Wang}, \citenamefont {Chen}, \citenamefont {Lin}, \citenamefont {Lee},\ and\
  \citenamefont {Ni}}]{DataCode}%
  \BibitemOpen
  \bibfield  {author} {\bibinfo {author} {\bibfnamefont {L.}~\bibnamefont
  {Li}}, \bibinfo {author} {\bibfnamefont {Y.}~\bibnamefont {Wei}}, \bibinfo
  {author} {\bibfnamefont {Y.}~\bibnamefont {Ruan}}, \bibinfo {author}
  {\bibfnamefont {G.}~\bibnamefont {Wu}}, \bibinfo {author} {\bibfnamefont
  {J.}~\bibnamefont {Wang}}, \bibinfo {author} {\bibfnamefont {S.}~\bibnamefont
  {Chen}}, \bibinfo {author} {\bibfnamefont {T.}~\bibnamefont {Lin}}, \bibinfo
  {author} {\bibfnamefont {C.~H.}\ \bibnamefont {Lee}}, \ and\ \bibinfo
  {author} {\bibfnamefont {Z.}~\bibnamefont {Ni}},\ }\href@noop {} {\enquote
  {\bibinfo {title} {Research data and codes supporting {Non-Hermitian Topology
  Driven by an Identity Term: An Exactly Solvable Paradigm}},}\ }\bibinfo
  {howpublished}
  {\url{https://github.com/Lingfang-Li/Identity-Term-Driven-Topology}}
  (\bibinfo {year} {2026})\BibitemShut {NoStop}%
\end{thebibliography}%

\onecolumngrid
\section*{End Matter}
\vspace{0.5cm}
\twocolumngrid

\appendix

{\it Topological phase transition}---In the conventional non-Hermitian SSH model, the bulk spectrum under OBC consists of two horizontally separated arcs in the complex plane, which are symmetric about $E=0$ due to chiral symmetry. The topological zero mode, when present, resides in the open gap between these two arcs. The topological phase transition occurs when the two arcs touch each other at $E=0$, closing the line gap and causing the zero mode to merge into the bulk continuum at the phase boundary.

\begin{figure}[H]
\centering
\includegraphics[width=8.7cm]{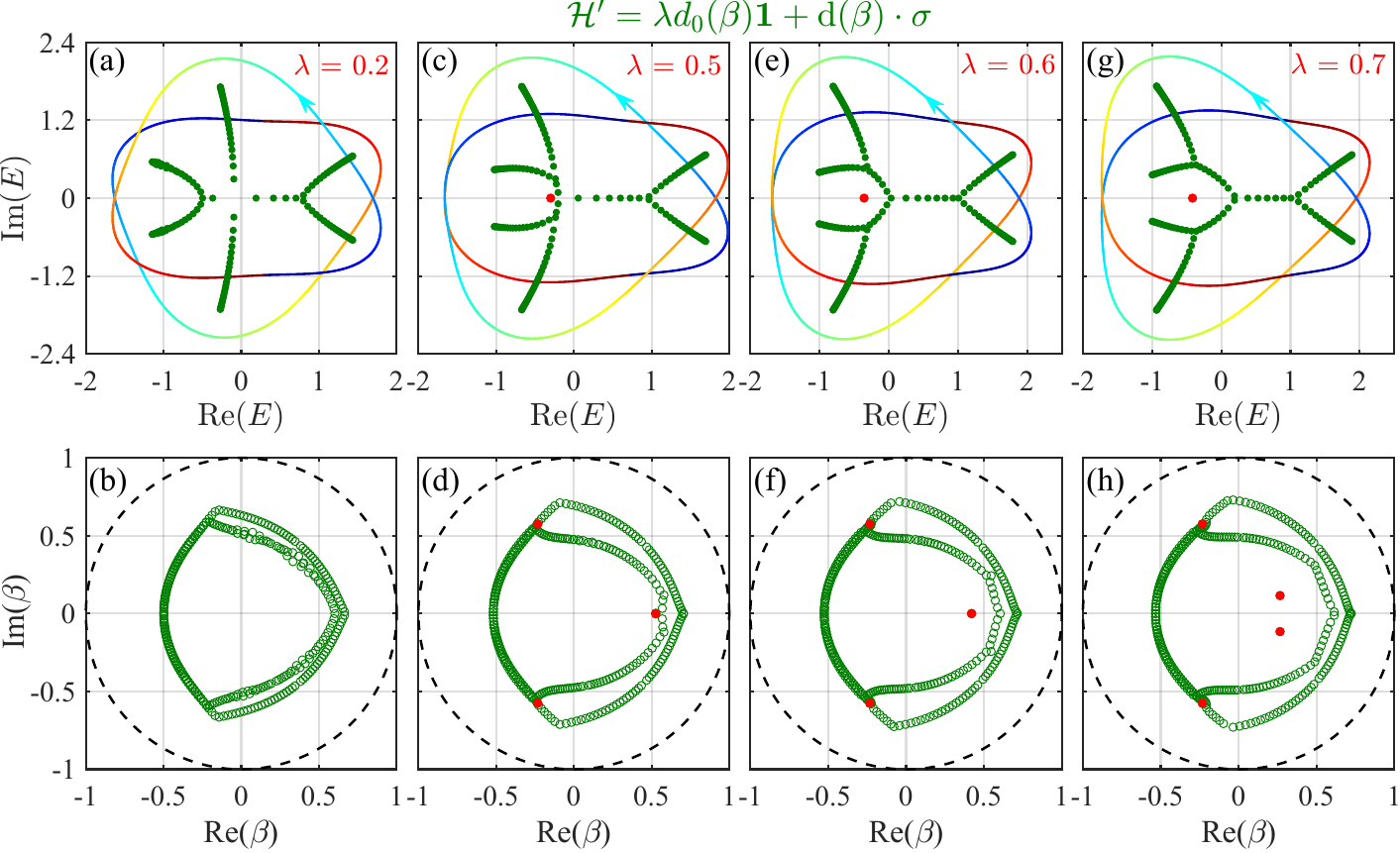}
\caption{Evolution of OBC spectrum (top row) and GBZ (bottom row) as a function of $\lambda$ in $\mathcal{H}^\prime(\beta,\lambda) = \lambda d_0(\beta)\mathds{1} + \mathbf{d}(\beta)\cdot\boldsymbol{\sigma}$, with other parameters identical to Fig.~\ref{fig1}(c) (where $\lambda=0$ and $\lambda=1$ are shown). (a,b) $\lambda=0.2$; (c,d) $\lambda=0.5$; (e,f) $\lambda=0.6$; (g,h) $\lambda=0.7$. The topological edge state (red dot) emerges at $\lambda\approx0.5$.}
\label{fig5}
\end{figure}

However, in our model with an identity term, the situation is fundamentally different. As demonstrated in Fig.~\ref{fig5}, even with all other parameters fixed, continuously tuning the strength $\lambda$ of the identity term from $\lambda=0$ [the reduced model in Fig.~\ref{fig1}(c)] to $\lambda=1$ (the full model) deforms the GBZ and the OBC spectrum. At $\lambda \approx 0.5$, a topological edge state emerges from the bulk continuum, showing that the identity term is an active driver of the transition.

\begin{figure}[H]
\centering
\includegraphics[width=8.7cm]{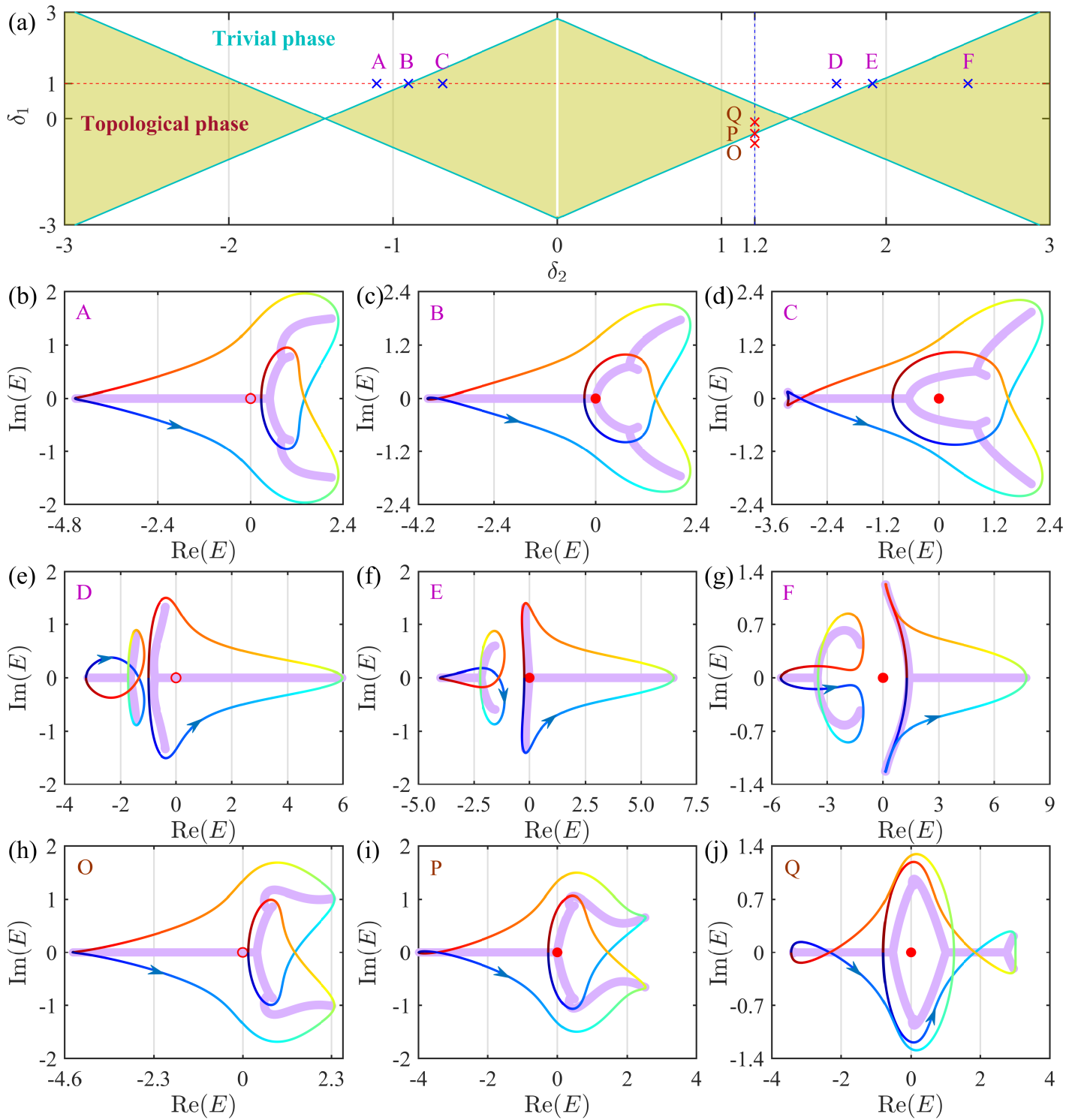}
\caption{Topological phase transition for three distinct spectral morphologies. (a) Zoomed-in zero-mode phase diagram reproduced from Fig.~\ref{fig3}(d), with points A--C, D--F, and O--Q marking three sets of parameters crossing the phase boundaries. (b)-(d) OBC (pink thick curves) and PBC (colored loops) spectra for points A ($\delta_2=-1.1$), B ($\delta_2=-0.91$), C ($\delta_2=-0.7$), showing the zero mode (red dots) emerging from the multi-branch continuum. (e)-(g) Same for D ($\delta_2=1.7$), E ($\delta_2=1.92$), F ($\delta_2=2.5$), showing the zero mode emerging from the right arc. (h)-(j) Same for O ($\delta_1=-0.7$), P ($\delta_1=-0.422$), Q ($\delta_1=-0.1$), showing the zero mode emerging from a closed-loop continuum. For finite-size OBC spectra at C, F, and Q points, see Supplemental Fig.~4(b), Fig.~\ref{fig3}(e), and Supplemental Fig.~4(e), respectively.}
\label{fig6}
\end{figure}

With the identity term present, the phase diagram defined by Eq.~(\ref{Eq10}) supports three distinct spectral morphologies: well-separated arcs [see Figs.~\ref{fig2}(n) and \ref{fig3}(e)], multi-branch spectra [see Figs.~\ref{fig2}(e), \ref{fig2}(k), and Supplemental Fig.~4(b)], and closed-loop spectra [see Supplemental Fig.~4(e)]. These diverse morphologies are a direct consequence of the interplay between the NHSE (driven by point-gap topology) and the wavefunction topology on the GBZ: the NHSE can deform the OBC bulk spectrum into arcs or closed loops, thereby concealing a generalized line gap defined in the parameter space. To examine the topological phase transition, we zoom in on the zero-mode phase diagram of Fig.~\ref{fig3}(d) [reproduced as Fig.~\ref{fig6}(a) with a tighter ($\delta_2,~\delta_1$) range] and select three sets of points crossing the phase boundaries: A--C ($\delta_1=1$) across the first phase boundary, D--F ($\delta_1=1$) across the second, and O-Q ($\delta_2=1.2$) across the third. The corresponding OBC spectra at $N=\infty$ (together with PBC spectra) are shown in Figs.~\ref{fig6}(b)--\ref{fig6}(j), with the zero modes denoted by red solid circles. As illustrated, due to the absence of ($E,~-E$) symmetry, the topological transition in all three cases occurs not by the meeting of two arcs as in the SSH model, but by the continuous deformation of the bulk spectrum---a consequence of the GBZ deformation driven by the identity term (in synergy with $\delta_2$). This deformation causes the generalized line gap in parameter space to close or open, which physically manifests as the zero mode being gradually uncovered from or enclosed into the bulk continuum.

This topological phase transition can be captured by the topological invariant of Ref.~\cite{Zhong25}, which is defined as
\begin{equation*}
W = W_1 + W_2 ,
\end{equation*}
\begin{equation*}
W_j = \left(1 + \frac{1}{2\pi i} \oint_{M(\mathcal{C}_{\mathrm{GBZ}})} dM \frac{d}{dM} \ln\frac{M - M_{\mathrm{deg},j}}{M - M_{\mathrm{branch}}}\right) \mod 2,
\end{equation*}
where the integration contour \(M(\mathcal{C}_{\mathrm{GBZ}})\) is the image of the GBZ on the complex $M$-plane.
The two eigenvector degeneracy points $M_{\mathrm{deg},j}$ ($j=1,2$) and the four branch points $M_{\mathrm{branch}}$ of the Riemann surface can be obtained from Eqs. (S69) and (S70) in Supplemental Material \cite{SuppMat}, respectively. We find that, at the phase boundary, both $M_{\mathrm{deg},j}$ points lie exactly on the $M(\mathcal{C}_{\mathrm{GBZ}})$ contour [see Supplemental Fig.~7(f)], rendering $W_1$ and $W_2$ undefined simultaneously. This critical behavior of the invariant precisely signifies the underlying topological phase transition.

\end{document}

\title{Supplemental Material for ``Non-Hermitian Topology Driven by an Identity Term: An Exactly Solvable Paradigm''}

\author{Lingfang Li}
\affiliation{Key Laboratory of Quantum Materials and Devices of Ministry of Education, School of Physics, Frontiers Science Center for Mobile Information Communication and Security, Southeast University, Nanjing 211189, China}

\author{Yating Wei}
\affiliation{Key Laboratory of Quantum Materials and Devices of Ministry of Education, School of Physics, Frontiers Science Center for Mobile Information Communication and Security, Southeast University, Nanjing 211189, China}

\author{Yang Ruan}
\affiliation{Key Laboratory of Quantum Materials and Devices of Ministry of Education, School of Physics, Frontiers Science Center for Mobile Information Communication and Security, Southeast University, Nanjing 211189, China}

\author{Gangzhou Wu}
\affiliation{Key Laboratory of Quantum Materials and Devices of Ministry of Education, School of Physics, Frontiers Science Center for Mobile Information Communication and Security, Southeast University, Nanjing 211189, China}

\author{Jun Wang}
\affiliation{Key Laboratory of Quantum Materials and Devices of Ministry of Education, School of Physics, Frontiers Science Center for Mobile Information Communication and Security, Southeast University, Nanjing 211189, China}

\author{Shihua Chen}
\affiliation{Key Laboratory of Quantum Materials and Devices of Ministry of Education, School of Physics, Frontiers Science Center for Mobile Information Communication and Security, Southeast University, Nanjing 211189, China}
\affiliation{Purple Mountain Laboratories, Nanjing 211111, China}

\author{Tong Lin}
\affiliation{Advanced Photonics Center, School of Electronic Science and Engineering, Southeast University, Nanjing 210096, China}

\author{Ching Hua Lee}
\affiliation{Department of Physics, National University of Singapore, Singapore 117551, Republic of Singapore}

\author{Zhenhua Ni}
\affiliation{Key Laboratory of Quantum Materials and Devices of Ministry of Education, School of Physics, Frontiers Science Center for Mobile Information Communication and Security, Southeast University, Nanjing 211189, China}
\affiliation{Purple Mountain Laboratories, Nanjing 211111, China}
\affiliation{Advanced Photonics Center, School of Electronic Science and Engineering, Southeast University, Nanjing 210096, China}

%\date{\today}
\maketitle

 %     \vspace{1.0cm}

The Supplemental Material is organized as follows. Section \ref{Sec1} presents exact solutions of the non-Hermitian spin-orbit (SO) coupled Hatano-Nelson (HN) model and an analysis of the role of the momentum-dependent identity term. In Sec.~\ref{Sec2}, we provide a detailed derivation of the full eigensystem under open boundary conditions (OBC). Section \ref{Sec3} derives the energy and parameter conditions for topological edge states in the thermodynamic limit. Section \ref{Sec4} discusses the non-Hermitian skin effect and its connection to point-gap topology. In Sec.~\ref{Sec5}, we introduce the concept of topological islands in the phase diagrams and perform a direct test of the symmetry-free topological invariant. Section \ref{Sec6} provides robustness tests of the topological edge states against generic disorders. Section \ref{Sec7} presents the analytical construction of the generalized Brillouin zone (GBZ). Finally, in Sec.~\ref{Sec8}, we extend the analytical framework to the most general SO-coupled HN model that retains no conventional symmetry.

\section{Exact Solutions and the Role of the Identity Term} \label{Sec1}

In this section, we present exact solutions of the non-Hermitian SO-coupled HN model and elucidate the decisive role played by the momentum-dependent identity term. We begin by writing the Hamiltonian under OBC as
\begin{align}
\begin{split}
\hat{H}=\sum_{n=1}^{N}\left\{t_{L}e^{i\alpha_L\sigma_z}\hat{c}_{n}^{\dagger}\hat{c}_{n+1}
+t_{R}e^{i\alpha_R\sigma_z}\hat{c}_{n+1}^{\dagger}\hat{c}_{n}
+\delta_1\hat{c}_{n}^{\dagger}(\boldsymbol{b}\cdot \boldsymbol{\sigma})\hat{c}_{n}+\left[\delta_2\hat{c}_{n}^{\dagger}(\boldsymbol{b}\cdot \boldsymbol{\sigma})\hat{c}_{n+1}+h.c.\right]\right\},
\end{split}\tag{S1}\label{S1}
\end{align}
where $\hat{c}_{n}^{\dagger}=(\hat{c}_{n,\uparrow}^{\dagger},\hat{c}_{n,\downarrow}^{\dagger})$ creates an electron with spin up ($\uparrow$) or down ($\downarrow$) on the $n$th unit cell (with $\hat{c}_{N+1}\equiv0$), $\boldsymbol{\sigma}=(\sigma_x,\sigma_y,\sigma_z)$ are the Pauli matrices, and $\boldsymbol{b}=(1,0,0)$ sets the magnetic field axis. The parameters $\delta_1$ and $\delta_2$ denote the intra-cell and inter-cell SO strengths, respectively. The hopping amplitudes carry SU(2) phases $e^{i\alpha_{L,R}\sigma_z}$ with real $t_{L,R}$ and $\alpha_{L,R}$. It is easy to show that this SO-coupled Hamiltonian does not respect chiral symmetry \cite{Kawa19}.

In momentum space ($\beta\equiv e^{ik}$), the Hamiltonian (via Fourier transform) becomes
\begin{align}
\mathcal{H}(\beta)=d_0(\beta)\mathds{1}+\boldsymbol{\mathrm{d}}(\beta)\cdot \boldsymbol{\sigma}=\left[\begin{array}{cc}G_{\uparrow}(\beta) &\Delta(\beta)\\\Delta(\beta)&G_{\downarrow}(\beta)
	\end{array}\right], \tag{S2}\label{S2}
\end{align}
where $\mathds{1}$ is the $2\times2$ identity matrix and
\begin{align}
d_0(\beta)=\frac{G_{\uparrow}(\beta)+G_{\downarrow}(\beta)}{2}=t_L\cos(\alpha_L)\beta
+\frac{t_R\cos(\alpha_R)}{\beta},~~~~
\boldsymbol{\mathrm{d}}(\beta)=\left(\Delta(\beta), ~0,~\frac{G_{\uparrow}(\beta)-G_{\downarrow}(\beta)}{2}\right).\tag{S3}\label{S3}
\end{align}
Here, $G_{\uparrow,\downarrow}(\beta)$ describe the spin-conserving hoppings, while $\Delta(\beta)$ encodes the spin-flip coupling induced by the SO interaction:
\begin{align}
G_{\uparrow}(\beta) =t_{L}e^{i\alpha_L}\beta+\frac{t_{R}e^{i\alpha_R}}{\beta},~~~~G_{\downarrow}(\beta) = t_{L}e^{-i\alpha_L}\beta+\frac{t_{R}e^{-i\alpha_R}}{\beta},~~~~\Delta(\beta)=\delta_{1}
+\delta_{2}\left(\beta+\frac{1}{\beta}\right).\tag{S4}\label{S4}
\end{align}
The term $d_0(\beta)\mathds{1}$ acts as a momentum-dependent diagonal shift. Unlike a constant energy offset, this $\beta$-dependent term actively deforms the GBZ and thereby alters the topological classification. Its presence moves the model outside the standard symmetry-based topological frameworks and necessitates the dedicated analytical treatment developed in this work.

Subsequently, we present the exact solution of the Schr{\"o}dinger equation $\hat{H}|\psi\rangle=E|\psi\rangle$ under OBC. The energy spectrum $E$ must satisfy $\mathrm{det}[\mathcal{H}(\beta)-E]=0$, which simplifies to
\begin{align}
\left[G_{\uparrow}(\beta)-E\right]\left[G_{\downarrow}(\beta)-E\right]
=[\delta_{1}+\delta_{2}(\beta+1/\beta)]^2. \tag{S5}\label{S5}
\end{align}
For a given $E$, Eq.~(\ref{S5}) yields four complex roots $\beta_i\ (i=1,\ldots,4)$. Because the bulk equations couple each site to its nearest neighbors, the general solution can be expressed as a linear combination of functions of the form $\beta_i^n$. Writing the eigenvector in the component form $|\psi\rangle=(\psi^{\uparrow}_{1}, \psi^{\downarrow}_{1},\psi^{\uparrow}_{2}, \psi^{\downarrow}_{2},\cdots,\psi^{\uparrow}_{N}, \psi^{\downarrow}_{N})^\mathrm{T}$ (with $\mathrm{T}$ denoting transpose), we have
\begin{align}
\psi^{\uparrow}_{n}=\sum_{i=1}^{4}c_i\phi(\beta_{i})\beta_i^n,~~~~
\psi^{\downarrow}_{n}=\sum_{i=1}^{4}c_i\beta_i^n, \tag{S6}\label{S6}
\end{align}
where the spinor component $\phi(\beta_{i})$ is determined from the bulk equations,
\begin{align}
\phi(\beta_{i})=\frac{\delta_{1}+\delta_{2}(\beta_i+1/\beta_i)}
{E-G_{\uparrow}(\beta_i)},\tag{S7}\label{S7}
\end{align}
and the coefficients $c_i$ are fixed (up to an overall factor) by the boundary conditions,
\begin{align}
c_{i}=\frac{1}{2}\sum_{j,k,l=1}^{4}\epsilon_{ijkl}\beta_j^{N+1}[\phi(\beta_k)-\phi(\beta_l)], \tag{S8}\label{S8}
\end{align}
with $\epsilon_{ijkl}$ being the four-dimensional Levi-Civita symbol.

The eigenenergies $E$ are ultimately determined by requiring that the wavefunction (\ref{S6}) satisfies all boundary equations simultaneously. This condition can be condensed, after some algebra, into a single $2N$-degree polynomial equation:
\begin{align}
p_1\lambda_{N+1}^{[1]}+p_2\lambda_{N+1}^{[2]}=0, \tag{S9}\label{S9}
\end{align}
where
\begin{align}
p_1=\sqrt{(t_R^2-\delta_{2}^2)(t_L^2-\delta_{2}^2)},~~~~
p_2=\delta_{2}^2-t_Lt_R\cos(\alpha_L-\alpha_R), \tag{S10}\label{S10}
\end{align}
and the polynomials $\lambda_{N+1}^{[i]}$ ($i=1,2$) are generated by the coupled recurrence relations
\begin{align}
\lambda_{n+1}^{[i]}=2A\lambda_n^{[i]}+2\mu_n^{[i]}-\lambda_{n-1}^{[i]},~~~~
\mu_{n+1}^{[i]}=-2B\lambda_n^{[i]}+2\gamma_n^{[i]}-\mu_{n-1}^{[i]},~~~~
\gamma_{n+1}^{[i]}=2C\lambda_n^{[i]}-\gamma_{n-1}^{[i]}, \tag{S11}\label{S11}
\end{align}
with the initial conditions
\begin{align}
\lambda_{0}^{[1]}=0,\mu_{0}^{[1]}=1,\gamma_{0}^{[1]}=0,
\lambda_{1}^{[1]}=1,\mu_{1}^{[1]}=0,\gamma_{1}^{[1]}=0,\tag{S12}\label{S12}
\end{align}
\begin{align}
\lambda_{0}^{[2]}=0,\mu_{0}^{[2]}=0,\gamma_{0}^{[2]}=1,
\lambda_{1}^{[2]}=0,\mu_{1}^{[2]}=1,\gamma_{1}^{[2]}=0.\tag{S13}\label{S13}
\end{align}
Here the coefficients $A$, $B$, and $C$ in Eqs. (\ref{S11}) are given by
\begin{align}
A=\frac{E^2-\delta_1^2-2p_2}{2p_1},  \tag{S14}\label{S14}
\end{align}
\begin{align}
B=\frac{(\delta_{1}\delta_{2}+Et_L\cos\alpha_L)(\delta_{1}\delta_{2}+Et_R\cos\alpha_R)}{p_1^2}-1, \tag{S15}\label{S15}
\end{align}
\begin{align}
C=\frac{(\kappa_2-1)E^2}{2p_1}+\frac{\kappa_1\delta_{1}\delta_{2}E}{p_1}
+\frac{2p_2+\delta_1^2+\kappa_0\delta_{1}^2\delta_{2}^2}{2p_1}, \tag{S16}\label{S16}
\end{align}
with
\begin{align}
\kappa_m=\frac{t_L^m\cos^m\alpha_L}{t_L^2-\delta^2_{2}}+ \frac{t_R^m\cos^m\alpha_R}{t_R^2-\delta^2_{2}},~~~(m=0,1,2). \tag{S17}\label{S17}
\end{align}
Hence, the full energy spectrum is obtained by solving Eq.~(\ref{S9}). Each eigenvalue $E$ then determines the corresponding $\beta_i$ via Eq.~(\ref{S5}) and, finally, the eigenfunction $|\psi\rangle$ via Eqs.~(\ref{S6})--(\ref{S8}).

\begin{figure}[h!]
\begin{center}
\includegraphics[width=17cm]{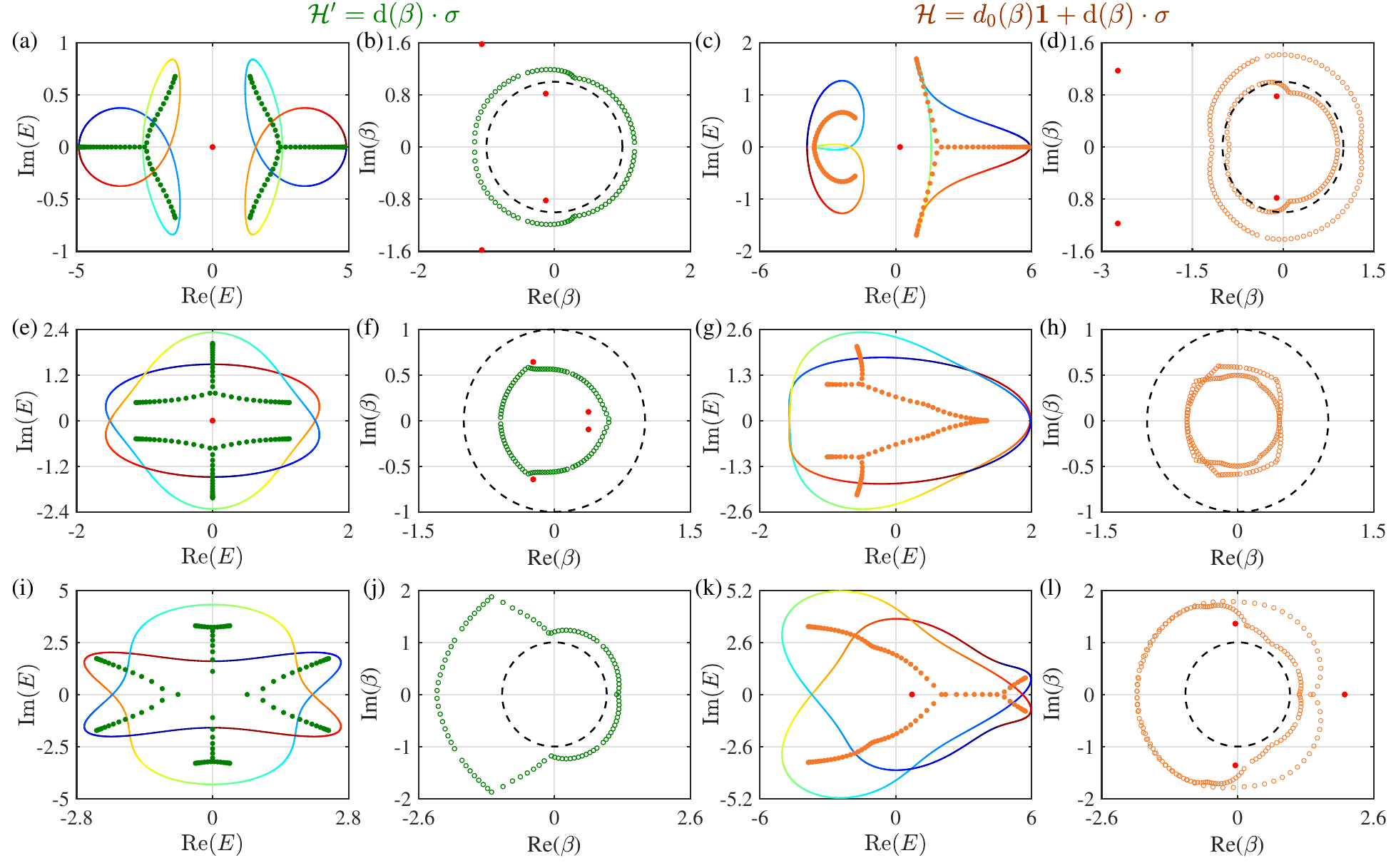}
\caption{\hbadness=10000 \vbadness=10000
  \setlength{\leftskip}{0pt}%
  \setlength{\rightskip}{0pt}%
  \setlength{\parfillskip}{0pt plus 1fil} Comparison between the reduced model $\mathcal{H}^\prime(\beta)=\mathbf{d}(\beta)\cdot\boldsymbol{\sigma}$ (left two columns) and the full model $\mathcal{H}(\beta)=d_0(\beta)\mathds{1}+\mathbf{d}(\beta)\cdot\boldsymbol{\sigma}$ (right two columns) for three distinct parameter sets, with $N=50$. Each row uses the same parameters for both models, isolating the effect of the identity term $d_0(\beta)\mathds{1}$. Reduced model (left): (a,e,i) Complex energy spectra under periodic boundary condition (PBC) (colored loops) and OBC (green dots); (b,f,j) Corresponding GBZ (green open circles) and BZ (dashed unit circle). Full model (right): (c,g,k) and (d,h,l) show the same quantities, where the OBC spectra and GBZ are marked by yellow dots and yellow open circles, respectively. In each panel, the topological edge state, when present, is highlighted by red markers.
The parameter sets used in rows 1--3 are specified by: (1) $t_L=2, t_R=0.5, \alpha_L=\pi/3, \alpha_R=\pi/2, \delta_1=1, \delta_2=2.2$; (2) $t_L=2$, $t_R=0.6$, $\alpha_L=\pi/3$, $\alpha_R=\pi/2$, $\delta_1=1$, $\delta_2=0.4$;
(3) $t_L=2, t_R=5, \alpha_L=\pi/3, \alpha_R=\pi/4, \delta_1=1, \delta_2=2.01$.} \label{figS1}
\end{center}
\end{figure}

To illustrate the decisive role of the identity term $d_0(\beta)\mathds{1}$, we compare in Supplemental Fig.~\ref{figS1} the reduced model $\mathcal{H}^\prime(\beta)=\mathbf{d}(\beta)\cdot\boldsymbol{\sigma}$ (lacking the identity term) with the full model $\mathcal{H}(\beta)=d_0(\beta)\mathds{1}+\mathbf{d}(\beta)\cdot\boldsymbol{\sigma}$. For each row, the same set of parameters is used for both models, revealing how $d_0(\beta)$ actively reshapes both the GBZ and the topological properties. Obviously, while the GBZ in the reduced model typically forms a single loop, its counterpart in the full model can be deformed into either two separated loops or a single twisted contour. Despite this universal GBZ deformation, the topological outcome varies: in rows~2 and~3, the identity term respectively annihilates and creates topological edge states, thereby driving a topological phase transition; in row~1, it does not alter the presence of edge states but qualitatively modifies their topological nature---from zero-energy edge states protected by a hidden chiral symmetry that enforces $(E,-E)$ spectral pairing in the reduced model, to finite-energy edge states that survive in a fully non-chiral setting when the identity term explicitly breaks this symmetry. These comparative analyses underscore that the momentum-dependent identity term is not a passive spectral offset but an active agent that can create, destroy, or qualitatively alter topological edge states, and dramatically reconfigure the GBZ geometry. This positions our model beyond conventional symmetry-based topological classifications.

\section{Derivation of Exact Eigensystem under OBC} \label{Sec2}

This section provides a detailed derivation of the exact open-boundary eigensystem summarized in Sec.~\ref{Sec1}. Starting from the real-space Hamiltonian, we derive the bulk and boundary equations and show the algebraic steps that lead to the compact polynomial equation for the spectrum, Eq.~(\ref{S9}). To proceed, we first write the Hamiltonian \eqref{S1} in a $2N \times 2N$ matrix form:
\begin{align}
	H=	\left[\begin{array}{ccccccccccccc}
		0 & \delta_{1}     &t_{L}e^{i\alpha_L}   &\delta_{2}           &0           & 0            &0           &   0          &  \cdots     &0     & 0     &0     & 0         \\
		\delta_{1}       &0 & \delta_{2}       & t_{L}e^{-i\alpha_L}   & 0          & 0            &0           &     0         &\cdots        &0 & 0    & 0     &0        \\
		t_{R}e^{i\alpha_R}    & \delta_{2}      & 0  & \delta_{1}      &t_{L}e^{i\alpha_L}   &\delta_{2}             &0           &0             & \cdots      &0  & 0   &0  & 0       \\
		\delta_{2}            &t_{R}e^{-i\alpha_R}   & \delta_{1}       &0  &\delta_{2}      &t_{L}e^{-i\alpha_L}     &0           &0             & \cdots      &0  & 0    &0  & 0         \\
		0            & 0           &t_{R}e^{i\alpha_R}     &\delta_{2}       &0  &\delta_{1}        &t_{L}e^{i\alpha_L}  &\delta_{2}          & \cdots       &0  & 0  &0  & 0        \\
		0            & 0           &\delta_{2}             &t_{R}e^{-i\alpha_R}   &\delta_{1}      & 0 & \delta_{2}     &t_{L}e^{-i\alpha_L}      & \cdots       &0 & 0    &0 & 0          \\
		0            & 0           &0             &0            &t_{R}e^{i\alpha_R}  &\delta_{2}        &0 & \delta_{1}        &\cdots       &0    &0  &0    &0        \\
		0            & 0           & 0            & 0       & \delta_{2}      &t_{R}e^{-i\alpha_R}           &\delta_{1}     &0      &\cdots     &0      &0     & 0    & 0        \\
\vdots       &\vdots       &\vdots        &\vdots       &\vdots      &\vdots        &\vdots      &   \vdots       &\ddots     & \vdots       & \vdots & \vdots       & \vdots      \\
	0            &0            &0             &0            &0           &0             &0           &0              &\cdots       & 0  &\delta_{1}   & t_{L}e^{i\alpha_L}  &\delta_{2}  \\
		0            & 0           & 0            & 0          & 0           & 0            &0           &0              & \cdots        &\delta_{1}        &0  & \delta_{2}    &t_{L}e^{-i\alpha_L}   \\
		0            &0            &0             &0            &0           &0             &0           &0              &\cdots       & t_{R}e^{i\alpha_R}  &\delta_{2}   & 0  &\delta_{1}  \\
		0            & 0           & 0            & 0          & 0           & 0            &0           &0             &  \cdots     &\delta_{2}        &t_{R}e^{-i\alpha_R}  & \delta_{1}   &0  \\
	\end{array}\right]_{2N\times2N}. \tag{S18}\label{S18}
\end{align}

Acting on the eigenvector $|\psi\rangle=(\psi^{\uparrow}_{1}, \psi^{\downarrow}_{1},\psi^{\uparrow}_{2}, \psi^{\downarrow}_{2},\cdots,\psi^{\uparrow}_{N}, \psi^{\downarrow}_{N})^\mathrm{T}$, the Schr\"{o}dinger equation $\hat{H}|\psi\rangle=E|\psi\rangle$ generates the bulk equations for interior sites $2\leqslant n\leqslant N-1$:
\begin{align}
	 t_{L}e^{i\alpha_L}\psi^{\uparrow}_{n+1} + t_{R}e^{i\alpha_R}\psi^{\uparrow}_{n-1} + \delta_{2}(\psi^{\downarrow}_{n-1} +\psi^{\downarrow}_{n+1}) +\delta_{1}\psi^{\downarrow}_{n } =E\psi^{\uparrow}_{n }, \tag{S19}\label{S19}
\end{align}
\begin{align}
	  t_{L}e^{-i\alpha_L}\psi^{\downarrow}_{n+1} + t_{R}e^{-i\alpha_R}\psi^{\downarrow}_{n-1} + \delta_{2}(\psi^{\uparrow}_{n-1} +\psi^{\uparrow}_{n+1}) +\delta_{1}\psi^{\uparrow}_{n } =E\psi^{\downarrow}_{n }, \tag{S20}\label{S20}
\end{align}
and the four boundary equations at the chain ends:
\begin{align}
	 t_{L}e^{i\alpha_L}\psi^{\uparrow}_{2}  + \delta_{2}\psi^{\downarrow}_{2} +\delta_{1}\psi^{\downarrow}_{1 } =E\psi^{\uparrow}_{1}, \tag{S21}\label{S21}
\end{align}
\begin{align}
	 t_{L}e^{-i\alpha_L}\psi^{\downarrow}_{2}  + \delta_{2}\psi^{\uparrow}_{2} +\delta_{1}\psi^{\uparrow}_{1 } =E\psi^{\downarrow}_{1 }, \tag{S22}\label{S22}
\end{align}
\begin{align}
	 	t_{R}e^{i\alpha_R}\psi^{\uparrow}_{N-1} + \delta_{2}\psi^{\downarrow}_{N-1} +\delta_{1}\psi^{\downarrow}_{N } =E\psi^{\uparrow}_{N }, \tag{S23}\label{S23}
\end{align}
\begin{align}
t_{R}e^{-i\alpha_R}\psi^{\downarrow}_{N-1} + \delta_{2}\psi^{\uparrow}_{N-1} +\delta_{1}\psi^{\uparrow}_{N} =E\psi^{\downarrow}_{N }.\tag{S24}\label{S24}
\end{align}

Given the translation-invariant structure of the bulk equations, we employ the ansatz presented in Eq.~(\ref{S6}), namely,
\begin{align*}
\psi^{\uparrow}_n = \sum_{i=1}^{4} c_i \phi(\beta_i) \beta_i^n, \qquad
\psi^{\downarrow}_n = \sum_{i=1}^{4} c_i \beta_i^n.
\end{align*}
Substituting this form into Eqs.~(\ref{S19}) and (\ref{S20}) leads directly to the expression for the spinor component $\phi(\beta_i)$ given in Eq.~(\ref{S7}), or equivalently,
\begin{align}
\phi(\beta_{i})
=\frac{E-G_{\downarrow}(\beta_i)}{\delta_{1}+\delta_{2}(\beta_i+1/\beta_i)}, \tag{S25}\label{S25}
\end{align}
and confirms that $\beta_i$ must satisfy the characteristic Eq. (\ref{S5}).

The crucial step is to impose the OBC. Substituting the ansatz (\ref{S6}) into the four boundary Eqs. (\ref{S21})--(\ref{S24}) yields a homogeneous linear system for the coefficients $c_i$:
\begin{align}
\left[\begin{array}{cccc}
    1&1&1&1\\
	\phi(\beta_{1})&\phi(\beta_{2})&\phi(\beta_{3})&\phi(\beta_{4})\\
	\beta_{1}^{N+1}&
	\beta_{2}^{N+1}&
	\beta_{3}^{N+1}&
	\beta_{4}^{N+1}\\
   \phi(\beta_{1})\beta_{1}^{N+1}&\phi(\beta_{2})\beta_{2}^{N+1}&\phi(\beta_{3})\beta_{3}^{N+1}&\phi(\beta_{4})\beta_{4}^{N+1}
\end{array}\right]\left[\begin{array}{c}
	c_1\\
	c_2\\
	c_3\\
    c_4
\end{array}\right]=0, \tag{S26}\label{S26}
\end{align}
where $\phi(\beta_i)$ is given by Eq. (\ref{S7}) or (\ref{S25}). This linear system expresses the physical condition that the wavefunction in Eq.~(\ref{S6}) must vanish at the would-be sites $n=0$ and $n=N+1$. For a non-zero solution to exist, the determinant of the coefficient matrix in Eq.~(\ref{S26}) must be zero. This condition generates an OBC equation relating the four roots $\beta_i$:
\begin{align}
\beta_{12}\beta_{34}[(\beta_1\beta_2)^{N+1}+(\beta_3\beta_4)^{N+1}]
-\beta_{13}\beta_{24}[(\beta_1\beta_3)^{N+1}+(\beta_2\beta_4)^{N+1}]
+\beta_{14}\beta_{23}[(\beta_1\beta_4)^{N+1}+(\beta_2\beta_3)^{N+1}]=0,  \tag{S27}\label{S27}
\end{align}
where
\begin{align}
\beta_{ij}=(\beta_{i}-\beta_{j})\left\{\delta_{1}(t_{L}e^{i\alpha_L}\beta_i\beta_j - t_{R}e^{i\alpha_R})+\delta_2[(\beta_i+\beta_j)(t_{L}e^{i\alpha_L}-t_{R}e^{i\alpha_R})+E(\beta_i\beta_j-1)]\right\}. \tag{S28}\label{S28}
\end{align}
Once the OBC Eq.~(\ref{S27}) is satisfied, the linear system (\ref{S26}) can be solved for the coefficients $c_i$. Their explicit expressions are
\begin{align}
c_1=[\phi(\beta_3) -\phi(\beta_4)]\beta_2^{N + 1} +[\phi(\beta_4) -\phi(\beta_2)]\beta_3^{N + 1} +[\phi(\beta_2) -\phi(\beta_3)]\beta_4^{N + 1},   \tag{S29}\label{S29}
\end{align}
\begin{align}
c_2=[\phi(\beta_4) -\phi(\beta_3)]\beta_1^{N + 1} +[\phi(\beta_1) -\phi(\beta_4)]\beta_3^{N + 1} +[\phi(\beta_3) -\phi(\beta_1)]\beta_4^{N + 1},  \tag{S30}\label{S30}
\end{align}
\begin{align}
c_3=[\phi(\beta_2) -\phi(\beta_4)]\beta_1^{N + 1} +[\phi(\beta_4) -\phi(\beta_1)]\beta_2^{N + 1} +[\phi(\beta_1) -\phi(\beta_2)]\beta_4^{N + 1}, \tag{S31}\label{S31}
\end{align}
\begin{align}
c_4=[\phi(\beta_3) -\phi(\beta_2)]\beta_1^{N + 1} +[\phi(\beta_1) -\phi(\beta_3)]\beta_2^{N + 1} +[\phi(\beta_2) -\phi(\beta_1)]\beta_3^{N + 1}.\tag{S32}\label{S32}
\end{align}
These four formulas can be combined into a single compact and fully antisymmetric expression:
 \begin{equation}
c_i=\frac{1}{2}\sum_{j,k,l=1}^4\epsilon_{ijkl}
\beta_j^{N+1}
\bigl[\phi(\beta_k)-\phi(\beta_l)\bigr],
\tag{S33}\label{S33}
\end{equation}
which is identical to Eq.~\eqref{S8} in Sec.~\ref{Sec1}.
Here $\epsilon_{ijkl}$ is the four-dimensional Levi-Civita symbol, and the summation over
$j,k,l$ enforces a completely antisymmetric combination with respect to the four roots $\beta_i$.

Subsequently, we simplify the cumbersome boundary condition (\ref{S27}) by exploiting the relations among the roots $\beta_i$ enforced by the characteristic Eq.~(\ref{S5}). Applying the Vieta's theorem to Eq. (\ref{S5}) gives
\begin{align}
\beta_{1}\beta_{2}\beta_{3}\beta_{4}=\frac{t_R^2-\delta_{2}^2}{t_L^2-\delta_{2}^2}, \tag{S34}\label{S34}
\end{align}
\begin{align}
\beta_{1}+\beta_{2}+\beta_{3}+\beta_{4}=\frac{2E t_L \cos\alpha_L+2\delta_{1}\delta_{2}}{t_L^2-\delta_{2}^2}, \tag{S35}\label{S35}
\end{align}
\begin{align}
\beta_1\beta_2+\beta_1\beta_3+\beta_1\beta_4+\beta_2\beta_3+\beta_2\beta_4+\beta_3\beta_4
=\frac{E^2+2t_L t_R\cos(\alpha_L-\alpha_R)-\delta_{1}^2-2\delta_{2}^2}{t_L^2-\delta_{2}^2},\tag{S36}\label{S36}
\end{align}
\begin{align}
\beta_1\beta_2\beta_3+\beta_1\beta_2\beta_4+\beta_1\beta_3\beta_4+\beta_2\beta_3\beta_4 = \frac{2E t_R \cos\alpha_R+2\delta_{1}\delta_{2}}{t_L^2-\delta_{2}^2}. \tag{S37}\label{S37}
\end{align}
A key step is to adopt a symmetric parametrization of the four roots:
\begin{align}
\beta_1 = re^{\frac{i}{2}(\Theta+\Phi-\Psi)},~~~~\beta_2 = re^{\frac{i}{2}(\Theta-\Phi+\Psi)}, ~~~~
\beta_3 = re^{\frac{i}{2}(-\Theta+\Phi+\Psi)},~~~~\beta_4= re^{-\frac{i}{2}(\Theta+\Phi+\Psi)},  \tag{S38}\label{S38}
\end{align}
where $r$ is the amplitude, and $\Theta$, $\Phi$, $\Psi$ are complex phase angles. Substituting Eq.~\eqref{S38} into Eqs.~\eqref{S34}--\eqref{S37} yields expressions for the new variables. From Eq.~\eqref{S34} we immediately obtain
\begin{align}
r=\left(\frac{t_R^2-\delta_{2}^2}{t_L^2-\delta_{2}^2}\right)^{1/4}. \tag{S39}\label{S39}
\end{align}
The remaining Vieta relations then give
\begin{align}
E=\pm\sqrt{\delta_{1}^2+2p_2+ 2p_1(\cos\Theta+\cos\Phi+\cos\Psi)}, \tag{S40}\label{S40}
\end{align}
\begin{align}
\cos\left(\frac{\Theta}{2}\right)\cos\left(\frac{\Phi}{2}\right) \cos\left(\frac{\Psi}{2}\right)= \frac{E(r^2t_L\cos\alpha_L+t_R\cos\alpha_R)+\delta_{1}\delta_{2}(r^2+1)}{4r^3(t_L^2-\delta_{2}^2)}, \tag{S41}\label{S41}
\end{align}
\begin{align}
\sin\left(\frac{\Theta}{2}\right)\sin\left(\frac{\Phi}{2}\right)\sin\left(\frac{\Psi}{2}\right) = \frac{-i\left[E(r^2t_L\cos\alpha_L-t_R\cos\alpha_R)+\delta_{1}\delta_{2}(r^2-1)\right]}{4r^3(t_L^2-\delta_{2}^2)}.   \tag{S42}\label{S42}
\end{align}
Most importantly, the original boundary condition (\ref{S27}) simplifies dramatically when expressed in terms of $\Theta$, $\Phi$, and $\Psi$. After substituting Eq.~\eqref{S38} and using the relations above, Eq.~\eqref{S27} reduces to the following symmetric form:
\begin{align}
f(\cos\Theta)\cos[(N+1)\Theta](\cos\Phi-\cos\Psi)+f(\cos\Phi)\cos[(N+1)\Phi](\cos\Psi-\cos\Theta) +f(\cos\Psi)\cos[(N+1)\Psi](\cos\Theta-\cos\Phi)=0,
 \tag{S43}\label{S43}
\end{align}
where $f(x)=p_1x+p_2$ with $p_1$ and $p_2$ being defined in Eq. (\ref{S10}).

Next, defining $X=\cos\Theta$, $Y=\cos\Phi$, and $Z=\cos\Psi$, and using the identity $\cos[(N+1)\arccos(x)]=T_{N+1}(x)$ (where $T_{N+1}(x)$ is the Chebyshev polynomial of the first kind), the OBC Eq. (\ref{S43}) becomes
\begin{align}
\frac{(p_1X+p_2)T_{N+1}(X)}{(X-Y)(Z-X)}+\frac{(p_1Y+p_2)T_{N+1}(Y)}{(X-Y)(Y-Z)} +\frac{(p_1Z+p_2)T_{N+1}(Z)}{(Y-Z)(Z-X)}=0. \tag{S44}\label{S44}
\end{align}
In the new variables $X$, $Y$, and $Z$, the relations (\ref{S40})-(\ref{S42}) take a more compact form. From (\ref{S40}) we obtain
\begin{align}
E=\pm\sqrt{\delta_{1}^2+2p_2+ 2p_1(X+Y+Z)}, \tag{S45}\label{S45}
\end{align}
while Eqs.~(\ref{S41}) and (\ref{S42}) can be rewritten as
\begin{align}
(1+X)(1+Y)(1+Z)= \frac{\left[E(r^2t_L\cos\alpha_L+t_R\cos\alpha_R)+
\delta_{1}\delta_{2}(r^2+1)\right]^2}{2r^6(t_L^2-\delta_{2}^2)^2}, \tag{S46}\label{S46}
\end{align}
\begin{align}
(1-X)(1-Y)(1-Z)=\frac{\left[E(r^2t_L\cos\alpha_L-t_R\cos\alpha_R)+
\delta_{1}\delta_{2}(r^2-1)\right]^2}{2r^6(t_L^2-\delta_{2}^2)^2}.   \tag{S47}\label{S47}
\end{align}

A direct algebraic manipulation of Eqs.~(\ref{S45})-(\ref{S47}) yields three symmetric combinations of $X$, $Y$, and $Z$ that depend only on the energy $E$ and the system parameters:
\begin{align}
X+Y+Z=\frac{E^2-\delta_1^2-2p_2}{2p_1}\equiv A. \tag{S48}\label{S48}
\end{align}
\begin{align}
XY+XZ+YZ=\frac{(Et_L\cos\alpha_L+\delta_{1}\delta_{2})
(Et_R\cos\alpha_R+\delta_{1}\delta_{2})}{p_1^2}-1\equiv B. \tag{S49}\label{S49}
\end{align}
\begin{align}
XYZ=\frac{(\kappa_2-1)E^2}{2p_1}+\frac{\kappa_1\delta_{1}\delta_{2}E}{p_1}
+\frac{2p_2+\delta_1^2+\kappa_0\delta_{1}^2\delta_{2}^2}{2p_1}\equiv C, \tag{S50}\label{S50}
\end{align}
with $\kappa_m$ ($m=0,1,2$) given by Eq.~(\ref{S17}).
Equations (\ref{S48})-(\ref{S50}) imply that $X$, $Y$, and $Z$ are actually the three roots of the cubic equation:
\begin{align}
z^3-Az^2+Bz-C=0. \tag{S51}\label{S51}
\end{align}

Now, Eq.~(\ref{S44}) can be further recast in a more systematic form. Introducing the notation
\begin{align}
\mathds{M}=\frac{XT_{N+1}(X)}{(X-Y)(Z-X)}+\frac{YT_{N+1}(Y)}{(Y-Z)(X-Y)}
+\frac{ZT_{N+1}(Z)}{(Y-Z)(Z-X)}, \tag{S52}\label{S52}
\end{align}
\begin{align}
\mathds{N}=\frac{T_{N+1}(X)}{(X-Y)(Z-X)}+\frac{T_{N+1}(Y)}{(Y-Z)(X-Y)}
+\frac{T_{N+1}(Z)}{(Y-Z)(Z-X)}, \tag{S53}\label{S53}
\end{align}
allows us to write Eq. (\ref{S44}) as
\begin{align}
p_1\mathds{M}+p_2\mathds{N}=0. \tag{S54}\label{S54}
\end{align}
Since $X,Y,Z$ are the three roots of the cubic Eq. (\ref{S51}), the quantities $\mathds{M}$ and $\mathds{N}$ are second-order divided differences of the polynomials $xT_{N+1}(x)$ and $T_{N+1}(x)$, respectively.  These divided differences can be expressed directly through the polynomials $\lambda_{N+1}^{[1]}$ and $\lambda_{N+1}^{[2]}$ generated by the following recurrence relation:
\begin{align}
G_{n+1}^{[i]}=Q G_{n}^{[i]}-G_{n-1}^{[i]}, ~~(i=1,2),   \tag{S55}\label{S55}
\end{align}
where
\begin{align}
Q=\left[\begin{array}{ccc}
	2A&2&0\\
	-2B&0&2\\
	2C&0&0
\end{array}\right],~~~G_{n}^{[i]}=\left[\begin{array}{c}
	\lambda_n^{[i]}\\
	\mu_n^{[i]}\\
	\gamma_n^{[i]}\\
\end{array}\right], \tag{S56}\label{S56}
\end{align}
with initial conditions
\begin{align}
G_{0}^{[1]}=\left[\begin{array}{c}
	0\\
	1\\
	0\\
\end{array}\right],~~~~G_{1}^{[1]}=\left[\begin{array}{c}
	1\\
	0\\
	0\\
\end{array}\right],~~~~G_{0}^{[2]}=\left[\begin{array}{c}
	0\\
	0\\
	1\\
\end{array}\right],~~~~G_{1}^{[2]}=\left[\begin{array}{c}
	0\\
	1\\
	0\\
\end{array}\right]. \tag{S57}\label{S57}
\end{align}
This matrix formulation is fully equivalent to the three scalar recurrences (\ref{S11})--(\ref{S13}) in Sec.~\ref{Sec1}. We note that such a matrix recurrence structure is typical for the exact solution of open-boundary spectra in the generic two-band Hamiltonian $\mathcal{H}(\beta)=d_0(\beta)\mathds{1}+\mathbf{d}(\beta)\cdot\boldsymbol{\sigma}$,  irrespective of the presence of chiral symmetry.  Indeed, one finds
\begin{align}
\mathds{M}=-\lambda_{N+1}^{[1]},\qquad\mathds{N}=-\lambda_{N+1}^{[2]}. \tag{S58}\label{S58}
\end{align}
Substituting (\ref{S58}) into (\ref{S54}) reproduces the polynomial equation of degree $2N$ for the energy $E$ already presented in Eq.~(\ref{S9}):
\begin{align*}
p_1\lambda_{N+1}^{[1]}+p_2\lambda_{N+1}^{[2]}=0.
\end{align*}
For any finite size $N$, this equation can be solved (e.g., by a standard root-finding routine) to obtain the complete set of $2N$ OBC eigenvalues. Each eigenvalue $E$ then determines the corresponding $\beta_i$ via Eq.~(\ref{S5}) and, through Eqs.~(\ref{S6})-(\ref{S8}), the full eigenfunction. Thus the derivation closes, yielding the exact eigensystem announced in Sec.~\ref{Sec1}.

\begin{figure}[h!]
\begin{center}
\includegraphics[width=17cm]{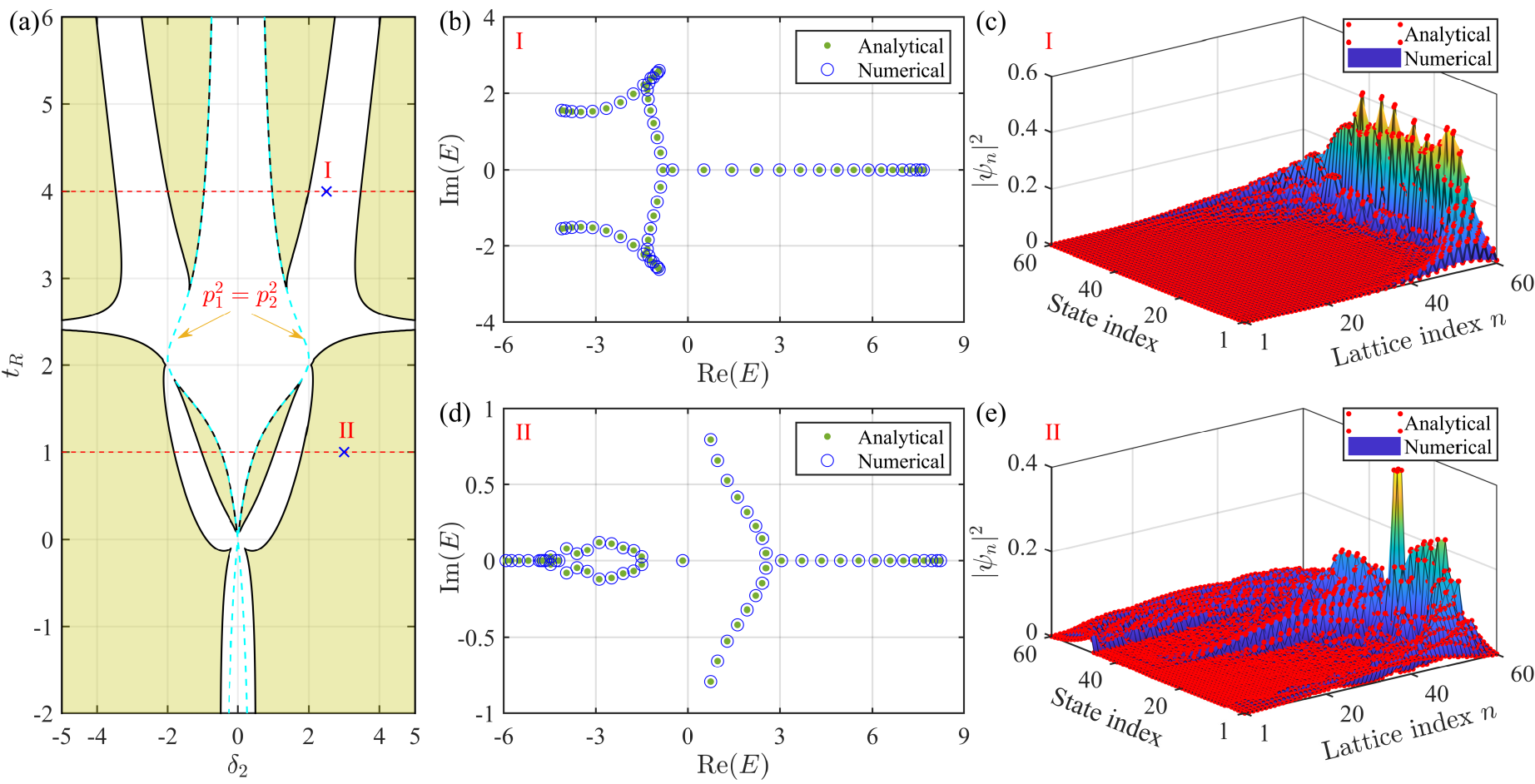}
\caption{\hbadness=10000 \vbadness=10000
  \setlength{\leftskip}{0pt}%
  \setlength{\rightskip}{0pt}%
  \setlength{\parfillskip}{0pt plus 1fil} Numerical verification of the exact solutions ($N=30$). (a) Phase diagram in the $(\delta_2, t_R)$ plane for $t_L=2$, $\alpha_L=\pi/3$, $\alpha_R=\pi/4$, $\delta_1=1$, reproduced from Fig.~2(a) of the main text. (b,c) Results for the parameter point I ($\delta_2=2.5$, $t_R=4$): (b) analytical energy spectrum [green dots, from Eq.~(\ref{S9})] compared to numerical one (open circles); (c) analytical eigenstates [red dots, from Eq.~(\ref{S6})] compared to numerical wavefunctions (solid lines). (d,e) Results for the parameter point II ($\delta_2=3$, $t_R=1$), presented in the same manner as in (b,c). } \label{figS2}
\end{center}
\end{figure}

To validate the analytical derivation, we compare its predictions with direct numerical diagonalization. Supplemental Fig. \ref{figS2} displays the excellent agreement between the analytical energy spectrum obtained from Eq.~(\ref{S9}) and the numerical OBC eigenvalues, as well as between the analytical eigenstates constructed via Eq.~(\ref{S6}) and their numerical counterparts, for two representative parameter sets marked by blue crosses in the phase diagram.

\section{Derivation of the Energy and Existence Condition for Topological Edge States} \label{Sec3}

In Sec.~\ref{Sec2}, we reduce the eigenvalue problem to solving a polynomial equation of degree
$2N$ for the energy $E$. While a closed-form solution for finite large $N$ is generally impossible (based on Abel's theorem), in the thermodynamic limit $N\rightarrow\infty$, we can obtain exact expressions for the topological edge states and for the parameter regions where they appear. This section presents that asymptotic derivation.

Topological edge states are isolated eigenvalues in the complex energy plane that do not merge with the bulk continuum even when $N \to \infty$. In our formalism, such an isolated state corresponds to a special solution of the OBC Eq. (\ref{S44}) for which one of the three variables $X,Y,Z$ (call it $X_e$) dominates the growth of the Chebyshev polynomials as $N$ increases. Concretely, an edge state requires $|T_{N}(X_e)| > |T_{N}(Y)|,~|T_{N}(Z)|$ ($N \to \infty$) \cite{Li2024}. Taking this dominance condition in Eq.~(\ref{S44}), we obtain
\begin{align}
f(X_e)T_{N+1}(X_e)(Y-Z)=0.
 \tag{S59}\label{S59}
\end{align}
As the factor $Y-Z$ cannot vanish for a genuine edge state, Eq. (\ref{S59}) gives immediately $f(X_e)=p_1X_e+p_2=0$, namely,
\begin{align}
X_e=-\frac{p_2}{p_1}=\frac{t_Lt_R\cos(\alpha_L-\alpha_R)
-\delta_{2}^2}{\sqrt{(t_R^2-\delta_{2}^2)(t_L^2-\delta_{2}^2)}}.
 \tag{S60}\label{S60}
\end{align}
It is clear that the quantity $X_e$ is real when $\delta_2$ lies outside the interval between $t_R$ and $t_L$; otherwise it becomes complex, reflecting a non-trivial deformation of the GBZ.

Once $X_e$ is fixed, the remaining two variables follow from the cubic relations (\ref{S48})-(\ref{S50}). Inserting $X=X_e$  into those equations gives
\begin{align}
Y=\frac{BX_e-C+\sqrt{(BX_e-C)^2-4CX_e^3}}{2X_e^2},
 \tag{S61}\label{S61}
\end{align}
\begin{align}
Z=\frac{BX_e-C-\sqrt{(BX_e-C)^2-4CX_e^3}}{2X_e^2}.
 \tag{S62}\label{S62}
\end{align}

\begin{figure}[b!]
\begin{center}
\includegraphics[width=17cm]{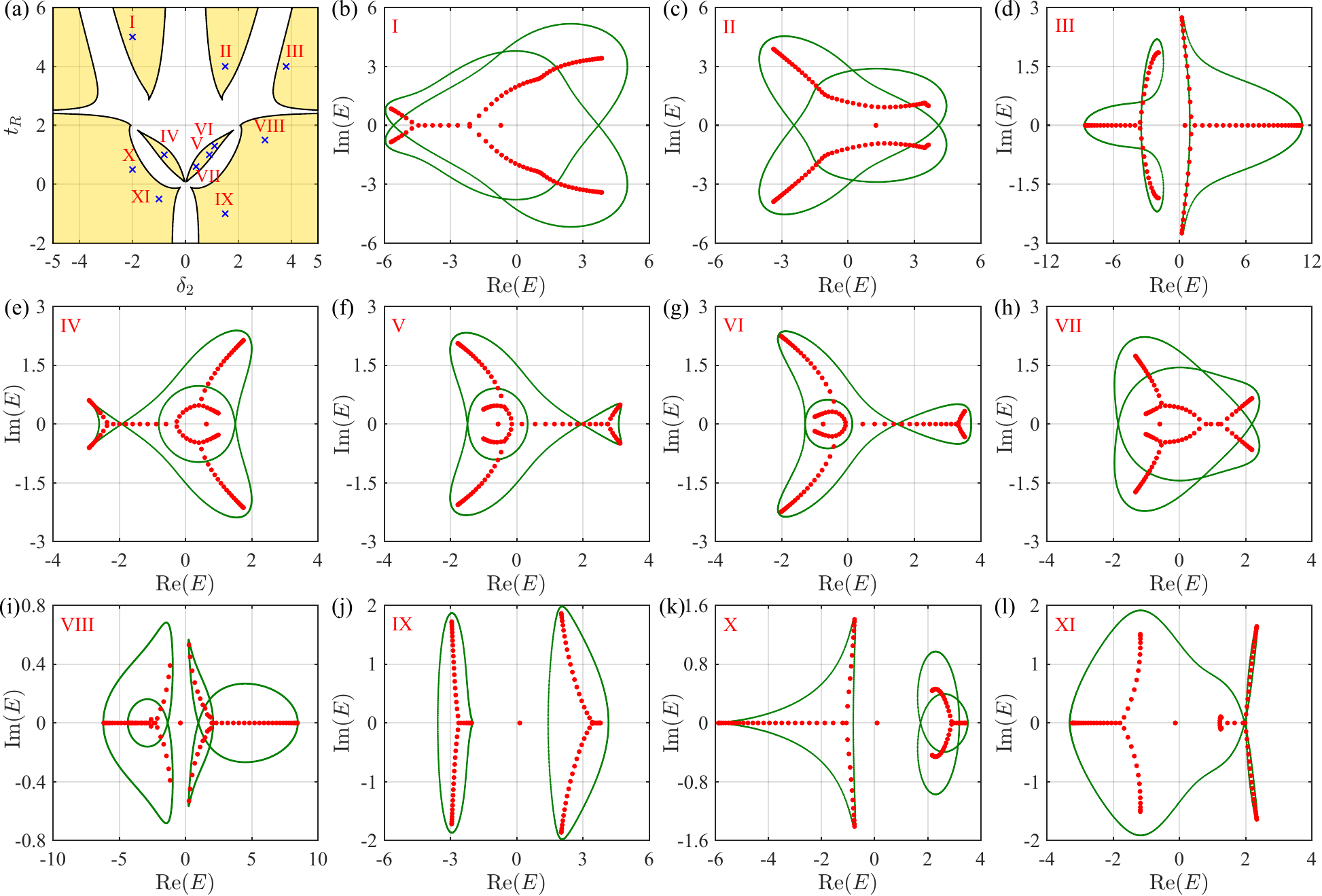}
\caption{\hbadness=10000 \vbadness=10000
  \setlength{\leftskip}{0pt}%
  \setlength{\rightskip}{0pt}%
  \setlength{\parfillskip}{0pt plus 1fil} Distinct spectral morphology within topological regimes. (a) Phase diagram in the $(\delta_2, t_R)$ plane for $t_L=2$, $\alpha_L=\pi/3$, $\alpha_R=\pi/4$, $\delta_1=1$, showing isolated topological regimes (yellow regions), identical to Fig.~2(a) in the main text. (b)-(l) OBC energy spectra (red dots) and PBC spectra (colored loops) for selected parameter points inside [panels (b), (c), (e--h)] and outside [panels (d), (i--l)] the topological islands. The spectra within the islands exhibit a characteristic multi-branch structure.
Parameters for the labeled points are given by:
I $(\delta_2,t_R)=(-2,5)$,
II $(1.5,4)$,
III $(3.8,4)$,
IV $(-0.8,1)$,
V $(0.9,1)$,
VI $(1.1,1.3)$,
VII $(0.4,0.6)$,
VIII $(3,1.5)$,
IX $(1.5,-1)$,
X $(-2,0.5)$,
and XI $(-1,-0.5)$.} \label{figS3}
\end{center}
\end{figure}

Finally, substitution of Eqs. (\ref{S60})--(\ref{S62}) into Eq. (\ref{S45}) yields the exact energy of the topological edge states (which are degenerate) in the thermodynamic limit:
\begin{align}
E_e=- \frac{\delta_{1}t_Lt_R\sin(\alpha_L-\alpha_R)}{\delta_{2}(t_L\sin\alpha_L-t_R\sin\alpha_R)}. \tag{S63}\label{S63}
\end{align}
Equation (\ref{S63}) is one of the central results of this work. Remarkably, setting $\delta_{1} = 0$ or $\alpha_L=\alpha_R$ gives $E_e=0$, i.e. a topological zero mode that survives even in the absence of chiral symmetry.

The existence of the topological edge state is guaranteed only when the assumed dominance condition ($|T_{N}(X_e)| > |T_{N}(Y)|,~|T_{N}(Z)|$ ) holds. Using the identity
\begin{align}
T_{N}(x)=\frac{1}{2}W_1^N(x)+\frac{1}{2}W_2^N(x),~~~W_{1,2}(x)=x\pm \sqrt{x^{2}-1}, \tag{S64}\label{S64}
\end{align}
and defining $W(x)=\mathrm{max}(|W_{1}(x)|,|W_{2}(x)|)$, the condition simplifies to
\begin{align}
|W(Y)|\leqslant|W(X_e)|,~~~~
|W(Z)|\leqslant|W(X_e)|. \tag{S65}\label{S65}
\end{align}
The inequalities (\ref{S65}), which are exactly equivalent to the phase condition Eq.~(10) in the main text, delineate the topological phase boundaries in parameter space. They are asymptotically exact and, in practice, already give accurate predictions for chains as short as $N>6$ (generally with errors below $1\%$).

One can also obtain the energy formula (\ref{S63}) for the topological edge states by a leading-term analysis of the original OBC Eq.~(\ref{S27}) \cite{Zhong25}. Since edge states do not satisfy the GBZ condition $|\beta_2|=|\beta_3|$, the eigenvalue Eq.~(\ref{S5}) allows four $\beta$ roots ordered as $|\beta_1|\leqslant|\beta_2|<|\beta_3|\leqslant|\beta_4|$. In this case, the dominant term of Eq.~(\ref{S27}) in the thermodynamic limit is $\beta_{12}\beta_{34}(\beta_3\beta_4)^{N+1}$, which must vanish. This requires $\beta_{12}=0$ or $\beta_{34}=0$, i.e., $\phi(\beta_1)=\phi(\beta_2)=M_{\mathrm{deg},1}$ or $\phi(\beta_3)=\phi(\beta_4)=M_{\mathrm{deg},2}$. Using the definition of $\phi(\beta)$ in Eq.~(\ref{S7}), these two conditions can be expressed as
\begin{align}
\frac{\delta_{1}+\delta_{2}(\beta+1/\beta)}{E-G_{\uparrow}(\beta)}=M_{\mathrm{deg}},~~~~
\frac{E-G_{\downarrow}(\beta)}{\delta_{1}+\delta_{2}(\beta+1/\beta)}=M_{\mathrm{deg}}. \tag{S66}\label{S66}
\end{align}
For a given $E_e$ and $M_{\mathrm{deg}}$, each equation in (\ref{S66}) simplifies, after cross-multiplication, to a quadratic equation for $\beta$. Since these two quadratic equations must be the same, their coefficients are proportional to each other, leading to
\begin{align}
\frac{t_Le^{i\alpha_L}M_{\mathrm{deg}}+\delta_2}{\delta_2M_{\mathrm{deg}}+t_Le^{-i\alpha_L}} = \frac{t_Re^{i\alpha_R}M_{\mathrm{deg}}+\delta_2}{\delta_2M_{\mathrm{deg}}+t_Re^{-i\alpha_R}}
=\frac{-E_eM_{\mathrm{deg}}+\delta_1}{\delta_1M_{\mathrm{deg}}-E_e}. \tag{S67}\label{S67}
\end{align}
The former equality in Eq.~(\ref{S67}) yields the two degeneracy points $M_{\mathrm{deg},1,2}$. Substituting these into the latter equality reproduces the edge-state energy formula (\ref{S63}). We note, however, that obtaining the compact phase boundary condition (\ref{S65}) from this route is not straightforward: although substituting $E_e$ back into the characteristic equation (\ref{S5}) would in principle give the explicit $\beta$ roots, and the inequality $|\beta_2|\leqslant|\beta_3|$ then determines the phase boundaries, the resulting algebraic expressions are substantially more cumbersome and less direct than those obtained via the Chebyshev-polynomial approach.

Here, we demonstrate in Supplemental Fig.~\ref{figS3} the distinct spectral morphology for parameter points selected from both the isolated topological ``islands'' (a detailed discussion of which is provided in Sec.~\ref{Sec5}) and the connected topological region in the phase diagram shown in Fig.~2(a) of the main text. A clear dichotomy emerges between these two subregions: within the islands (points I, II, IV, V, VI, VII), the bulk OBC spectra generally form continuous, multi-branch structures, which may evolve or split into vertically separated arcs under specific parameters, as shown in panel (c). In stark contrast, in the non-island topological regimes (points III, VIII, IX, X, XI), the OBC bulk spectra split into horizontally separated arcs, a unique characteristic that never evolves into the multi-branch form [see panels (d), and (i--l)]. This fundamental difference in spectral connectivity signals that the islands and non-island regimes correspond to distinct GBZ topologies, consistent with their segregation in the phase diagram. This dichotomy is also reflected in the evolution of spectral morphologies across the phase boundaries, as detailed in Fig.~6 of the End Matter for the topological phase transition processes. While the trivial regime can also exhibit multi-branch features when adjacent to the topological islands (as shown by point I in Supplemental Fig.~\ref{figS2}), it is the absence of isolated in-gap edge states that distinguishes it from the topological islands. This underscores that the spectral morphology is not the phase-defining criterion; rather, the emergence of edge states is entirely governed by our exact phase condition Eq.~(10).

\begin{figure}[b!]
\begin{center}
\includegraphics[width=17cm]{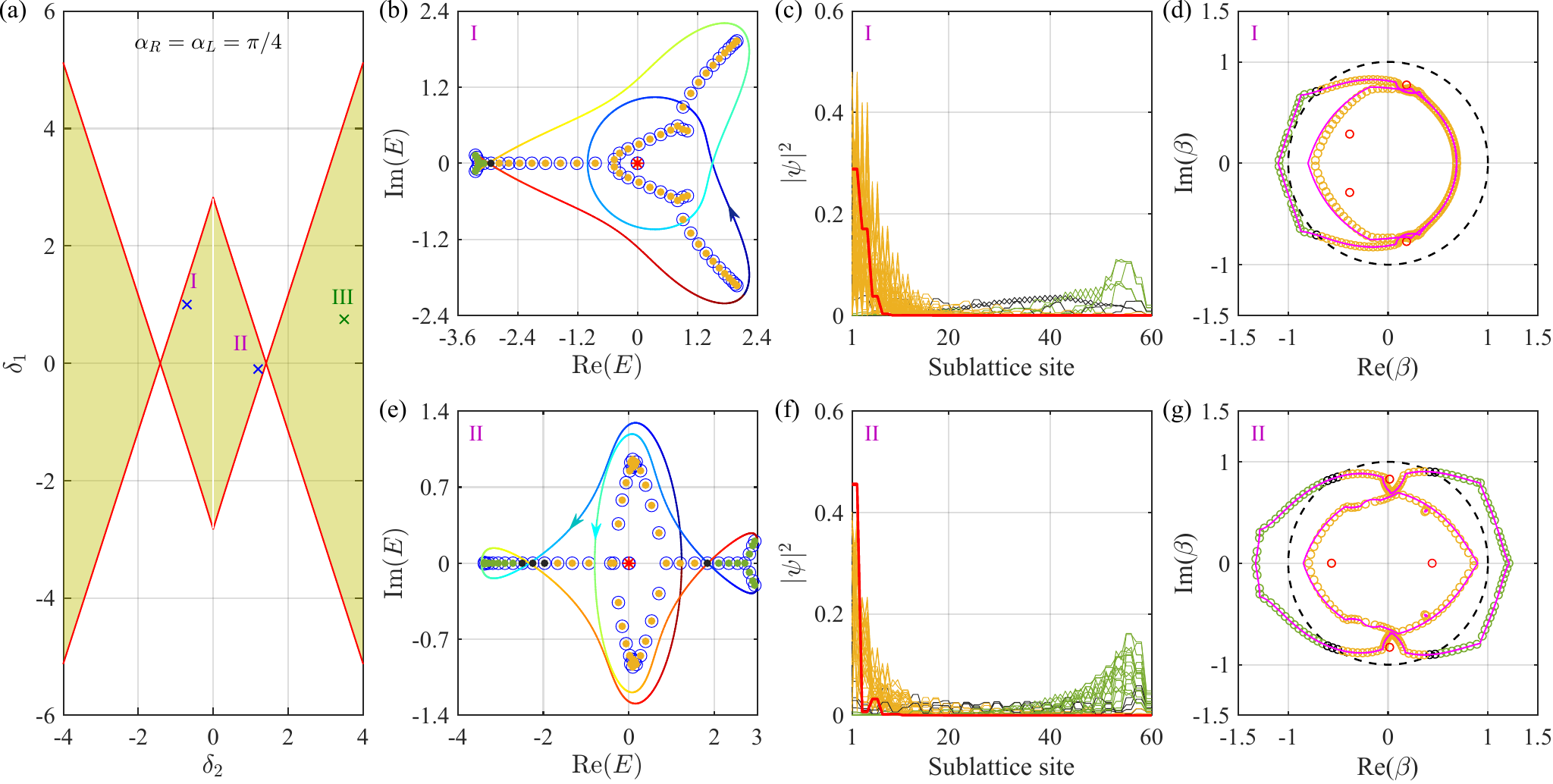}
\caption{\hbadness=10000 \vbadness=10000
  \setlength{\leftskip}{0pt}%
  \setlength{\rightskip}{0pt}%
  \setlength{\parfillskip}{0pt plus 1fil} Topological zero modes in systems without $(E,-E)$ spectral symmetry. (a) Zero-mode phase diagram copied from Fig.~3(d) in the main text, with representative points I, II, and III indicated. (b--d) Data for the marked point I $(\delta_2=-0.7, \delta_1=1)$ in the phase diagram: (b) OBC energy spectrum partially enclosing the zero mode, (c) spatial profiles of eigenstates, and (d) corresponding GBZ, which forms a single twisted contour. The topological zero mode is highlighted in red (asterisk for eigenvalue, curve for wavefunction, circles for GBZ points). (e--g) Data for point II $(\delta_2=1.2, \delta_1=-0.1)$, displayed in the same manner; its GBZ consists of two separate loops. In both cases, the OBC spectra lack $E \leftrightarrow -E$ pairing, confirming that topological zero modes can exist even in the absence of symmetries that enforce such spectral symmetry.  }\label{figS4}
\end{center}
\end{figure}

Furthermore, Supplemental Fig.~\ref{figS4} explicitly shows that the topological zero modes, available for the specific parameter condition $\alpha_R=\alpha_L$ here, are not contingent upon the spectral $(E,-E)$ symmetry, which is typically associated with chiral, particle-hole, or the weaker $\pi$-translation symmetry. We select two representative points (labeled I and II) inside the topological islands of the zero-mode phase diagram. Their OBC spectra are symmetric about the real axis but exhibit no $E \leftrightarrow -E$ pairing about the imaginary axis: for point I, the spectrum forms an asymmetric continuum that partially encloses the zero mode [see Supplemental Figs.~\ref{figS4}(b-d)]; for point II, the zero mode is completely enclosed by the bulk spectral continuum [see Supplemental Figs.~\ref{figS4}(e-g)]. These examples provide direct evidence that robust zero modes can emerge in systems devoid of any conventional symmetry that mandates $(E,-E)$ pairing. This highlights the symmetry-agnostic nature of the topological mechanism, which is fundamentally driven by the synergy between the momentum-dependent identity term and the SO coupling.

\section{Non-Hermitian skin effect and point-gap topology} \label{Sec4}

While the topological edge states studied in the previous section are protected by wavefunction topology on the GBZ, our model also hosts a distinct topological phenomenon---the non-Hermitian skin effect (NHSE), which stems from an entirely different topological origin, namely, the point-gap topology of the PBC spectrum \cite{Oku2020,Zha2020}. For a 1D non-Hermitian system described by a non-Bloch Hamiltonian $\mathcal{H}(\beta)$ with $\beta \equiv e^{ik}$, this point-gap topology is characterized by the spectral winding number with respect to a reference energy $E_b \in \mathbb{C}$:
\begin{align}
W(E_b) = \frac{1}{2\pi i} \oint_{\mathrm{BZ}} dk \frac{d}{dk} \ln \det[\mathcal{H}(e^{ik}) - E_b], \tag{S68}\label{S68}
\end{align}
where the integration is performed along the Brillouin zone (BZ), i.e., the unit circle $|\beta| = 1$.
In the thermodynamic limit, a non-zero $W(E_b)$, meaning that the PBC spectrum winds around $E_b$, has been rigorously established as both necessary and sufficient for the existence of skin modes \cite{Oku2020,Zha2020}. Specifically, the sign of $W(E_b)$ determines the direction of skin-mode localization: clockwise (counterclockwise) winding corresponds to right-localized (left-localized) skin modes.

This global criterion, however, is rigorously established only in the thermodynamic limit ($N \to \infty$), where the OBC spectrum forms a continuous set fully enclosed by the PBC spectral loops \cite{Yok19,Yang20}. For finite systems, the OBC spectrum consists of discrete points that do not necessarily obey the same enclosure property. This deviation is particularly pronounced in the present model, where the momentum-dependent identity term $d_0(\beta)\mathds{1}$ deforms the GBZ and consequently the OBC spectrum significantly.

\begin{figure}[h!]
\begin{center}
\includegraphics[width=17cm]{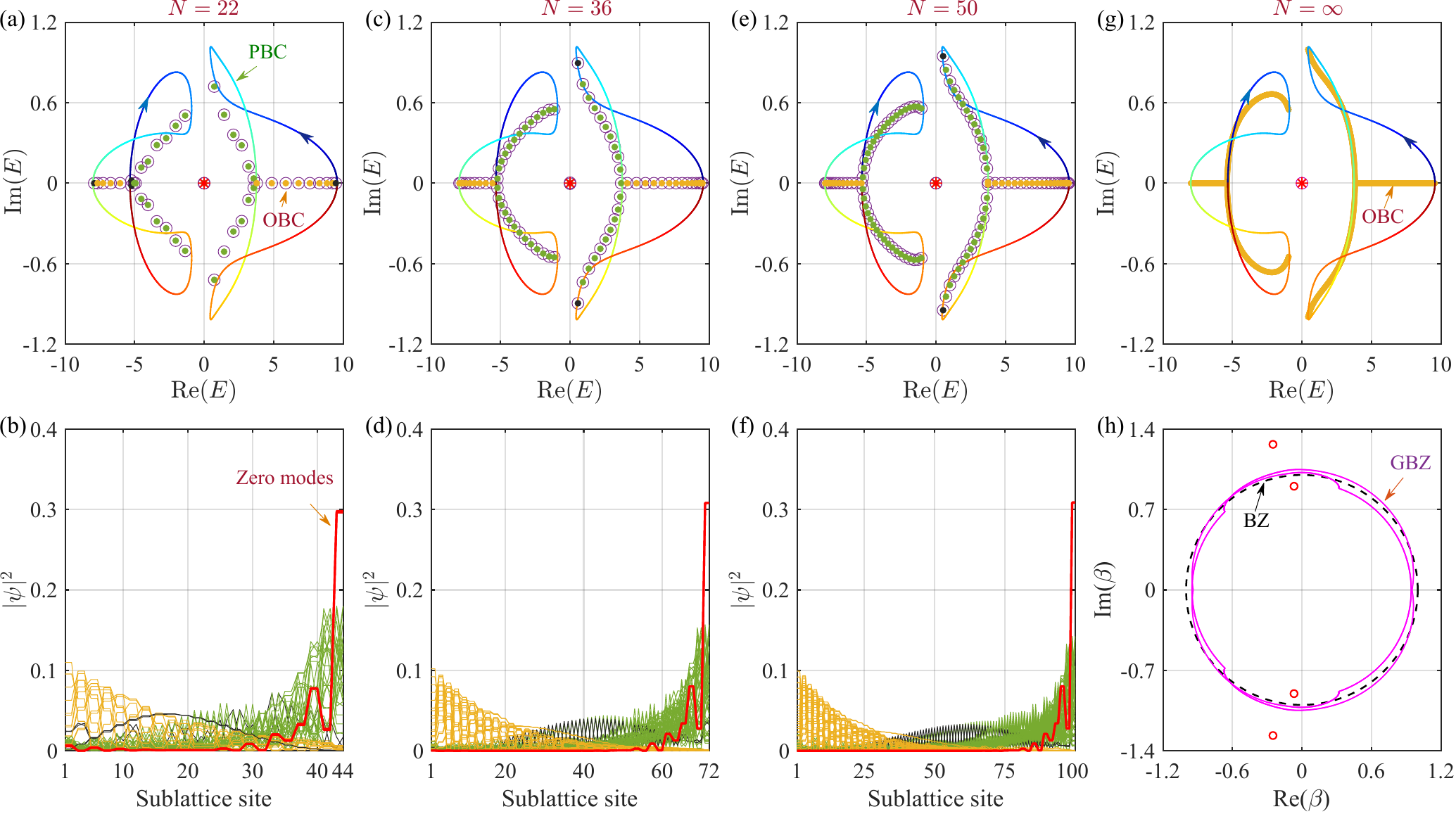}
\caption{\hbadness=10000 \vbadness=10000
  \setlength{\leftskip}{0pt}%
  \setlength{\rightskip}{0pt}%
  \setlength{\parfillskip}{0pt plus 1fil} Finite-size evolution of the OBC spectrum and its thermodynamic limit.
(a,c,e) Complex OBC energy spectrum, with PBC spectrum as colored loops, and (b,d,f) the corresponding spatial profiles of all eigenstates, obtained for $N=22$, $36$, and $50$, respectively. (g) The OBC spectrum in the thermodynamic limit $N=\infty$ (yellow thick curves) obtained from the GBZ, fully enclosed by the PBC loops. (h) GBZ (magenta curves) and BZ (dashed unit circle). The system parameters used correspond to point III in Supplemental Fig. \ref{figS4}(a) where $\delta_2=3.5$ and $\delta_1=0.75$. The topological zero modes (red asterisks) remain pinned at $E_e=0$ for all finite $N$ and persist in the thermodynamic limit.}\label{figS5}
\end{center}
\end{figure}

Supplemental Fig.~\ref{figS5} illustrates this behavior for a representative point $(\delta_2 = 3.5, \delta_1 = 0.75)$ inside the topological regime [point III in Supplemental Fig.~\ref{figS4}(a)], where the GBZ splits into two separate loops [panel (h)]. For small system sizes ($N = 22, 36$), a significant fraction of the OBC eigenvalues (green dots) lie \emph{outside} the PBC spectral loops [panels (a), (c)], yet they remain robust right-localized skin modes [panels (b), (d)]. These ``overflow'' eigenvalues are not directly captured by the spectral winding number $W(E_b)$, as the latter is formulated in the thermodynamic limit. This does not indicate a failure of the underlying theorems, but simply reflects their range of validity. As $N$ increases, the OBC spectrum gradually retracts and converges precisely onto the continuous spectrum calculated from the GBZ [panels (e), (f)], which at $N \to \infty$ is fully enclosed by the PBC loops [panel (g)], in complete agreement with the thermodynamic-limit theorems of Refs.~\cite{Oku2020, Zha2020}. This convergence demonstrates that the point-gap topological criteria correctly predict the asymptotic behavior, while our exact solution presented in Sec.~\ref{Sec1} fills a crucial gap by providing a quantitative description of the finite-size approach to this limit.

Crucially, the topological zero modes (red asterisks) remain exactly pinned at $E_e = 0$ for all $N$, completely insensitive to whether the surrounding OBC bulk spectrum lies inside or outside the PBC loops. This stark contrast highlights the physical independence of two distinct topological phenomena: the NHSE, governed by point-gap topology on the BZ, and the topological edge states, protected by wavefunction topology on the GBZ. This distinction is fully consistent with the framework of Ref.~\cite{Zhong25}.

It should be emphasized that even in the thermodynamic limit, the spectral winding number $W(E)$ is a purely topological invariant: it faithfully diagnoses whether an OBC eigenstate at energy $E$ is a skin mode and determines its localization direction, but it does not quantify the skin-mode intensity (i.e., the localization length). The latter is governed by the deviation of the GBZ from the BZ, specifically by the magnitude of the corresponding $\beta$ roots $|\beta(E)|$ that satisfy the GBZ condition. In our model with self-intersecting PBC spectra, some OBC energies may lie in regions with winding number $|W(E_{\rm OBC})|=2$, yet their skin-mode localization can be weaker than that of states in regions with $|W(E_{\rm OBC})|=1$, as observed in Fig.~3 in the main text. This reflects the fact that the PBC spectral winding is a global topological classification, whereas the localization length is a local property determined by the analytic structure of the non-Bloch Hamiltonian.

\section{Topological islands in phase diagrams and topological invariant test}
\label{Sec5}

In the main text, we have presented phase diagrams in various two-dimensional (2D) parameter slices, revealing that the regions supporting topological edge states can appear as disconnected components [see the isolated yellow regions in Figs.~2(a) and 3(d)]. We refer to these components as topological islands. Within a given 2D slice, an island is defined as a connected set of parameters where topological edge states exist, which may be separated from other such sets by a finite region of trivial parameters or connected to them only at isolated points, as visualized in Supplemental Fig.~\ref{figS3}(a) and Fig.~3(d) in the main text. This phenomenon is absent in models without the momentum-dependent identity term, where the topological region is simply connected (see, e.g., Fig. 2(a) in Ref.~\cite{Li2024}).

\begin{figure}[h!]
\begin{center}
\includegraphics[width=17cm]{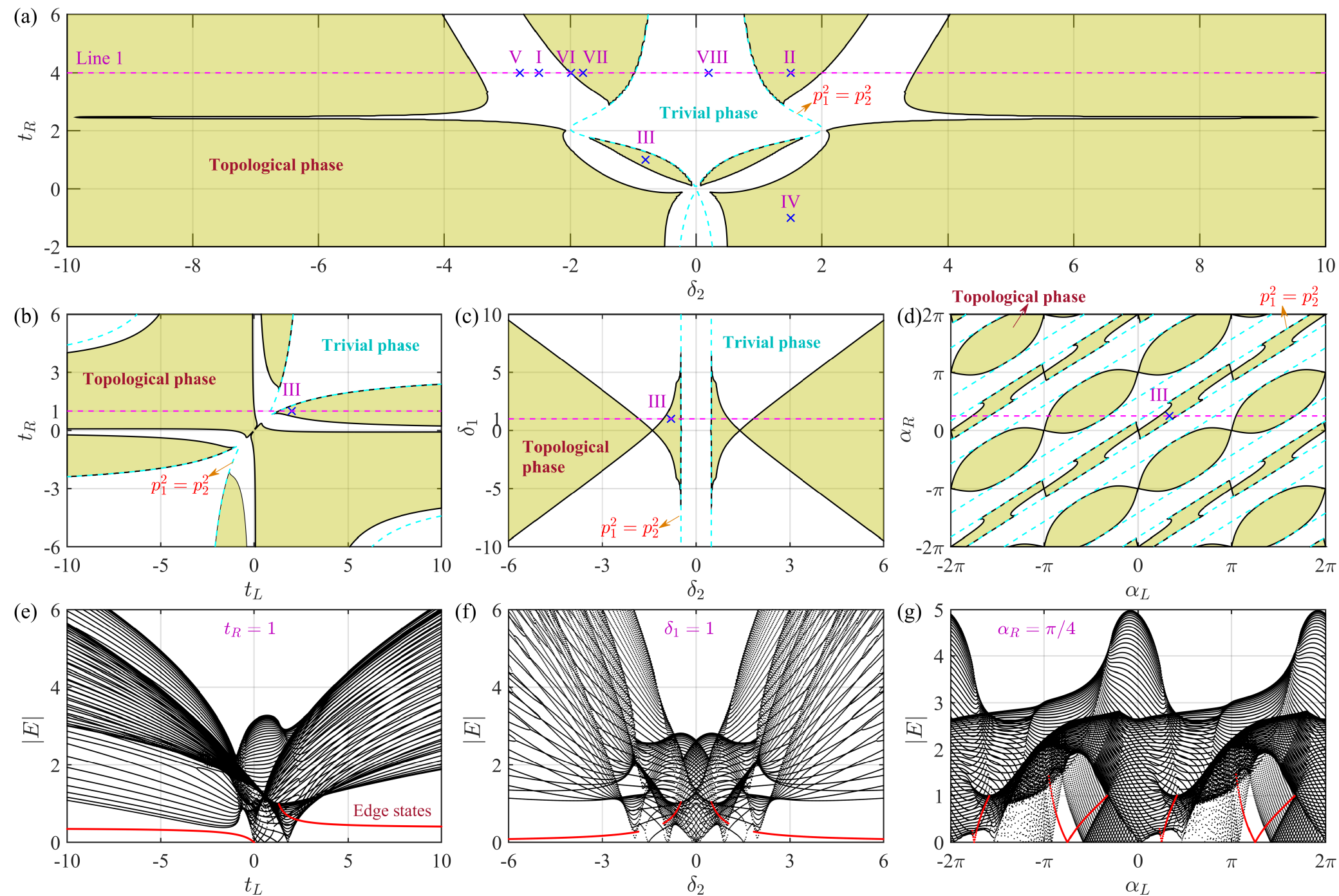}
\caption{\hbadness=10000 \vbadness=10000
  \setlength{\leftskip}{0pt}%
  \setlength{\rightskip}{0pt}%
  \setlength{\parfillskip}{0pt plus 1fil} Phase diagrams in different parameter slices and band structures.
(a) Phase diagram in the $(\delta_2, t_R)$ plane, identical to Fig.~2(a) in the main text. Points I--VIII are marked, with coordinates I $(-2.5,4)$, II $(1.5,4)$, III $(-0.8,1)$, IV $(1.5,-1)$, V $(-2.8,4)$, VI $(-1.99,4)$, VII $(-1.8,4)$, and VIII $(0.2,4)$. (b), (c), and (d) show the phase diagrams in the $(t_L, t_R)$, $(\delta_2, \delta_1)$, and $(\alpha_L, \alpha_R)$ planes, respectively, all passing through point III which corresponds to $\delta_2=-0.8$, $\delta_1=1$, $t_L=2$, $t_R=1$, $\alpha_L=\pi/3$, $\alpha_R=\pi/4$. In (a)--(d), yellow regions denote topological edge states, and cyan dashed curves represent the $p_1^2 = p_2^2$ boundary. (e)--(g) Energy band diagrams $|E|$ along the cross-sections indicated by the magenta dashed lines in (b)--(d), respectively. Red curves highlight topological edge states, confirming consistency with the phase boundaries.}
\label{figS6}
\end{center}
\end{figure}

Supplemental Fig.~\ref{figS6}(a) reproduces the phase diagram of Fig.~2(a) in the $(\delta_2, t_R)$ plane. In addition to the four representative points (I--IV) discussed in the main text, we introduce four extra points (V--VIII) along line~1 (see the caption for detailed coordinates) to facilitate the topological invariant test described below. The topological island containing point III is a genuine ``island'' in the sense that it is completely surrounded by trivial regions. The region containing point II, by contrast, is a ``peninsula'', yet it exhibits the same multi-branch spectral features as an island and is therefore also referred to as an island (see Supplemental Fig.~\ref{figS3}). Supplemental Figs.~\ref{figS6}(b)--\ref{figS6}(d) show additional phase diagrams in different parameter slices, all taken through point III. In each slice, topological islands are clearly visible, though their shapes differ. The cyan dashed curves mark the boundary of the simple inequality $p_2^2 \geqslant p_1^2$, which defines a large simply connected region. The actual topological phase (yellow), however, is a non-trivial subset of this region, carved into islands and peninsulas by the momentum-dependent identity term. Supplemental Figs.~\ref{figS6}(e)--\ref{figS6}(g) display the corresponding energy bands $|E|$ along the cross-sections through point III, with edge states highlighted by red curves, confirming consistency with the phase boundaries.

To explicitly verify the topological nature of the edge states, we compute the topological invariant recently proposed by Zhong \textit{et al.}~\cite{Zhong25} for points I--VIII in Supplemental Fig.~\ref{figS6}(a). This invariant is constructed from the eigenvector ratio $M(\beta)$ defined in Eq.~(5) of Ref.~\cite{Zhong25}, which coincides with the spinor ratio $\phi(\beta)$ given by Eq.~(\ref{S7}) or (\ref{S25}) in our model. The key quantities are the two bulk eigenvector degeneracy points $M_{\mathrm{deg},j}$ ($j=1,2$), which follow readily from Eq. (\ref{S67}) as
\begin{align}
M_{\mathrm{deg},1}&=\frac{-it_Lt_R\sin(\alpha_L-\alpha_R)+\sqrt{p_2^2-p_1^2}}
{\delta_2\bigl(t_Le^{i\alpha_L}-t_Re^{i\alpha_R}\bigr)},~~~~
M_{\mathrm{deg},2}=\frac{-it_Lt_R\sin(\alpha_L-\alpha_R)-\sqrt{p_2^2-p_1^2}}
{\delta_2\bigl(t_Le^{i\alpha_L}-t_Re^{i\alpha_R}\bigr)}, \tag{S69}\label{S69}
\end{align}
and the four branch points $M_{\mathrm{branch}}$ of the Riemann surface, which are obtained from the discriminant of $g(M,z)=0$ (see Ref.~\cite{Zhong25} for details). For our model parameters, these branch points are the roots of the quartic equation:
\begin{align}
M^4 + \frac{8i\delta_2(t_L\sin\alpha_L+t_R\sin\alpha_R)}{\delta_1^2-4\delta_2^2} M(M^2-1)
+ \left[\frac{16 t_L t_R \sin\alpha_L \sin\alpha_R}{\delta_1^2-4\delta_2^2} - 2\right] M^2 + 1 = 0. \tag{S70}\label{S70}
\end{align}
Equation (\ref{S70}) generally yields four $M_{\mathrm{branch}}$ points. Since their contributions to the winding number are identical when taken modulo 2, it suffices to consider any one of them.

The invariant for each degeneracy point is then defined as
\begin{align}
W_j = \left(1 + \frac{1}{2\pi i} \oint_{M(\mathcal{C}_{\mathrm{GBZ}})} dM \frac{d}{dM} \ln\frac{M - M_{\mathrm{deg},j}}{M - M_{\mathrm{branch}}}\right) \mod 2, \tag{S71}\label{S71}
\end{align}
where the integration contour $M(\mathcal{C}_{\mathrm{GBZ}})$ is the image of the GBZ on the complex $M$-plane. Equivalently, $W_j$ can be computed by evaluating the winding numbers of $M_{\mathrm{deg},j}$ and $M_{\mathrm{branch}}$ with respect to this contour:
\begin{align}
W_j = \left(1 + W_{\mathrm{deg},j} - W_{\mathrm{branch}}\right) \mod 2, \tag{S72}\label{S72}
\end{align}
where $W_{\mathrm{deg},j}$ and $W_{\mathrm{branch}}$ count how many times the contour encloses the respective point. The total invariant is $W = W_1 + W_2$, and it predicts the number of topological edge states.

The results of this calculation are summarized in Supplemental Fig.~\ref{figS7}. For points inside the topological regimes (II, III, IV, VII), we obtain $W = 2$, correctly predicting the two degenerate topological edge states [panels (b), (c), (d), (g)]. For point VIII inside the central trivial regime, we obtain $W = 0$, also consistent with our analytical results [panel (h)]. For point VI, which lies on the phase boundary, both $W_1$ and $W_2$ are undefined because the corresponding $M_{\mathrm{deg},1}$ and $M_{\mathrm{deg},2}$ lie exactly on the integration contour [panel (f)]; this is expected at a topological phase transition.

However, for points I and V, which lie in the trivial region away from any phase boundary, we find that \(M_{\mathrm{deg},2}\) lies on a self-intersection line of the \(M(\mathcal{C}_{\mathrm{GBZ}})\) curve, rendering \(W_2\) undefined [panels (a) and (e)]. Consequently, the invariant of Ref.~\cite{Zhong25} in its current form does not provide a definite value for these points. While setting \(W = W_1\) when \(W_2\) is undefined would coincidentally yield \(W=0\) at points I and V, this does not follow from the established prescription. Indeed, when the invariant is well-defined, it correctly yields \(W=0\) for a trivial point, as shown in panel (h). The self-intersecting, multi-loop, or fragmented structures of the GBZ induced by the identity term \(d_0(\beta)\mathds{1}\) therefore present a scenario not covered in the examples studied in Ref.~\cite{Zhong25}. Whether the invariant could be further extended to cover such cases remains to be explored. In contrast, our analytical phase boundary Eq.~(10) correctly predicts the absence of edge states in the entire trivial region, including points I and V.

Thus, our exactly solvable model serves as a rigorous benchmark for testing the range of applicability of symmetry-free topological invariants. Where the invariant of Ref.~\cite{Zhong25} is well-defined, it agrees perfectly with our analytical phase boundary; where it becomes undefined, our exact solution continues to provide a reliable criterion. This agreement, wherever a direct comparison is possible, provides strong evidence for the topological origin of the edge states reported in this work.

\begin{figure}[h!]
\begin{center}
\includegraphics[width=17cm]{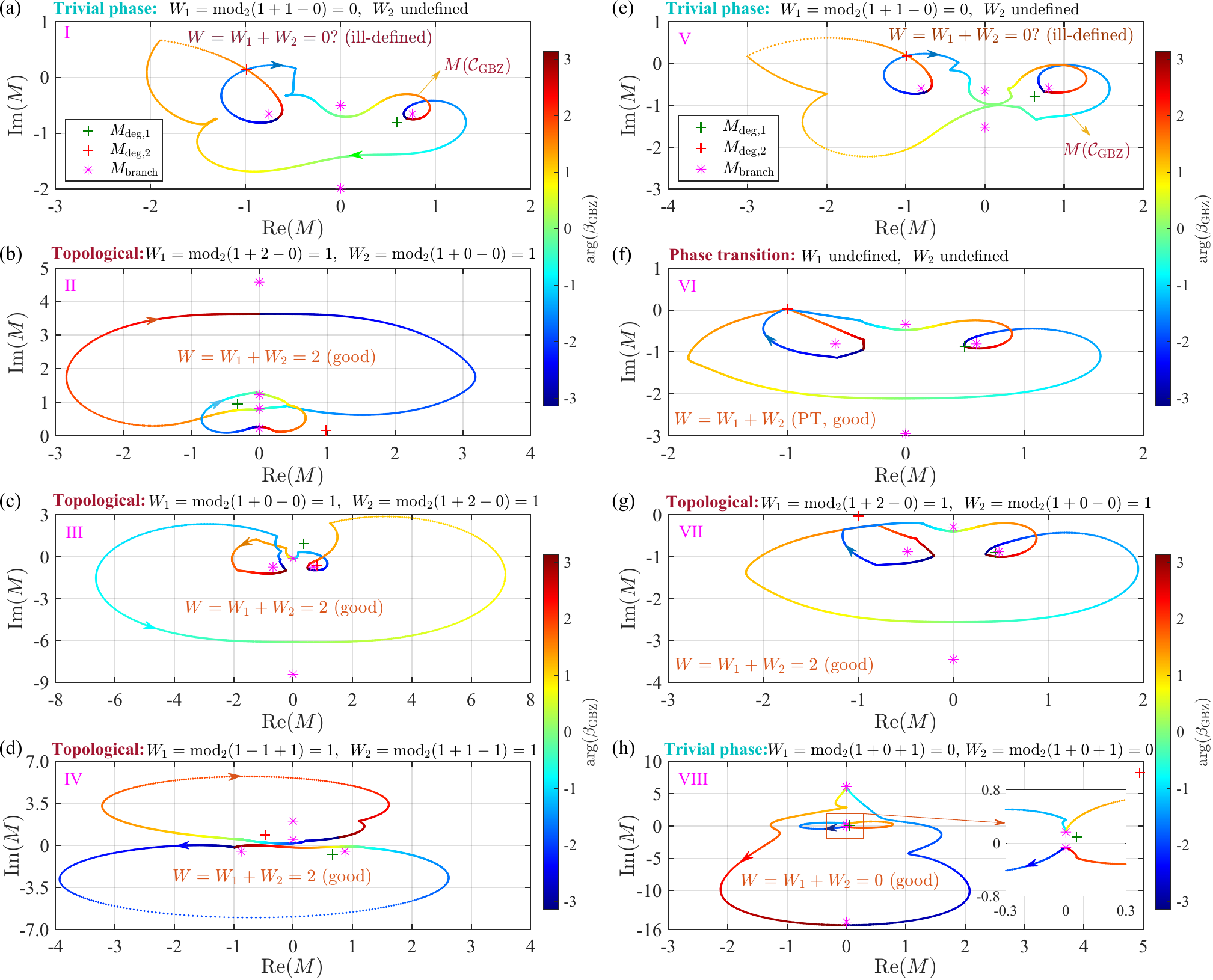}
\caption{\hbadness=10000 \vbadness=10000
  \setlength{\leftskip}{0pt}%
  \setlength{\rightskip}{0pt}%
  \setlength{\parfillskip}{0pt plus 1fil} Topological invariant test at representative points. (a)--(h) correspond to points I--VIII marked in Supplemental Fig.~\ref{figS6}(a), respectively. In each panel, the solid curves represent the closed contour $M(\mathcal{C}_{\mathrm{GBZ}})$, the image of the GBZ; the green and red markers denote the two $M_{\mathrm{deg},j}$ points ($j=1,2$), and the magenta asterisks denote the four $M_{\mathrm{branch}}$ points. The computed winding numbers $W_1 = (1 + W_{\mathrm{deg},1} - W_{\mathrm{branch}}) \mod 2$, $W_2 = (1 + W_{\mathrm{deg},2} - W_{\mathrm{branch}}) \mod 2$, and the total $W = W_1 + W_2$ (where defined) are indicated. Panels (b), (c), (d), (g) yield $W=2$, correctly predicting the topological edge states; panel (h) yields $W=0$, consistent with the trivial phase; panel (f) lies on the phase boundary, where both $W_1$ and $W_2$ are undefined because $M_{\mathrm{deg},1}$ and $M_{\mathrm{deg},2}$ lie exactly on the contour; panels (a) and (e) have $W_2$ undefined because $M_{\mathrm{deg},2}$ lies on a self-intersection line of $M(\mathcal{C}_{\mathrm{GBZ}})$, rendering $W$ also undefined for some trivial phases.}
\label{figS7}
\end{center}
\end{figure}

\section{Robustness of topological edge states against generic disorders}
\label{Sec6}

In the main text (Fig.~4), we have demonstrated that the topological edge states are robust against a local perturbation that modifies the hopping phase at a single unit cell. To further assess their stability under more generic and spatially uncorrelated perturbations, we now test the effect of random on-site potentials and random hopping amplitudes.

\begin{figure}[h!]
\begin{center}
\includegraphics[width=17cm]{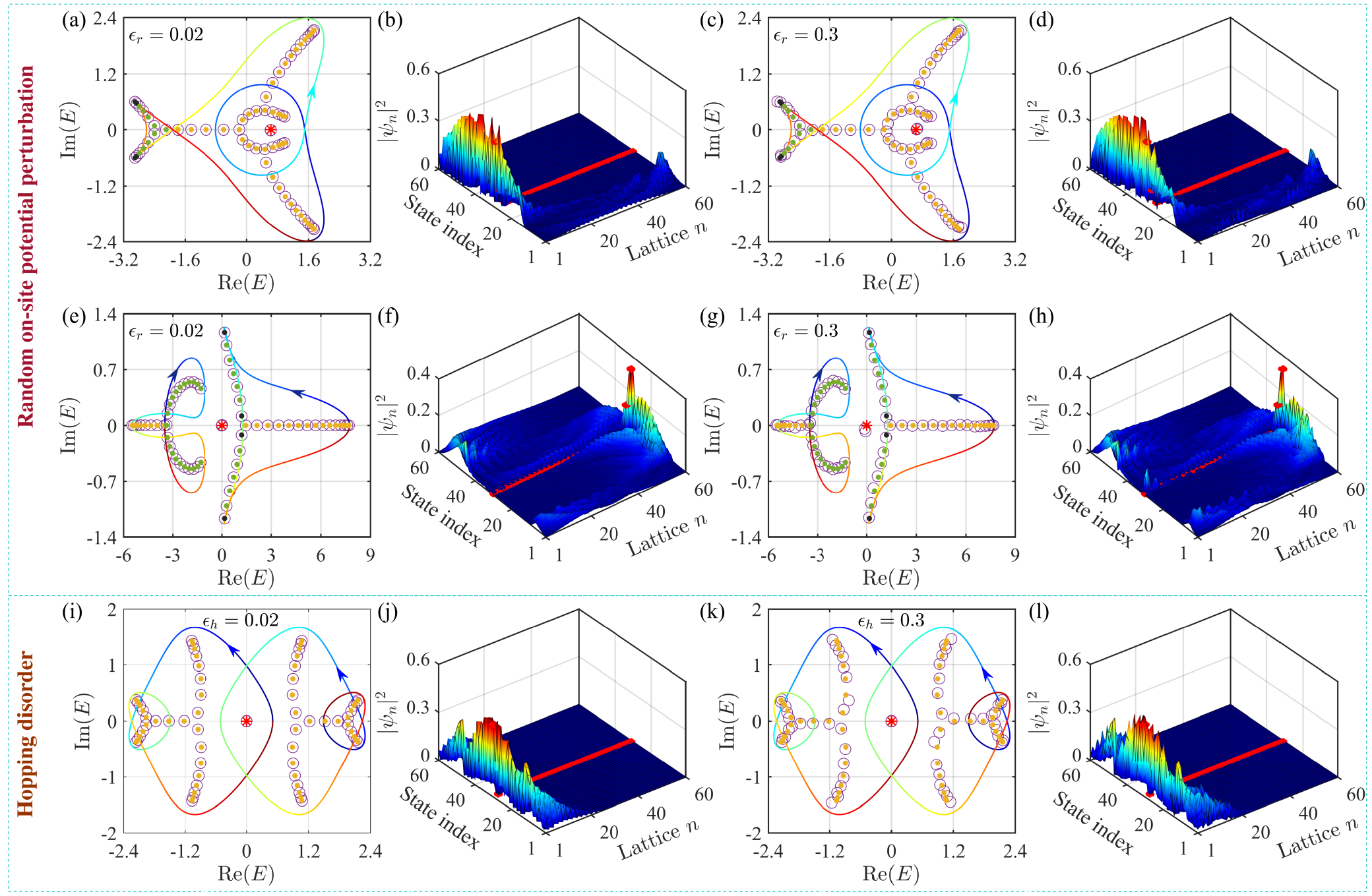}
\caption{\hbadness=10000 \vbadness=10000
  \setlength{\leftskip}{0pt}%
  \setlength{\rightskip}{0pt}%
  \setlength{\parfillskip}{0pt plus 1fil} Robustness of topological edge states against random on-site and hopping disorders. (a)--(d) Finite-energy topological edge states (point III in Fig.~2 of the main text) and (e)--(h) topological zero modes [Fig.~3(e) of the main text] under random on-site disorder: (a,b,e,f) $\epsilon_r = 0.02$; (c,d,g,h) $\epsilon_r = 0.3$. (i)--(l) Topological zero modes [Fig.~3(b) of the main text] under random hopping disorders: (i,j) $\epsilon_h = 0.02$; (k,l) $\epsilon_h = 0.3$. In each case, the left panels show the OBC energy spectrum (green or yellow dots: unperturbed; open purple circles: perturbed), with the analytical edge state highlighted by a red asterisk, and the right panels display the spatial profiles of all eigenstates (red dots for analytical edge states). Even under strong disorder, the topological edge states remain well isolated and localized at the boundary, confirming their topological robustness.}
\label{figS8}
\end{center}
\end{figure}

The random on-site potential is implemented by adding a diagonal matrix $V_{\mathrm{diag}}$ to the real-space Hamiltonian (\ref{S18}), whose entries are independent random variables uniformly distributed in $[-\epsilon_r, \epsilon_r]$ with zero mean, where $\epsilon_r$ measures the disorder strength relative to the bulk gap. We begin with the finite-energy topological edge states from Fig.~2 (point III) of the main text. Under weak on-site disorder ($\epsilon_r = 0.02$), the edge state remains clearly isolated in the OBC spectrum and its wavefunction stays sharply localized at the boundary [Supplemental Figs.~\ref{figS8}(a) and \ref{figS8}(b)]. When the disorder strength is increased to $\epsilon_r = 0.3$, the bulk states shift significantly, yet the edge state persists at essentially the same energy and retains its boundary localization; only a tiny splitting appears, which remains well within the gap and very close to the unperturbed value $E_e$ [Supplemental Figs.~\ref{figS8}(c) and \ref{figS8}(d)]. We next turn to the topological zero modes from Fig.~3(e) (the $\alpha_R = \alpha_L$ case with strictly zero energy). Under the same random on-site disorder perturbations, the zero mode remains exactly pinned at $E = 0$ and its wavefunction stays localized at the right boundary [Supplemental Figs.~\ref{figS8}(e)--\ref{figS8}(h)]. In neither case do the finite-energy edge states or the zero modes merge into the bulk continuum, confirming their exceptional robustness against on-site disorders.

We also test the effect of random hopping disorder on the topological zero modes of Fig.~3(b) in the main text. The hopping amplitudes $t_L$ and $t_R$ are independently perturbed as $t_L \to t_L (1 + \eta_{n})$ and $t_R \to t_R (1 + \eta'_{n})$, where each $\eta_{n}$ and $\eta'_{n}$ is an independent random variable drawn from a uniform distribution $[-\epsilon_h, \epsilon_h]$, with $\epsilon_h$ being the disorder strength. Under weak disorder ($\epsilon_h = 0.02$), the zero mode remains at $E = 0$ and the bulk spectrum is only slightly modified [Supplemental Figs.~\ref{figS8}(i) and \ref{figS8}(j)]. For strong disorder ($\epsilon_h = 0.3$), the bulk eigenvalues deviate considerably from the unperturbed analytical results, yet the zero mode remains perfectly pinned at zero energy and localized at the boundary [Supplemental Figs.~\ref{figS8}(k) and \ref{figS8}(l)].

These extensive numerical tests confirm that both the finite-energy edge states and the zero-energy topological modes are robust against generic, spatially uncorrelated disorders. The slight splitting or broadening observed under strong disorder does not destroy the edge states; they persist as well-defined isolated eigenvalues inside the gap, in complete agreement with the analytical predictions of our exactly solvable model.

\section{Derivation of Exact GBZ in the Thermodynamic Limit} \label{Sec7}

In this section, we derive the exact GBZ in the thermodynamic limit, which is essential for understanding the non-Hermitian bulk topology. Previous studies~\cite{Yok19,Yang20} have shown that, in the continuum limit, the two middle solutions among the four allowed $\beta$ values satisfy an equal modulus condition (say, $|\beta_2|=|\beta_3|$ here), and together constitute the GBZ of the bulk modes. This GBZ forms a closed loop encircling the origin in the complex plane. This criterion, first proposed in Ref.~\cite{Yok19}, provides a direct way to determine the GBZ numerically. To illustrate the procedure, we take the Hamiltonian~(\ref{S18}) as an example. One first chooses a sufficiently large lattice size $N$ and solves Hamiltonian~(\ref{S18}) to obtain all eigenvalues $E$. For each eigenvalue, substituting $E$ into Eq.~(\ref{S5}) yields four corresponding roots $\beta_i$, which are then sorted according to their moduli. The middle two roots, $\beta_2$ and $\beta_3$, are subsequently selected. By plotting all $\beta_2$ and $\beta_3$ values in the complex plane, one obtains the numerical GBZ. Naturally, increasing the lattice size improves the accuracy of the resulting GBZ, though at higher computational cost.

To obtain the exact GBZ (i.e., in the limit $N \to \infty$), we start from the characteristic equation $\det[\mathcal{H}(\beta)-E]=0$, expanded as
\begin{align}
 \beta^4-\frac{2Et_L\cos\alpha_L + 2\delta_{1}\delta_{2}}{t_L^2-\delta_{2}^2} \beta^3+\frac{E^2-\delta_1^2-2p_2}{t_L^2-\delta_{2}^2}\beta^2-\frac{2Et_R\cos\alpha_R + 2\delta_{1}\delta_{2}}{t_L^2-\delta_{2}^2} \beta+\frac{t_R^2-\delta_{2}^2}{t_L^2-\delta_{2}^2}=0.\tag{S73}\label{S73}
\end{align}
Specifically, because of $|\beta_2|=|\beta_3|$, one can set $\beta_2=z$ and $\beta_3=ze^{i\theta}$, where $\theta\in[0,~2\pi]$, and insert them into Eq. (\ref{S73}), respectively. Then, subtracting the resultant equations to eliminate $E$, one obtains an eighth-degree polynomial in $z$:
\begin{align}
 a_8z^8+a_7z^7+a_6z^6+a_5z^5+a_4z^4+a_3z^3+a_2z^2+a_1z+a_0=0, \tag{S74}\label{S74}
\end{align}
whose coefficients are given by
\begin{align*}
a_0 &= (e^{-i\theta} - 1)^2\left[4t_R^2 \cos^2(\alpha_{R})e^{-i\theta}- (t_R^2 - \delta_{2}^2)(e^{-i\theta} + 1)^2\right](t_R^2 - \delta_{2}^2),\\
a_1 &=4\delta_{1}\delta_{2}(t_R^2 - \delta_{2}^2)(e^{-i\theta} - 1)(e^{-2i\theta} - 1), \\
a_2 &=4t_Lt_R\cos(\alpha_{L})\cos(\alpha_{R})(t_R^2 - \delta_{2}^2)(e^{-2i\theta} + 1)(e^{-i\theta} - 1)^2e^{i\theta}+4(e^{-i\theta} - 1)^2\left[(\delta_1^2+2p_2)t_R^2\cos^2(\alpha_{R})-\delta_{1}^2\delta_{2}^2\right],\\
a_3&=4\delta_{1}\delta_{2}(2i\sin\theta+e^{-2i\theta}-1)
\left[-2t_Lt_R\cos(\alpha_{L})\cos(\alpha_{R})+t_R^2\cos(2\alpha_{R})+\delta_{2}^2\right],\\
a_4&=16\left[\delta_{1}^2\delta_{2}^2-(\delta_1^2+2p_2)t_Lt_R\cos(\alpha_{L})\cos(\alpha_{R})
\right](\cos\theta-1)-8(t_L^2 - \delta_{2}^2)(t_R^2 - \delta_{2}^2)\sin^2{\theta}\\
&~~~~-8\left[t_L^2 \cos^2(\alpha_{L})(t_R^2 - \delta_{2}^2)+t_R^2 \cos^2(\alpha_{R})(t_L^2 - \delta_{2}^2)\right](2\cos\theta+1)(\cos\theta-1),\\
a_5&=4\delta_{1}\delta_{2}(2i\sin\theta-e^{2i\theta}+1)\left[2t_Lt_R\cos(\alpha_{L})\cos(\alpha_{R})
-t_L^2\cos(2\alpha_{L})-\delta_{2}^2\right],\\
a_6 &=4t_Lt_R\cos(\alpha_{L})\cos(\alpha_{R})(t_L^2 - \delta_{2}^2)(e^{2i\theta} +1)(e^{i\theta} - 1)^2e^{-i\theta}+4(e^{i\theta} - 1)^2\left[(\delta_1^2+2p_2)t_L^2\cos^2(\alpha_{L})-\delta_{1}^2\delta_{2}^2\right],\\
a_7 &=4\delta_{1}\delta_{2}(t_L^2 - \delta_{2}^2)(e^{i\theta} - 1)(e^{2i\theta} - 1), \\
a_8 &= (e^{i\theta} - 1)^2\left[4t_L^2 \cos^2(\alpha_{L})e^{i\theta}- (t_L^2 - \delta_{2}^2)(e^{i\theta} + 1)^2\right](t_L^2 - \delta_{2}^2).
\tag{S75}\label{S75}
\end{align*}

The exact GBZ in the thermodynamic limit is obtained as follows. For each \(\theta\in[0,2\pi]\), we solve the eighth-degree polynomial Eq. (\ref{S74}) to obtain eight candidate \(z\) values. For each candidate \(z\), we substitute it into the characteristic Eq. (\ref{S5}) to compute the corresponding energy \(E\). Substituting this \(E\) back into Eq. (\ref{S5}) yields four roots \(\beta_{j}\) (\(j=1,\cdots,4\)), which are sorted by their moduli as \(|\beta_1|\leqslant|\beta_2|\leqslant|\beta_3|\leqslant|\beta_4|\). The candidate \(z\) constitutes a valid GBZ point if and only if it coincides with either \(\beta_2\) or \(\beta_3\); otherwise, it is discarded. By sweeping \(\theta\) and collecting all valid \(z\) points (and their corresponding energies \(E\)), we obtain the exact GBZ and the OBC bulk spectrum. This analytical construction corresponds to the thermodynamic limit (\(N\to\infty\)) of the numerical GBZ obtained from finite-size matrix diagonalization.

As an example, we demonstrate in Supplemental Fig. \ref{figS9} the comparison between the numerical GBZ (finite $N=120$) and the exact GBZ derived above for two representative parameter sets [corresponding to Figs.~2(g) and 2(m) in the main text].  The excellent agreement validates our analytical derivation. Moreover, the figure illustrates how the synergy between the identity term $d_0(\beta)$ and the SO couplings ($\delta_{1,2}$) deforms the GBZ into complex shapes, including separated loops or twisted contours, highlighting the nontrivial impact of the momentum-dependent identity term on the non-Hermitian band topology.

\begin{figure}[t!]
\begin{center}
\includegraphics[width=17cm]{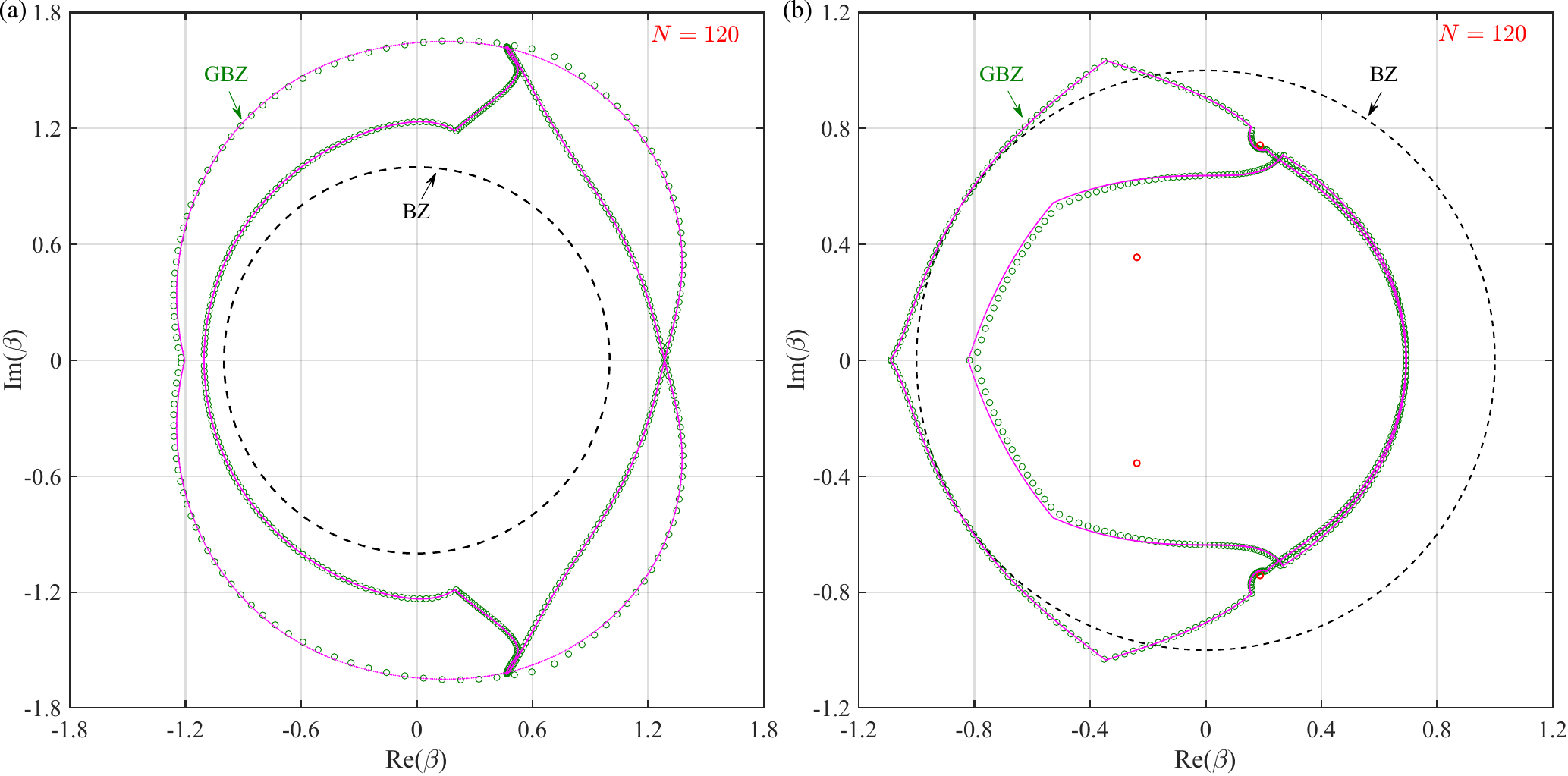}
\caption{\hbadness=10000 \vbadness=10000
  \setlength{\leftskip}{0pt}%
  \setlength{\rightskip}{0pt}%
  \setlength{\parfillskip}{0pt plus 1fil} Comparison between the numerically obtained GBZ (green circles, $N=120$) and the exact GBZ (magenta curves) derived in the thermodynamic limit. (a) and (b) correspond to the parameter sets of panels (g) and (m) in Fig.~2 of the main text, respectively. The close agreement confirms the accuracy of the analytical GBZ solution. }\label{figS9}
\end{center}
\end{figure}

\section{The Most General Two-Band Model and Exact Solutions} \label{Sec8}

In this section, we extend our approach to a completely generic setting that retains no conventional symmetry. By introducing additional degrees of freedom in the on-site and spin-flip terms, we arrive at the most general model that breaks every usual protecting symmetry (e.g., chiral, particle-hole, and even the weaker $\pi$-translation symmetry), while still preserving lattice translation invariance. This generalized model, which is equivalent to the one studied in Ref. \cite{Zhong25}, falls entirely within the scope of our analytical framework, thereby demonstrating the universality of our method.

\begin{figure}[b!]
\begin{center}
\includegraphics[width=17cm]{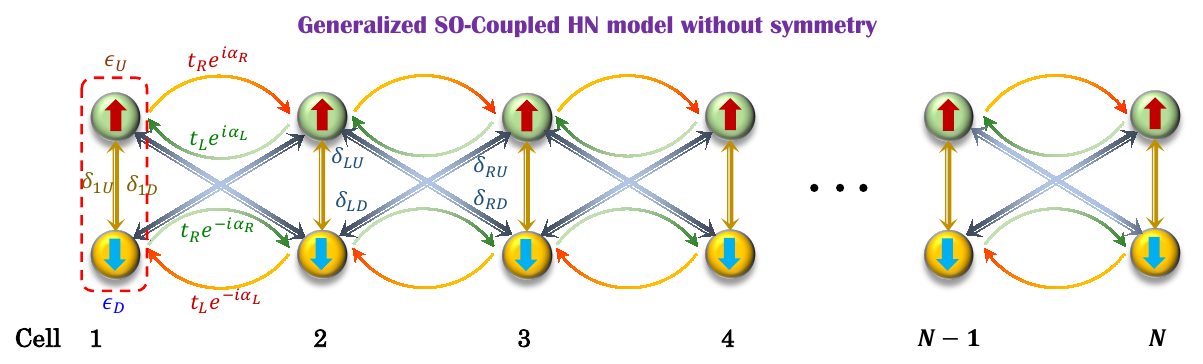}
\caption{\hbadness=10000 \vbadness=10000
  \setlength{\leftskip}{0pt}%
  \setlength{\rightskip}{0pt}%
  \setlength{\parfillskip}{0pt plus 1fil} Schematic of the most general non-Hermitian SO-coupled HN chain without any protecting symmetry. The model is defined by the twelve free parameters shown: spin-dependent on-site potentials $\epsilon$, non-reciprocal intra-cell ($\delta_1$) and inter-cell ($\delta_2,\delta^\prime_2$) SO couplings, as well as the spin-conserving hoppings $t_{L,R}e^{i\alpha_{L,R}\sigma_z}$, as detailed in the text. }\label{figS10}
\end{center}
\end{figure}

As sketched in Supplemental Fig. \ref{figS10}, the Hamiltonian under OBC can be written as
\begin{align}
	\begin{split}
		\hat{H}=\sum_{n=1}^{N}\left[t_{L}e^{i\alpha_L\sigma_z}\hat{c}_{n}^{\dagger}\hat{c}_{n+1}
		+t_{R}e^{i\alpha_R\sigma_z}\hat{c}_{n+1}^{\dagger}\hat{c}_{n}+\hat{c}_{n}^{\dagger}\epsilon\hat{c}_{n}
		+\hat{c}_{n}^{\dagger}(\delta_1\cdot \sigma_x)\hat{c}_{n}+\hat{c}_{n}^{\dagger}(\delta_2\cdot \sigma_x)\hat{c}_{n+1}+\hat{c}_{n+1}^{\dagger}(\delta^\prime_2\cdot \sigma_x)\hat{c}_{n}\right],
	\end{split}\tag{S76}\label{S76}
\end{align}
where $\sigma_x$ and $\sigma_z$ are Pauli matrices, $t_{L,R}e^{i\alpha_{L,R}\sigma_z}$ are spin-conserving hoppings, and
\begin{align}
\epsilon=\left[\begin{array}{cc}\epsilon_{U} &0\\0&\epsilon_{D}
	\end{array}\right],~~~\delta_1=\left[\begin{array}{cc}\delta_{1U} &0\\0&\delta_{1D}
	\end{array}\right], ~~~\delta_2=\left[\begin{array}{cc}\delta_{LU} &0\\0&\delta_{LD}
	\end{array}\right], ~~~\delta^\prime_2=\left[\begin{array}{cc}\delta_{RU} &0\\0&\delta_{RD}
	\end{array}\right], \tag{S77}\label{S77}
\end{align}
denote the nonreciprocal on-site energy offsets and the intra-cell and inter-cell SO coupling strengths, respectively (giving twelve free parameters in total). Here, the subscripts denote the spin direction and hopping direction: $D$ (downward), $U$ (upward), $LU$ (left-upward), $LD$ (left-downward), $RU$ (right-upward), and $RD$ (right-downward). The exact solvability of this model rests on two structural pillars: nearest-neighbor hoppings only between unit cells, and spin-orbit coupling along a single axis (chosen as the $x$-direction, $\sigma_x$). The latter does not sacrifice generality, as one can always fix the spin quantization axis via an SU(2) rotation. Within these design principles, this is the most general symmetry-free two-band system. Then, the real-space $\hat{H}$ for a chain of $N$ unit cells can be expressed explicitly by a $2N\times2N$ matrix:
\begin{align}
	H=	\left[\begin{array}{ccccccccccccc}
		\epsilon_U & \delta_{1U}     &t_{L}e^{i\alpha_L}   &\delta_{LU}           &0           & 0            &0           &   0          &  \cdots     &0     & 0     &0     & 0         \\
		\delta_{1D}       &\epsilon_D & \delta_{LD}       & t_{L}e^{-i\alpha_L}   & 0          & 0            &0           &     0         &\cdots        &0 & 0    & 0     &0        \\
		t_{R}e^{i\alpha_R}    & \delta_{RU}      & \epsilon_U & \delta_{1U}      &t_{L}e^{i\alpha_L}   &\delta_{LU}             &0           &0             & \cdots      &0  & 0   &0  & 0       \\
		\delta_{RD}            &t_{R}e^{-i\alpha_R}   & \delta_{1D}       &\epsilon_D &\delta_{LD}      &t_{L}e^{-i\alpha_L}     &0           &0             & \cdots      &0  & 0    &0  & 0         \\
		0            & 0           &t_{R}e^{i\alpha_R}     &\delta_{RU}       &\epsilon_U  &\delta_{1U}        &t_{L}e^{i\alpha_L}  &\delta_{LU}          & \cdots       &0  & 0  &0  & 0        \\
		0            & 0           &\delta_{RD}             &t_{R}e^{-i\alpha_R}   &\delta_{1D}      & \epsilon_D & \delta_{LD}     &t_{L}e^{-i\alpha_L}      & \cdots       &0 & 0    &0 & 0          \\
		0            & 0           &0             &0            &t_{R}e^{i\alpha_R}  &\delta_{RU}        &\epsilon_U & \delta_{1U}        &\cdots       &0    &0  &0    &0        \\
		0            & 0           & 0            & 0       & \delta_{RD}      &t_{R}e^{-i\alpha_R}           &\delta_{1D}     &\epsilon_D      &\cdots     &0      &0     & 0    & 0        \\
\vdots       &\vdots       &\vdots        &\vdots       &\vdots      &\vdots        &\vdots      &   \vdots       &\ddots     & \vdots       & \vdots & \vdots       & \vdots      \\
	0            &0            &0             &0            &0           &0             &0           &0              &\cdots       & \epsilon_U  &\delta_{1U}   & t_{L}e^{i\alpha_L}  &\delta_{LU}  \\
		0            & 0           & 0            & 0          & 0           & 0            &0           &0              & \cdots        &\delta_{1D}        &\epsilon_D  & \delta_{LD}    &t_{L}e^{-i\alpha_L}   \\
		0            &0            &0             &0            &0           &0             &0           &0              &\cdots       & t_{R}e^{i\alpha_R}  &\delta_{RU}   & \epsilon_U  &\delta_{1U}  \\
		0            & 0           & 0            & 0          & 0           & 0            &0           &0             &  \cdots     &\delta_{RD}        &t_{R}e^{-i\alpha_R}  & \delta_{1D}   &\epsilon_D  \\
	\end{array}\right]_{2N\times2N}. \tag{S78}\label{S78}
\end{align}

In momentum space ($\beta=e^{ik}$), the generalized non-Bloch Hamiltonian reads
\begin{align}
\mathcal{H}(\beta)=d_0(\beta)\mathds{1}+\boldsymbol{\mathrm{d}}(\beta)\cdot \boldsymbol{\sigma}=\left[\begin{array}{cc}G_{\uparrow}(\beta) &\Delta_1(\beta)\\\Delta_2(\beta)&G_{\downarrow}(\beta)
	\end{array}\right], \tag{S79}\label{S79}
\end{align}
where the scalar coefficient
\begin{align}
d_0(\beta)=\frac{G_{\uparrow}(\beta)+G_{\downarrow}(\beta)}{2}=\frac{\epsilon_U+\epsilon_D}{2}+t_L\cos(\alpha_L)\beta
+\frac{t_R\cos(\alpha_R)}{\beta}, \tag{S80}\label{S80}
\end{align}
acts as a momentum-dependent diagonal shift, and the vector part $\boldsymbol{\mathrm{d}}(\beta)=(d_x, d_y, d_z)$ is given by
\begin{align}
\boldsymbol{\mathrm{d}}(\beta)=\left(\frac{\Delta_1(\beta)+\Delta_2(\beta)}{2}, ~\frac{i[\Delta_1(\beta)-\Delta_2(\beta)]}{2},
~\frac{G_{\uparrow}(\beta)-G_{\downarrow}(\beta)}{2}\right),\tag{S81}\label{S81}
\end{align}
with the spin-conserving and spin-flip components explicitly written as
\begin{align}
G_{\uparrow}(\beta) =\epsilon_U+t_L e^{i\alpha_L}\beta+\frac{t_Re^{i\alpha_R}}{\beta},~~~~G_{\downarrow}(\beta) = \epsilon_D+t_L e^{-i\alpha_L}\beta+\frac{t_Re^{-i\alpha_R}}{\beta}, \tag{S82}\label{S82}
\end{align}
\begin{align}
\Delta_1(\beta)=\delta_{1U}
+\delta_{LU}\beta+\frac{\delta_{RU}}{\beta},~~~~\Delta_2(\beta)=\delta_{1D}
+\delta_{LD}\beta+\frac{\delta_{RD}}{\beta}.\tag{S83}\label{S83}
\end{align}
Because the vector $\boldsymbol{\mathrm{d}}(\beta)$ does not lie in a fixed plane and, moreover, the identity term $d_0(\beta)\mathds{1}$ is non-vanishing and $\beta$-dependent, the Hamiltonian breaks chiral symmetry completely. One can further verify that it also violates time-reversal symmetry in its standard (Wigner) form. Consequently, the model described by Eqs. (\ref{S76})--(\ref{S83}) possesses no conventional protecting symmetry beyond lattice translation, thereby representing the most generic two-band non-Hermitian setting \cite{Zhong25}.

The absence of conventional symmetries does not hinder the exact solvability of the model. Following the same analytical blueprint established in Secs. \ref{Sec1} and \ref{Sec2}, we can obtain the complete eigensystem $\hat{H}|\psi\rangle=E|\psi\rangle$ under OBC. The eigenvalue $E$ must satisfy the characteristic equation $\mathrm{det}[\mathcal{H}(\beta)-E]=0$:
\begin{align}
\left[G_{\uparrow}(\beta)-E\right]\left[G_{\downarrow}(\beta)-E\right]
=\Delta_1(\beta)\Delta_2(\beta), \tag{S84}\label{S84}
\end{align}
which for a given $E$ yields four complex roots $\beta_i~(i=1,\ldots,4)$.
Adopting the ansatz consistent with the bulk equations, the eigenvector components can be written as a linear combination:
\begin{align}
\psi^{\uparrow}_{n}=\sum_{i=1}^{4}c_i\phi(\beta_{i})\beta_i^n,~~~~
\psi^{\downarrow}_{n}=\sum_{i=1}^{4}c_i\beta_i^n, \tag{S85}\label{S85}
\end{align}
where the spinor ratio $\phi(\beta_{i})$ and the coefficients $c_i$ are defined by
\begin{align}
\phi(\beta_{i})=\frac{\Delta_1(\beta_i)}
{E-G_{\uparrow}(\beta_i)}=\frac{E-G_{\downarrow}(\beta_i)}{\Delta_2(\beta_i)},
~~~c_{i}=\frac{1}{2}\sum_{j,k,l=1}^{4}\epsilon_{ijkl}\beta_j^{N+1}[\phi(\beta_k)-\phi(\beta_l)]. \tag{S86}\label{S86}
\end{align}
The spectral equation implementing the OBC retains the compact polynomial form
\begin{align}
p_1\lambda_{N+1}^{[1]}+p_2\lambda_{N+1}^{[2]}=0, \tag{S87}\label{S87}
\end{align}
with the generalized parameters
\begin{align}
p_1=\sqrt{(t_R^2-\delta_{RD}\delta_{RU})(t_L^2-\delta_{LD}\delta_{LU})},~~~~
p_2=\frac{\delta_{RD}\delta_{LU}+
\delta_{RU}\delta_{LD}}{2}-t_Lt_R\cos(\alpha_L-\alpha_R). \tag{S88}\label{S88}
\end{align}
The polynomials $\lambda_{N+1}^{[i]}$ ($i=1,2$) are generated by the same three-term matrix recurrence as in Eq. (\ref{S55}), namely,
\begin{align}
G_{n+1}^{[i]}=Q G_{n}^{[i]}-G_{n-1}^{[i]}, ~~(i=1,2),  \tag{S89}\label{S89}
\end{align}
where the coefficient matrix $Q$, defined in Eq. (\ref{S56}), depends on the energy through the three scalar quantities $A$, $B$, and $C$. For the present fully generic model these coefficients become
\begin{align}
A=\frac{(E-\epsilon_D)(E-\epsilon_U)-\delta_{1U}\delta_{1D}-2p_2}{2p_1},  \tag{S90}\label{S90}
\end{align}
\begin{align}
B=\frac{(\delta_{1D}\delta_{RU}+\delta_{1U}\delta_{RD}
+t_R\omega_R)(\delta_{1D}\delta_{LU}+\delta_{1U}\delta_{LD}
+t_L\omega_L)}{4p_1^2}-1, \tag{S91}\label{S91}
\end{align}
\begin{align}
C=\frac{(\kappa_2-1)(E-\epsilon_D)(E-\epsilon_U)}{2p_1}
+\frac{\kappa_1[E(\epsilon_D+\epsilon_U)-\epsilon_D\epsilon_U]}{4p_1}
+\frac{8p_2+\kappa_0+2\eta_1-\eta_2}{8p_1}, \tag{S92}\label{S92}
\end{align}
with the auxiliary functions
\begin{align}
\omega_R=(E-\epsilon_D)e^{i\alpha_R}
+(E-\epsilon_U)e^{-i\alpha_R}, ~~~~~
\omega_L=(E-\epsilon_D)e^{i\alpha_L}
+(E-\epsilon_U)e^{-i\alpha_L}, \tag{S93}\label{S93}
\end{align}
\begin{align}
\varpi_R=\epsilon_D(2E-\epsilon_D)e^{2i\alpha_R}
+\epsilon_U(2E-\epsilon_U)e^{-2i\alpha_R}, ~~~~~
\varpi_L=\epsilon_D(2E-\epsilon_D)e^{2i\alpha_L}
+\epsilon_U(2E-\epsilon_U)e^{-2i\alpha_L}, \tag{S94}\label{S94}
\end{align}
\begin{align}
\eta_1=\frac{t_L\omega_L(\delta_{1U}\delta_{LD}
+\delta_{1D}\delta_{LU})}{t_L^2-\delta_{LD}\delta_{LU}}+ \frac{t_R\omega_R(\delta_{1U}\delta_{RD}+\delta_{1D}\delta_{RU})}{t_R^2-\delta_{RD}\delta_{RU}}, ~~~~~
\eta_2=\frac{t_L^2\varpi_L}{t_L^2-\delta_{LD}\delta_{LU}}+ \frac{t_R^2\varpi_R}{t_R^2-\delta_{RD}\delta_{RU}}, \tag{S95}\label{S95}
\end{align}
and the parameter combinations
\begin{align}
\kappa_0=\frac{4t_L^2\delta_{1U}\delta_{1D}}{t_L^2-\delta_{LD}\delta_{LU}}+ \frac{(\delta_{1U}\delta_{LD}-\delta_{1D}
\delta_{LU})^2}{t_L^2-\delta_{LD}\delta_{LU}}+ \frac{(\delta_{1U}\delta_{RD}+\delta_{1D}\delta_{RU})^2}{t_R^2-\delta_{RD}\delta_{RU}}, \tag{S96}\label{S96}
\end{align}
\begin{align}
\kappa_1=\frac{t_L^2\cos(2\alpha_L)}{t_L^2-\delta_{LD}\delta_{LU}}+ \frac{t_R^2\cos(2\alpha_R)}{t_R^2-\delta_{RD}\delta_{RU}}, ~~~~
\kappa_2=\frac{t_L^2\cos^2\alpha_L}{t_L^2-\delta_{LD}\delta_{LU}}+ \frac{t_R^2\cos^2\alpha_R}{t_R^2-\delta_{RD}\delta_{RU}}. \tag{S97}\label{S97}
\end{align}
Thus, solving the polynomial equation (\ref{S87}) yields the complete OBC spectrum $\{E\}$. For each eigenvalue, the corresponding generalized momentum roots $\beta_i$ are obtained from Eq. (\ref{S84}), and the full eigenstate $|\psi\rangle$ is then constructed via Eqs. (\ref{S85}) with the coefficients $c_i$ given above. This explicit solution demonstrates that our analytical framework seamlessly encompasses the most general non-Hermitian two-band model without any protecting symmetry.

The same analytical structure also allows us to identify topological edge states in this generic setting. Following the asymptotic procedure outlined in Sec. \ref{Sec3}, the edge-state energy condition is still encoded in the relation $X_e + Y + Z = A$, where $A$ is given by Eq. (\ref{S90}), $X_e = -p_2/p_1$, and $Y,Z$ are determined from Eqs. (\ref{S61}) and (\ref{S62}). In contrast to the simpler cases discussed earlier, the topological edge states now generally separate into two distinct (non-degenerate) eigenvalues in the complex energy plane, reflecting the complete breaking of symmetry.

For illustration, we provide in Supplemental Fig. \ref{figS11} a direct numerical verification of the exact solution for the fully generic, symmetry-free model. Using a representative set of parameters that breaks all conventional symmetries, we obtain perfect agreement between the analytical and numerical results, both for the OBC energy spectrum [see Supplemental Fig. \ref{figS11}(a)] and the wavefunctions [see Supplemental Fig. \ref{figS11}(b)]. Notably, two non-degenerate topological edge states emerge within the spectral gap, exactly as predicted by the analytical condition discussed above. This point-for-point agreement not only validates our exact solution, but also demonstrates that the topological phenomena, including symmetry-agnostic edge modes, remain fully tractable within our analytical framework even when all conventional symmetries are absent.

\begin{figure}[h!]
\begin{center}
\includegraphics[width=17cm]{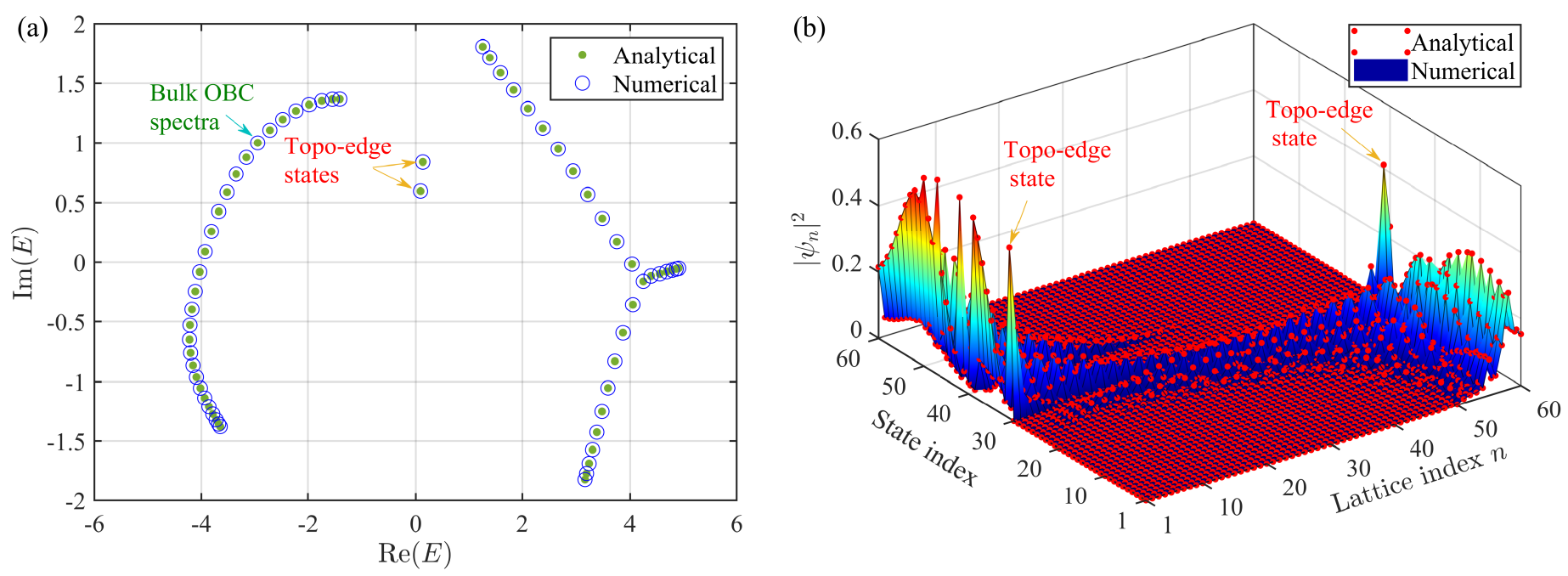}
\caption{\hbadness=10000 \vbadness=10000
  \setlength{\leftskip}{0pt}%
  \setlength{\rightskip}{0pt}%
  \setlength{\parfillskip}{0pt plus 1fil} Numerical verification of the exact solutions for the fully generic model ($N=30$). The parameters used are $t_L =2$, $t_R=-1$, $\alpha_L=\pi/3$, $\alpha_R = \pi/4$, $\epsilon_D = -1$, $\epsilon_U = 1$, $\delta_{1U} = 3/2$, $\delta_{1D} = 2/3$, $\delta_{LD} = 5/3$, $\delta_{LU} = 3$, $\delta_{RD} = 4/3$, and $\delta_{RU} = 5/2$. (a) Analytical OBC spectrum from Eq. (\ref{S87}) (green dots) coincides perfectly with the numerical diagonalization results (open circles), despite the complete absence of symmetry. (b) Analytical eigenstates constructed via Eqs. (\ref{S85}) (red dots) match the numerical wavefunctions (solid lines) point-for-point.} \label{figS11}
\end{center}
\end{figure}